\documentclass[referee,sn-basic]{sn-jnl}

\jyear{2021}%

\theoremstyle{thmstyleone}%
\theoremstyle{thmstyletwo}%

\theoremstyle{thmstylethree}%
\usepackage{subcaption}

\usepackage{statmath}

\usepackage[section]{placeins}

\begin{document}

\title[EmulART: Emulating Radiative Transfer]{EmulART: Emulating Radiative Transfer - A pilot study on autoencoder based dimensionality reduction for radiative transfer models\\ \small{(accepted in Neural Computing and Applications)}}


\author*[1]{\fnm{Jo\~ao} \sur{Rino-Silvestre}}\email{joao.silvestre@tecnico.ulisboa.pt}

\author[1]{\fnm{Santiago} \sur{Gonz\'alez-Gait\'an}}\email{gongsale@gmail.com}
\equalcont{These authors contributed equally to this work.}

\author[2,3]{\fnm{Marko} \sur{Stalevski}}\email{mstalevski@aob.rs}
\equalcont{These authors contributed equally to this work.}

\author[2]{\fnm{Majda} \sur{Smole}}\email{msmole@aob.rs}

\author[4]{\fnm{Pedro} \sur{Guilherme-Garcia}}\email{pguilhermegarcia@gmail.com}

\author[5]{\fnm{Jo\~ao} \sur{Paulo Carvalho}}\email{joao.carvalho@inesc-id.pt}

\author[1]{\fnm{Ana Maria} \sur{Mour\~ao}}\email{amourao@tecnico.ulisboa.pt}

\affil[1]{\orgdiv{CENTRA}, \orgname{Instituto Superior T\'ecnico}, \orgaddress{\street{Av. Rovisco Pais}, \city{Lisboa}, \postcode{1049-001}, \state{} \country{Portugal}}}

\affil[2]{\orgdiv{Astronomical Observatory}, \orgname{} \orgaddress{\street{Volgina 7}, \city{Belgrade}, \postcode{1060}, \state{} \country{Serbia}}}

\affil[3]{\orgdiv{Sterrenkundig Observatorium}, \orgname{Universiteit Gent}, \orgaddress{\street{Krijgslaan 281-S9}, \city{Gent}, \postcode{9000}, \state{} \country{Belgium}}}

\affil[4]{\orgdiv{CENTRA}, \orgname{Faculdade de Ci\^encias da Universidade de Lisboa}, \orgaddress{\street{} \city{Lisboa}, \postcode{1749-016}, \state{} \country{Portugal}}}

\affil[5]{\orgdiv{INESC-ID}, \orgname{Instituto Superior T\'ecnico}, \orgaddress{\street{Rua Alves Redol, 9}, \city{Lisboa}, \postcode{1000-029}, \state{} \country{Portugal}}}

\abstract{Dust is a major component of the interstellar medium. Through scattering, absorption and thermal re-emission, it can profoundly alter astrophysical observations. Models for dust composition and distribution are necessary to better understand and curb their impact on observations. A new approach for serial and computationally inexpensive production of such models is here presented.\par
Traditionally these models are studied with the help of radiative transfer modelling, a critical tool to understand the impact of dust attenuation and reddening on the observed properties of galaxies and active galactic nuclei. Such simulations present, however, an approximately linear computational cost increase with the desired information resolution.\par
Our new efficient model generator proposes a denoising variational autoencoder (or alternatively PCA), for spectral compression, combined with an approximate Bayesian method for spatial inference, to emulate high information radiative transfer models from low information models.\par
For a simple spherical dust shell model with anisotropic illumination, our proposed approach successfully emulates the reference simulation starting from less than 1\% of the information. Our emulations of the model at different viewing angles present median residuals below 15\% across the spectral dimension, and below 48\% across spatial and spectral dimensions. EmulART infers estimates for $\sim$85\% of information missing from the input, all within a total running time of around 20 minutes, estimated to be 6$\times$ faster than the present target high information resolution simulations, and up to 50$\times$ faster when applied to more complicated simulations.}




\maketitle
\section{Introduction}\label{sec:intro}
Cosmic dust is ubiquitous in the Universe, particularly present in the interstellar medium (ISM) \citep{ismdust} and in the line of sight towards astrophysical objects such as supernovae remnants \citep{snedust}, galaxies \citep{galdust} and active galactic nuclei (AGN) \citep[AGN,][]{agndust}. Dust grains absorb and scatter UV/optical radiation and re-emit that energy at infrared wavelengths, and are thus responsible for both attenuation and reddening of light in the line of sight, which also impact distance measurements for cosmology when using ``standard candles" such as supernovae \citep{Betoule}. Moreover, scattering on the dust grains and dichroic absorption in the dusty medium may lead to polarization of the light as it traverses the interstellar medium.\par
The effect these processes - of dust absorption, scattering and emission - have on the light detected from astronomical objects must be accounted when studying their intrinsic properties. Moreover, the nature of the dust can also inform us on physical and chemical processes related to its own history, from formation, variation in composition, growth and destruction in different astrophysical structures such as accretion disks, clouds and galaxies, as well as its interaction with magnetic fields through dust grain alignment. The analysis of the aforementioned interactions requires performing radiative transfer (RT) calculations \citep{RTreview}, allowing us to simulate the light path, from the source to the observer, depending on the physical properties of the emitting source embedded in dust structures of different geometries, sizes and composition. Comparing various RT models, based on different simulated properties, to global and pixel-by-pixel spectral energy distributions (SEDs) obtained in astronomical observations we can infer valuable information on the properties of the light sources, as well as the distribution and properties of the dust \citep{andre,cox,fritz}.\par
Observations of molecular clouds\citep{beech,falga} have shown that dust distributions are often inhomogeneous and complex, consequently the understanding of their intrinsic properties requires 3D radiative transfer calculations. This type of non-local and non-linear problem requires calculations which are computationally very costly \citep{RTreview}, this has prompted the search for alternative non-analytic approximate ways to address them. One of the most successful ways is Monte Carlo Radiative Transfer \citep[MCRT,][]{MCRT1,MCRT2}.\par
Monte Carlo Radiative Transfer methods simulate a large number of test photons, that propagate from their emission by a source through their journeys through the dusty medium. At every stage the characteristics that define their paths are determined by generating numbers from the probability density function most suited for each process they may undergo (absorption, scattering, re-emission, etc). At the end of the simulation, the radiation field is recovered from a statistical analysis of the photon paths. As an ensemble, the distribution of particles provides a good representation of the radiative transfer, as long as a sufficient number of photons is chosen.\par
Stellar Kinematics Including Radiative Transfer \citep[SKIRT,][]{SKIRT1,SKIRT2} is an MCRT suite that offers some built-in source templates, geometries, dust characterizations, spatial grids, and instruments, as well as an interface so that a user can easily describe a physical model. The user can in this way avoid coding the physics that describes both the source (e.g. AGN or galaxy type, observation perspective, emission spectrum) and environment (between the simulated source and observer, such as dust grain type and orientation, dust density distribution, etc) but instead design a model of modular complexity by following a Q\&A prompt (itself adaptable to the user expertise).\par
MCRT simulations suffer from computational limitations, namely the memory requirement scaling with the volume grid density, and the processing time scaling quasi-linearly with the amount of photons simulated \citep{skirt9}. Autoencoders \citep{Rabczuk-ae} together with collocation strategies \citep{Rabczuk} have been applied to solve complex interactions such as the ones described above, our approach differs by attempting to upscale the information density within a simulated data product instead. Considering that the objects and phenomena modeled by MCRT simulations present non-random spatial structures with heavily correlated spectral features, we tackle the computational cost issue through the development of an emulator that can achieve HPN-like MCRT models by exploring and implementing an autoencoder neural network in combination with Integrated Nested Laplace Approximation \citep[INLA,][]{inla}, an approximate method for Bayesian inference of spatial maps modeled with Gaussian Markov random fields, on LPN-like MCRT simulations. The results are then compared against an analogous implementation employing principal component analysis (PCA).\par
Section \ref{sec:meth} provides a brief highlight of the employed methods. Section \ref{sec:impl} describes our pipeline architecture and some of steps that lead to its development. Results and performance evaluation follow in Section \ref{sec:results}. Section \ref{sec:conc} presents our perspective on the significance of the obtained results and provides the steps to follow in order to both improve and generalize them in future developments.\par
All files concerning this work (SKIRT simulations, R scripts, neural network models, emulation products and performance statistics) can be obtained from our repository\footnote{\href{https://github.com/SN-CRISP/EmulART}{https://github.com/SN-CRISP/EmulART}}.\par

\section{Methods}\label{sec:meth}
To reduce the computational cost of SKIRT simulations, without compromising, as much as possible, the quality of the resulting models, an autoencoder, i.e. a dimensionality reduction neural network, is implemented to compress the spectral information within LPN spectroscopic data cubes. Then approximate Bayesian inference is performed with INLA on the spatial information of the compressed feature maps. Lastly, the reconstructed feature maps are decompressed to an HPN emulation.\par

\subsection{SKIRT}\label{sec:SKIRT}
As previously state, SKIRT allows the creation of models by prompting a Q\&A. Through it the user can configure any one- to three-dimensional geometry and combine multiple sources and media components, each with their own spatial distribution and physical properties, by either employing the built-in library or importing components from hydrodynamic simulations. Media types include dust, electrons and neutral hydrogen; the user can configure their own mixture, including optical properties and size distributions or simply choose from the available built-in mixtures. The included instruments allow the "observation" and recording of spectra, broad-band images or spectroscopic data cubes.\par
SKIRT uses Monte Carlo method for tracing photon packets through the spatial domain (both regular and adaptive spatial grids area available, as well as some optimized for 1D or 2D models), these packets are wavelength sampled from the source/medium spectrum (the wavelength grids can be separately configured for both storing the radiation field and for discretizing medium emission spectra). As they progress through the spatial grid cells, these photon packets can experience different physical interactions, such as multiple anisotropic scattering, absorption and (re-)emission by the transfer medium, Doppler shifts due to kinematics of sources and/or media, polarization caused by scattering off dust grains as well as polarized emission, among others.\par
The present application is intended for the combination of both spatial and spectral information. The outputs used here will be spectroscopic data cubes, these are in the flexible image transport system (FITS) format \citep{fits} and are composed of 2D spatial distributions at the desired wavelength bins.\par
The simulated spatial flux densities vary according to the amount of photons/photon packets simulated. The simulation starts by assigning some energy to those photons following the spectral energy distribution (SED) of a given astrophysical source, ensuring that no matter how many photons are simulated the spectral information is preserved. Simulations with a lower photon number (LPN) will consequently display fewer spatial positions with information (non-zero flux) and some of these pixels will have higher flux, some lower, i.e. the SED will have lower signal to noise ratio than in simulations with higher photon number (HPN). SKIRT has already been employed in the study of various galaxies \citep{delooze, viaene, verstocken}, AGN \citep{marko,marko2} and other objects\footnote{\href{https://skirt.ugent.be/root/_publications_gallery.html}{https://skirt.ugent.be/root/_publications_gallery.html}}. \par

\subsection{Dimensionality Reduction}\label{sec:DR}
Dimensionality reduction methods can more familiarly be called as compressors. These are methods that analyse and transform data from a high-dimensional space into a low-dimensional space while attempting to retain as much meaningful properties of the original data as possible. Working in high-dimensional spaces can be undesirable for many reasons; raw data are often sparse as a consequence of the curse of dimensionality\footnote{As the number of parameters increases (rapidly increasing the volume of the parameter-space) the density of the data quickly decreases.}, and analyzing the data usually becomes computationally intractable. Popular dimensionality reduction techniques in astronomy include PCA \citep{UMT, HC5} and Non-Negative Matrix Factorization \citep[NMF,][]{nmf1,nmf2}.\par
Autoencoder networks are an alternative method which has been gaining attention within the astronomy, astrophysics and cosmology community \citep{VAE3,CAE,VAE2,VAE1,DAE2,DAE1}.

\subsubsection{Denoising Variational Autoencoders}\label{sec:dvae}
Autoencoders (AEs) are a type of neural network architecture employed to learn compressed representations of data. In such architectures the input and output layers are equal, and the hidden layers display a funneling in and out scheme in regards to the number of neurons per layer, with the middle layer having the least amount of neurons and the input and output layers having the most. The models built this way can be seen as the coupling of a compressor/encoder and a decompressor/decoder, the first generating a more (ideally) fundamental representation of the data, and the second bringing it back to its initial feature space. After training, the encoder can be coupled at the beginning of other architectures, providing them with more efficient features from which to learn from. Interesting to note that AEs have shown promise as auxiliary tools in citizen science projects\footnote{Projects where non-scientists can participate either by collecting or processing data.}, such as the Radio Galaxy Zoo \citep{CAE}.\par 
Alternative ways of training this kind of network exist, such as:
\begin{itemize}
    \item Having multiple instances of each data point, resulting from the injection of noise or from a set of  transformations. Each instance of the same data point is then matched to the same output with the aim of making the model robust to noise and/or invariant to those transformations.
    \item Having the mid layer composed by two complementary layers of neurons (a mean layer and a standard deviation layer) instead of a single layer. Complemented with an appropriate loss function, the model will learn approximate distributions of values instead of single values, making it more robust and allowing for the decoder to also become a generator of different, yet statistically identical, examples.
\end{itemize}
Such strategies fall under different categories such as denoising autoencoders (DAEs) \citep{MoEVAE} and variational autoencoders (VAEs) \citep{OldVAE}  respectively. In this work we implement both, a denoising variational autoencoder (DVAE).\par
\citep{ImDVAE} showed that the DVAE, by introducing a corruption model, can lead to an encoder that covers a broader class of distributions when compared to a regular VAE (the corruption model may however remove the information necessary for reconstruction). In this context the loss function, $\mathcal{L}_{DVAE}$, to be minimized is given by the weighted sum of the Kullback-Leibler divergence (which influences the encoder) and the reconstruction error (see Eq. \ref{eq:lossDVAE}), similarly to a VAE, with the difference that the encoder now learns to approximate a prior $p(z)$ given corrupted examples and that in this case the reconstruction error can be interpreted as a denoising criterion: \par

\begin{equation}\label{eq:lossDVAE}
    \mathcal{L}_{DVAE} \propto a_1\mathbb{KL}(q(z\mid y')\mid p(z)) + a_2\ln{p(y \mid z)}\,\, ,
\end{equation}

\noindent where $\mathbb{KL}$ refers to the Kullback-Liebler divergence, $y'\sim q(y'\mid y)$ is a sample of the corruption model, $p(z)$ is the prior for the latent feature, $q(z\mid y')$ models the encoder, $p(y \mid z)$ models the decoder/generator and ($a_1$, $a_2$) are weights. For more details the reader is referred to \citep{ImDVAE,tutVAE}.\par  

\subsubsection{Principal Component Analysis}\label{sec:pca}

Principal Component Analysis (PCA) is a method that analyzes the feature space of the training data and creates orthogonal vectors (linear combinations of the initial variables) whose direction indicates the most variability. These new vectors in the transformed data set are called eigenvectors, or principal components, while the eigenvalues represent the coefficients attached to eigenvectors and give the relative amount of variance carried by each principal component.\par
PCA transforms the original space through rotation of its axes and re-scaling of the axes range. The first new PC is aligned with the direction of largest variance in the data. The second PC should also maximize the variance, while being orthogonal to the first, and respectively for the remaining PCs. Mathematically, these directions can be determined through the covariance matrix, as expressed in Eq. \ref{eq:cov}:\par

\begin{equation}\label{eq:cov}
    \Sigma_{ff'} = \sum_{j=1}^{j=N} \frac{(X_f^j - \Bar{X}_f)(X_{f'}^j - \Bar{X}_{f'})}{N} \,\, , 
\end{equation}

\noindent where $X_f$ is the mean of all values at feature $f$ and $N$ is the total number of data points (for more details and a modern review on PCA we suggest \citep{PCA3}). Once $\Sigma_{ff'}$ is diagonalized, the PCs are its eigenvectors, the first PC being the one with the largest associated eigenvalue and so on.\par
PCs are uncorrelated and frequently the information is compressed into the first $K$ components, with $K \ll M$ (where $M$ is the total number of features of the original space).\par
A data point from the original data set can then be reasonably recovered using those $K$ PCs,\par

\begin{equation}\label{eq:proj}
    \hat{\bfX} \approx \Bar{\bfX} + \sum_{m=1}^{m=K} c_m \bfP_m \,\, , 
\end{equation}

\noindent where $\bf\Bar{X}$ represents the mean of all data points, $\bfP_m$ is the m-th PC and $c_m$ is the projection of the data point on $\bfP_m$.\par
Using all the $M$ PCs the reconstruction becomes identical to the original data, but with a new basis that captures a large fraction of the variance in a small number of components $K$, dimensionality reduction is achieved. In this work we used two approaches to determine $K$ (see Section \ref{sec:pipePCA}).\par
Further discussion on the importance of PCA and its applications in astronomy can be found in e.g. \citep{astropca1,astropca2,astropca3}.


\subsection{Spatial Approximate Bayesian Inference}\label{sec:IB}
Bayesian inference (BI) refers to a family of methods of statistical inference where a hypothetical probabilistic model is updated, following Bayes theorem (Eq. \ref{eq:Bayes}), whenever new data is obtained. Bayes theorem allows to calculate a posterior distribution $p(\theta\mid y)$ (the conditional probability of $\theta$ occurring given $y$) by weighting in $p(y\mid \theta)$ (the likelihood of $y$ occurring given $\theta$), $p(\theta)$ (estimation of the probability distribution of $\theta$ before observing $y$, also designated as a prior), and $p(y)$ (the marginal probability of $y$, obtained from integrating $\theta$ out of $p(y\mid \theta)$):\par

\begin{equation}\label{eq:Bayes}
    p(\theta\mid y) = \frac{p(y\mid \theta)}{p(y)} p(\theta)\,\, ,
\end{equation}

\begin{equation}\label{eq:margin}
    p(y) = \int{p(y \mid\theta)p(\theta)\textrm{d}\theta}\,\, .
\end{equation}

This kind of update is of utmost relevance in the dynamical analysis of data streams or in the analysis of correlated data, and has been proposed in astronomy, from the study of variable stars \citep{B3} to 3-D mapping of the Milky Way \citep{B1}.\par 
Approximation techniques \citep{LaplaceApprox,VB,EP} have been developed over the years in order to help curb the very time-consuming process of sampling the whole likelihood $p(y\mid \theta)$. To this end, we here implement the Integrated Nested Laplace Approximation \citep[INLA,][]{inla}.\par 

\subsubsection{Integrated Nested Laplace Approximation}
INLA is an approximate BI method that accounts for spatial correlations between observed data points to recover an assumed Gaussian latent field and, in doing so, it is not only capable of predicting unobserved points of that field but also of correcting noisy observed ones, as well as associating a variance to those inferences.\par
Most techniques for calculating posterior distribution rely on Markov chain Monte Carlo \citep[MCMC,][]{MCMC} methods. In this class of sampling-based numerical methods, the posterior distribution is obtained after many iterations, which is often computationally expensive. INLA provides a novel approach for faster BI. While MCMC methods draw a sample from the joint posterior distribution, the Laplace approximation is a method that approximates posterior distributions of the model parameters to Gaussians, which is computationally more effective. Within the INLA framework, the posterior distribution of the latent Gaussian variables $\pmb{x}$ and hyper-parameters of the model $\pmb{\theta}$ is:

\begin{align}
   p(\pmb{x},\pmb{\theta} \mid \pmb{y}) = \frac{p(\pmb{y} \mid  \pmb{x},  \pmb{\theta})~ p(\pmb{x}, \pmb{\theta})}{p(\pmb{y})} \propto p(\pmb{y} \mid  \pmb{x},  \pmb{\theta})~ p(\pmb{x}, \pmb{\theta}),
    \label{joint_post}
\end{align}

\noindent where $\pmb{y}=(y_{1},...,y_{n})$ represents a set of observations. Each observation is treated with a latent Gaussian effect, with each $x_{i}$ (a Gaussian distribution of mean value $\mu_{i}$ and standard deviation $\sigma_{i}$) corresponding to an observation $y_{i}$, where $i\in[1,...,n]$. The observations are conditionally independent given the latent effect $\pmb{x}$ and the hyper-parameters $\pmb{\theta}$, the model likelihood is then:

\begin{align}
 p(\pmb{y} \mid  \pmb{x},  \pmb{\theta}) = \prod_{i} p(y_{i}  \mid x_{i}, \pmb{\theta})\,\, .
\end{align}

The joint distribution of the latent effects and the hyper-parameters, $p(\pmb{x}, \pmb{\theta})$ can be written as $p(\pmb{x}\mid\pmb{\theta})~p(\pmb{\theta})$, where $p(\pmb{\theta})$ represents the prior distribution of hyper-parameters $\pmb{\theta}$. 
It is assumed that the spatial information can be treated as a discrete sampling of an underlying continuous spatial field, a latent Gaussian Markov Random Field (GMRF), that takes into account the spatial correlations, and whose hyper-parameters are inferred in the process.
For a GMRF, the posterior distribution of the latent effects is:

\begin{equation}
 p(\pmb{x} \mid  \pmb{\theta}) \propto \mid\pmb{Q}(\pmb{\theta})\mid^{1/2} \exp {-\frac{1}{2} \pmb{x}^{T}\pmb{Q}(\pmb{\theta})~\pmb{x}}\,\, ,
\end{equation}

\noindent where $\pmb{Q}(\pmb{\theta})$ represents a precision matrix, or inverse of a covariance matrix, which depends on a vector of hyper-parameters $\pmb{\theta}$. This kernel matrix is what actually treats the spatial correlation between neighboring observations.
Using equation \ref{joint_post}, the joint posterior distribution of the latent effects and hyper-parameters can be written as:

\begin{align}
\begin{aligned}
  p(\pmb{x}, \pmb{\theta} \mid \pmb{y})\propto  p(\pmb{\theta}) \mid\pmb{Q}(\pmb{\theta})\mid^{1/2} \exp {-\frac{1}{2} \pmb{x}^{T}\pmb{Q}(\pmb{\theta})~\pmb{x}} \prod_{i} p(y_{i} \mid x_{i}, \pmb{\theta}) =\\ 
 =p(\pmb{\theta}) \mid\pmb{Q}(\pmb{\theta})\mid^{1/2} \exp{-\frac{1}{2} \pmb{x}^{T}\pmb{Q}(\pmb{\theta})~\pmb{x} + \sum_{i} \log (p(y_{i} \mid x_{i}, \pmb{\theta}))}\,\, .
 \label{post_joint}
 \end{aligned}
\end{align}

Instead of obtaining the exact posterior distribution from equation \ref{post_joint} INLA approximates the posterior marginals of the latent effects and hyper-parameters, its key methodological feature is to use appropriate approximations for the following integrals:

\begin{equation}
 p( x_{i} \mid \pmb{y})=\int p( x_{i} \mid \pmb{\theta}, \pmb{y})p( \pmb{\theta}\mid \pmb{y}) ~ d\pmb{\theta}
\end{equation}

\begin{equation}
 p( \theta_{j} \mid \pmb{y})=\int
 p( \pmb{\theta}\mid \pmb{y}) ~ d\pmb{\theta}_{-j}, 
\end{equation}

\noindent where $\pmb{\theta}_{-j}$ is a vector of hyper-parameters $\pmb{\theta}$  without element $\theta_{j}$.\par
INLA constructs nested approximations:

\begin{equation}
 \tilde{p}( x_{i} \mid \pmb{y})=\int \tilde{p}( x_{i} \mid \pmb{\theta}, \pmb{y})\tilde{p}( \pmb{\theta}\mid \pmb{y}) ~ d\pmb{\theta}\,\, ,
   \label{post_lat}
\end{equation}

\begin{equation}
 \tilde{p}( \theta_{j} \mid \pmb{y})=\int
 \tilde{p}( \pmb{\theta}\mid \pmb{y}) ~ d\pmb{\theta}_{-j}\,\, ,
\end{equation}

\noindent where $\tilde{p}(\cdot\mid\cdot)$  is an approximated posterior density. Using the Laplace approximation, the posterior marginals of hyper-parameters $p( \pmb{\theta}\mid \pmb{y})$ at a specific value  $\pmb{\theta}=\pmb{\theta}_{j}$ can be written as:

\begin{equation}
 \tilde{p}( \pmb{\theta}_{j}\mid \pmb{y})\propto \frac{p(\pmb{x},\pmb{\theta}_{j},\pmb{y})} {\tilde{p}_{G}(\pmb{x}\mid\pmb{\theta}_{j},\pmb{y})}  \propto  \frac{p(\pmb{y}\mid\pmb{x},\pmb{\theta}_{j}) p(\pmb{x}\mid\pmb{\theta}_{j}) p(\pmb{\theta}_{j})} {\tilde{p}_{G}(\pmb{x}\mid\pmb{\theta}_{j},\pmb{y})} \mid_{\pmb{x}=\pmb{x}^{*}(\pmb{\theta}_{j})}\,\, ,  
 \label{post_hyper}
\end{equation}

\begin{equation}
 \tilde{p}_{G}(\pmb{x}\mid\pmb{\theta},\pmb{y}) \propto \exp{-\frac{1}{2}\pmb{x}^{T}\pmb{Q}(\pmb{\theta})~\pmb{x} + \sum_{i} g_{i}(x_{i})}\,\, ,
\end{equation}

\noindent where $\tilde{p}_{G}(\pmb{x}\mid\pmb{\theta},\pmb{y})$ is the Gaussian approximation to the full conditional of $\pmb{x}$, and $\pmb{x^{*}}(\pmb{\theta}_{j})$ is the mode of the full conditional $\pmb{x}$  for given $\pmb{\theta}_{j}$. The posterior marginals of the latent effects are then numerically integrated as follows:

\begin{equation}
\tilde{p}(x_{i} \mid \pmb{y}) \backsimeq \sum_{j} \tilde{p}( x_{i} \mid \pmb{\theta}_{j}, \pmb{y})\tilde{p}( \pmb{\theta}_{j}\mid \pmb{y}) \Delta_{j}\,\, ,
\end{equation}

\noindent where $\Delta_{j}$ represents the integration step.\par
A good approximation for $\tilde{p}( x_{i} \mid \pmb{\theta}, \pmb{y})$ is required and INLA offers three different options: Gaussian approximation, Laplace approximation and simplified Laplace approximation \citep{inla}. In this work we used the simplified Laplace approximation, which represents a compromise between the accuracy of the Laplace approximation and the reduced computational cost achieved with the Gaussian approximation.\par
INLA has been shown \citep{inla} to greatly outperform MCMC sampling under limited computational power/time conditions, with the estimation error of INLAs results being invariably smaller than those of MCMC. Other approximated inference methods exist, such as variational Bayes \citep{VB} and expectation-propagation \citep{EP}, however these methods are not only slower, but they struggle with estimating the variance of the posterior since they execute iterative calculations instead of analytic approximations, unlike INLA \citep{inla}.\par
INLA suffers nonetheless from some limitations. The first, already mentioned above, is that to get meaningful results the latent field to be inferred must be Gaussian - which is not always the case -, and it must display conditional independence properties; the second is that for fast inference it is required that the number of hyper-parameters (characterizers of the parameter models) should be inferior to 6, and that the number of available observations of the field to infer be much smaller than the size of that field.\par
INLA is freely available as an \texttt{R} package \citep{inlaR}, and it has already been shown to: 1) be capable of recovering structures in scalar and vector fields, with great fidelity, out of sparse sets of observations, and even of inferring structures never seen before; 2) be robust to noise injections \citep{inlaApp1}. We refer to \citep{inla,inla_book} for more details on the mathematical background of INLA and the methods it employs.\par

\section{Implementation}\label{sec:impl}
This section describes both the dataset, the combination of a DVAE/PCA with INLA to enhance low information density SKIRT simulation data cubes and the tools to do so.\par
Our pilot pipeline aims to emulate radiative transfer models, as such we named it EmulART. All scripts were written and executed under R \citep{R} (version 3.6.3) and make use of the Keras API \citep{keras}.\par
The DVAE architecture was adapted from the one described in the Keras \href{documentation}{https://tensorflow.rstudio.com/guides/keras/making_new_layers_and_models_via_subclassing.html#putting-it-all-together-an-end-to-end-example}. The encoder block starts with an input layer of 64 features/neurons (the wavelengths of the SEDs) and each consecutive layer halves the amount of features until the latent space layer is reached. That layer, unlike the ones that precede it, is comprised of a vector doublet, each with 8 neurons. One vector corresponds to the mean value of the latent features and the other to their variance, together these vectors describe a value distribution for each of the latent features. The input of the decoder will be drawn from those distributions, and each subsequent layer will double the amount of features, decompressing the data, until the output, with the same number of features as the input layer of the encoder, 64, is reached.\par
Two pipelines using principal component models were used to compare against the DVAE pipeline. One pipeline makes use of the 8 PCs which explained the most variance (the same number of latent variables as available for the DVAE), while the other uses the number of PCs determined by the elbow method.  

\subsection{Dataset}\label{sec:dataset}
In this work 30 SKIRT simulations were used for separate purposes. All simulations model a spherical dust shell composed by silicates and graphites surrounding a bright point source with anisotropic emission \citep{marko}, as defined by Eq. \ref{eq:geo_sig} following \cite{netzer}:

\begin{equation}\label{eq:geo_sig}
    L(\theta) \propto \cos{\theta}(2\cos{\theta} + 1)\,\, ,
\end{equation}

\noindent where $\theta$ is the polar angle of the coordinate system. Each realization is a cube of 300 by 300 pixels maps at 103 distinct wavelength bins. The first 39 wavelength bins were discarded (leaving us with 64) for displaying very low signal of randomly scattered emission (less than 0.0001\% of the pixels at these wavelengths display flux density different than 0). The final dataset thus includes 90,000 spaxels\footnote{By spaxel we refer to the array across the spectral dimension of at a given pixel spatial coordinates.} per cube, each spaxel with 64 fluxes, or "features", at wavelength bins ranging from $\sim$1 $\mu$m to 1 mm\footnote{This wavelength range covers almost completely the infrared part of the light spectrum}.\par
\begin{figure}[ht]
    \centering
    \begin{subfigure}[b]{\textwidth}
        \centering
        \includegraphics[width =\textwidth]{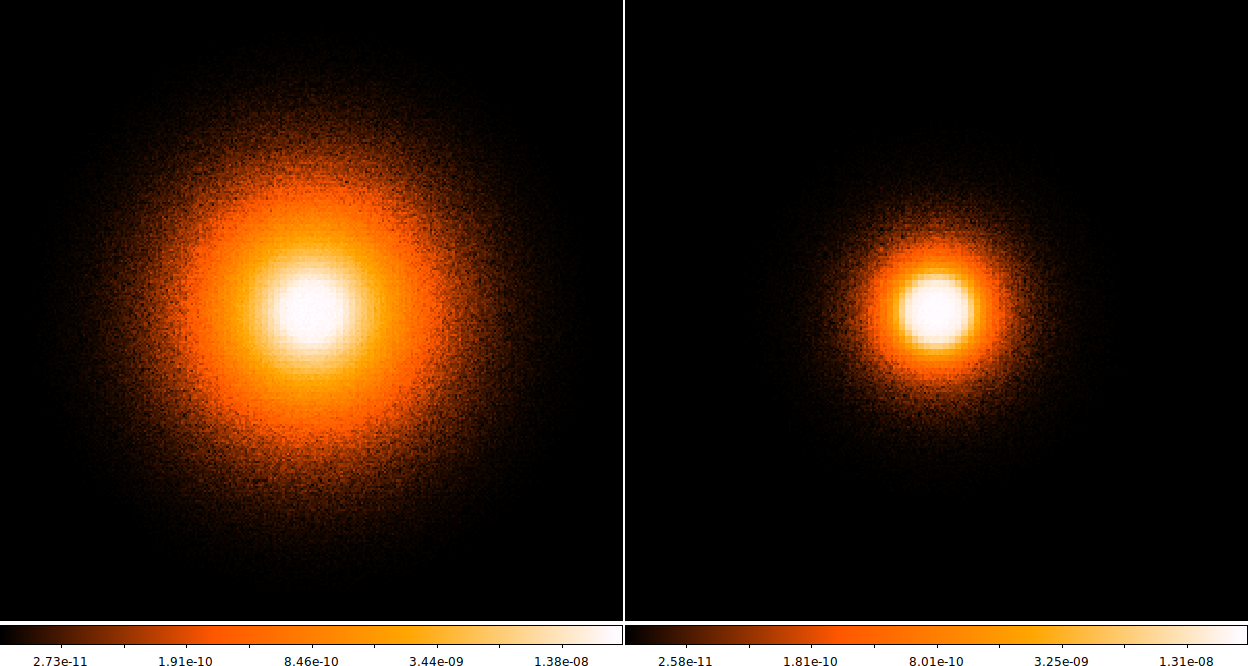}
        \caption{$\phi = 0^\circ$. Left: $\tau_{9.7} = 0.05$; Right: $\tau_{9.7} = 1.0$.}
        \label{fig:par1}
        \vspace{7mm}
    \end{subfigure}
    \begin{subfigure}[b]{\textwidth}
        \centering
        \includegraphics[width =\textwidth]{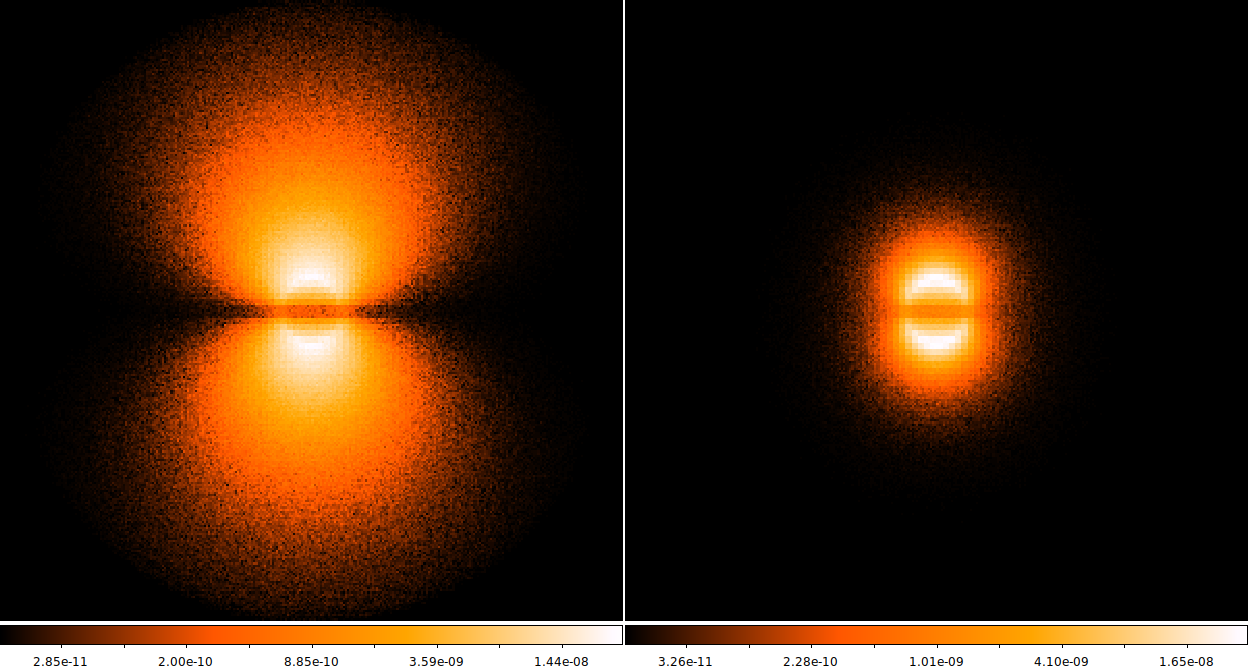}
        \caption{$\phi = 90^\circ$. Left: $\tau_{9.7} = 0.05$; Right: $\tau_{9.7} = 1.0$.}
        \label{fig:par2}
    \end{subfigure}
    \caption{High photon number references of a spherical dust shell composed of silicates and graphites surrounding a bright anisotropic point source. The present flux density maps represent the simulated observations at wavelength 1,85 $\mu$m, with $\phi = 0^\circ$ (Fig. \ref{fig:par1}) and $\phi = 90^\circ$ (Fig. \ref{fig:par2}) for $\tau_{9.7} \in \{0.05, 1.0\}$. Color indicates flux density in W/m$^2$.}
    \label{fig:parCombs}
\end{figure}
\begin{figure}[ht]
    \centering
    \begin{subfigure}[b]{\textwidth}
        \centering
        \includegraphics[width =\textwidth]{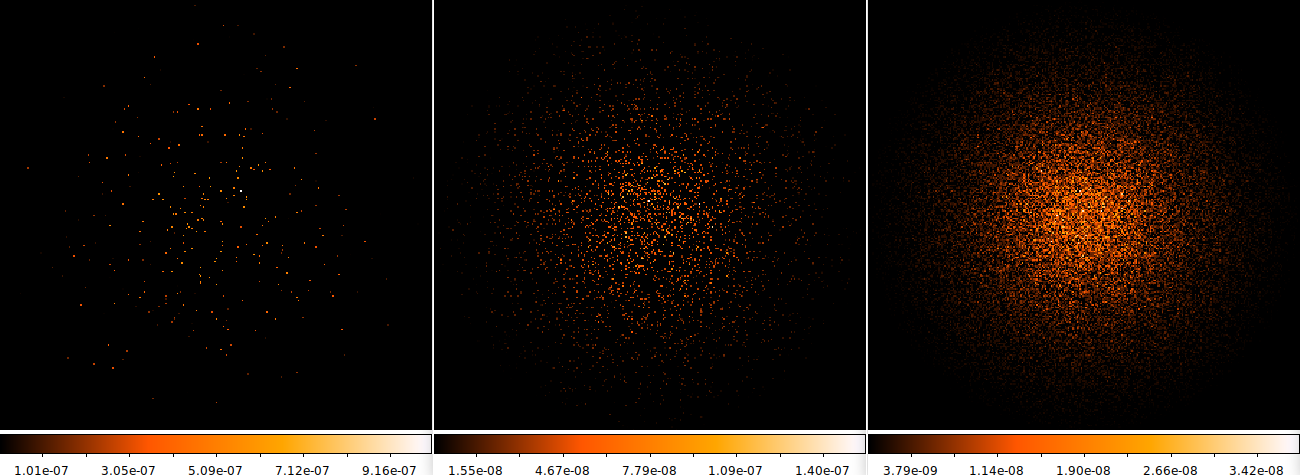}
        \caption{Left: $N_p = 10^4$; Middle: $N_p = 10^5$; Right: $N_p = 10^6$.}
        \label{fig:obj1}
        \vspace{7mm}
    \end{subfigure}
    \begin{subfigure}[b]{.66\textwidth}
        \centering
        \includegraphics[width =\textwidth]{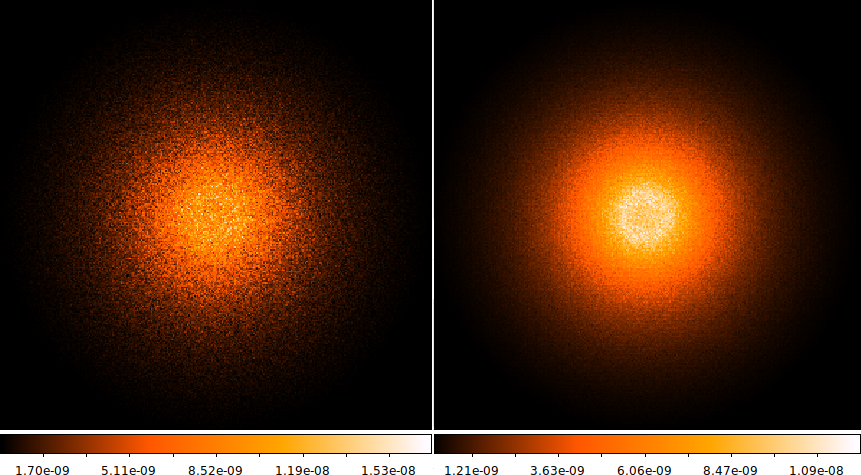}
        \caption{Left: $N_p = 10^7$; Right: $N_p = 10^8$.}
        \label{fig:obj2}
    \end{subfigure}
    \caption{Models of a spherical dust shell composed of silicates and graphites surrounding a bright anisotropic point source, with $\tau_{9.7} = 0.05$ and $\phi = 0^\circ$, realized by simulating different photon amounts. The present flux density maps represent the simulated observations at wavelength 9,28 $\mu$m. Fig. \ref{fig:obj1} presents the realization obtained by simulating $N_p \in \{10^4, 10^5, 10^6\}$, while Fig. \ref{fig:obj2} presents the realizations obtained by simulating $N_p \in \{10^7, 10^8\}$. Color indicates flux density in W/m$^2$.}
    \label{fig:obj}
\end{figure}

The 30 realizations differ from each other by up to three parameters: the tilt angle, $\phi$, of the object as seen by the observer ($0^\circ$, face-on, and $90^\circ$, edge-on\footnote{The face-on/edge-on terminology refers to the shape of a disk-like object as seen by an observer when tilt angle between the plane of the disk and the plane of the observer is, respectively, $0^\circ$ or $90^\circ$.}); the optical depth\footnote{The optical depth, $\tau$, describes the fraction of light that is transmitted through a material following the relation $\frac{I_0}{I_T}=e^{-\tau}$, where $I_0$ is the incoming light and $I_T$ the outgoing light.}, $\tau_{9.7}$\footnote{Optical depth at wavelength 9.7 $\mu$m corresponding to the peak emission wavelength of silicates.}, of the dust shell (0.05, 0.1 and 1.0); and the amount\footnote{Throughout this text the symbol $N_X$ will stand for "amount/number/quantity of X".} of photon packets simulated, $N_p \in \{10^4, 10^5, 10^6, 10^7, 10^8\}$. For each particular $\tau_{9.7}$ and $\phi$ combination the corresponding $N_p = 10^8$ realization was regarded as the HPN reference, or ``ground truth", for the purpose of evaluating the performance of our routines, since those yield the highest information density, while all other simulations, with $N_p \in \{10^4, 10^5, 10^6, 10^7\}$, were considered LPN simulations. Fig. \ref{fig:parCombs} illustrates the difference between HPN references through different $\tau_{9.7}$ and $\phi$ combinations, while Fig. \ref{fig:obj} shows LPN models with differing $N_p$, keeping $\tau_{9.7} = 0.05$ and $\phi = 0^\circ$. Tab. \ref{tab:tab_sources} displays the difference between the quality of the individual spaxels\footnote{We discriminate spaxels according to their completeness along the spectral dimension. Spaxels that have 0 flux at all wavelengths bins are classified as "null spaxel"; spaxels that have all wavelength bins with positive flux are classified as "full spaxel"; spaxels in between the two previous cases are classified as "partial spaxel"; finally, the "empty spaxel" information metric is the difference between the amount of null spaxels within the HPN reference and a given LPN realization.} that compose each LPN realization and HPN reference; the median, M, and mean absolute deviation (MAD) of the normalized residuals (see Eq. \ref{eq:normres}) of every pixel within each LPN realization; as well as the total information ratio (TIR) for all realizations, here defined as the ratio of the number of pixels with flux different than 0 of an LPN input or emulation, $N_{X' \neq 0}$, and that same number for the HPN reference, $N_{X \neq 0}$ (see Eq. \ref{eq:tir}), as an information metric to balance against the normalized residuals\footnote{Should the estimation, X’, be 0 the normalized residual will be 100\%, meaning there is no actual estimate for the reference value, X.}:\par

\begin{equation}\label{eq:normres}
    Residuals (\%) = \mid\frac{X' - X }{X}\mid \times 100\%\,\, ,
\end{equation}

\begin{equation}\label{eq:tir}
    TIR (\%) = \frac{N_{X' \neq 0}}{N_{X \neq 0}} \times 100\%\,\, .
\end{equation}.\par

\begin{table}[ht]
\centering
\begin{tabular}{cccccccc}
\hline
\multirow{5}{*}{\textbf{Sim.}} & \multicolumn{2}{c}{\multirow{2}{*}{\textbf{\begin{tabular}[c]{@{}c@{}}Normalized\\Residuals of Sim. (\%)\end{tabular}}}} & \multicolumn{5}{c}{\multirow{2}{*}{\textbf{Information of Simulation}}}      \\
    & \multicolumn{2}{c}{}  & \multicolumn{5}{c}{} \\ \cline{2-8}
    & \multirow{3}{*}{\textbf{M}}   & \multirow{3}{*}{\textbf{MAD}} & \multirow{3}{*}{\textbf{\begin{tabular}[c]{@{}c@{}}TIR\\ (\%)\end{tabular}}} & \multirow{3}{*}{\textbf{\begin{tabular}[c]{@{}c@{}}Null\\ Spxs\\ ($N$)\end{tabular}}} & \multirow{3}{*}{\textbf{\begin{tabular}[c]{@{}c@{}}Empty\\ Spxs\\ ($\Delta N$)\end{tabular}}} & \multirow{3}{*}{\textbf{\begin{tabular}[c]{@{}c@{}}Full\\ Spxs\\ ($N$)\end{tabular}}} & \multirow{3}{*}{\textbf{\begin{tabular}[c]{@{}c@{}}Partial\\ Spxs\\ ($N$)\end{tabular}}} \\
    &       &       &       &       &       &       &       \\
    &       &       &       &       &       &       &       \\ \cline{1-8}
    $10^4$  & 100   & 0     & 0.2   & 82,041& 64,185& 0     & 7,959 \\ \cline{1-8}
    $10^5$  & 100   & 0     & 1.9   & 47,964& 30,108& 0     & 42,036\\ \cline{1-8}
    $10^6$  & 100   & 0     & 15.4  & 19,115& 1,259 & 0     & 70,885\\ \cline{1-8}
    $10^7$  & 100   & 59    & 65.2  & 17,904& 48    & 2,784 & 69,312\\ \cline{1-8}
    $10^8$  & 0     & 0     & 100   & 17,856& 0     & 47,160& 24,984\\ \hline
\end{tabular}
\caption{Statistics regarding how the LPN realizations compare to the HPN reference. Median normalized residuals of 100\% indicate that missing information permeates the LPN realizations. The quantity and quality of the available information is further described by the total information ratio (TIR), and the amounts of different qualities of spaxels. Null spaxels are spaxels with zero information; empty spaxels are null spaxels that were not tagged as such for the HPN reference; full spaxels are spaxels that have information at every wavelength and, partial are those spaxels that are neither null nor full.}
\label{tab:tab_sources}
\end{table}

The realizations were split into two subsets: one to train the autoencoder, and perform the first batch of tests to the emulation pipeline, labeled AESet and described in Section \ref{sub:sub1}; and another, comprised exclusively by data the autoencoder did not see during training, to better assess EmulARTs performance, labeled EVASet and described in Section \ref{sub:sub2}.\par

\subsubsection{AESet}\label{sub:sub1}
This subset is comprised of 5 SKIRT simulation outputs of the same model, a spherical shell of dust composed by silicates and graphites, with optical depth $\tau_{9.7} = 0.05$, surrounding a bright point source with anisotropic emission  seen face-on, $\phi = 0^\circ$, making the emission appear isotropic. The only different parameter across the realizations in AESet was $N_p \in \{10^4, 10^5, 10^6, 10^7, 10^8\}$ (see Fig. \ref{fig:obj}). The $N_p = 10^8$ realization is the HPN and was used as the reference, or ``ground truth", for the purpose of both training the autoencoder model as well as evaluating the performance of EmulART.\par
\begin{figure}[ht]
    \centering
    \begin{subfigure}[b]{.48\textwidth}
        \centering
        \includegraphics[width =\textwidth]{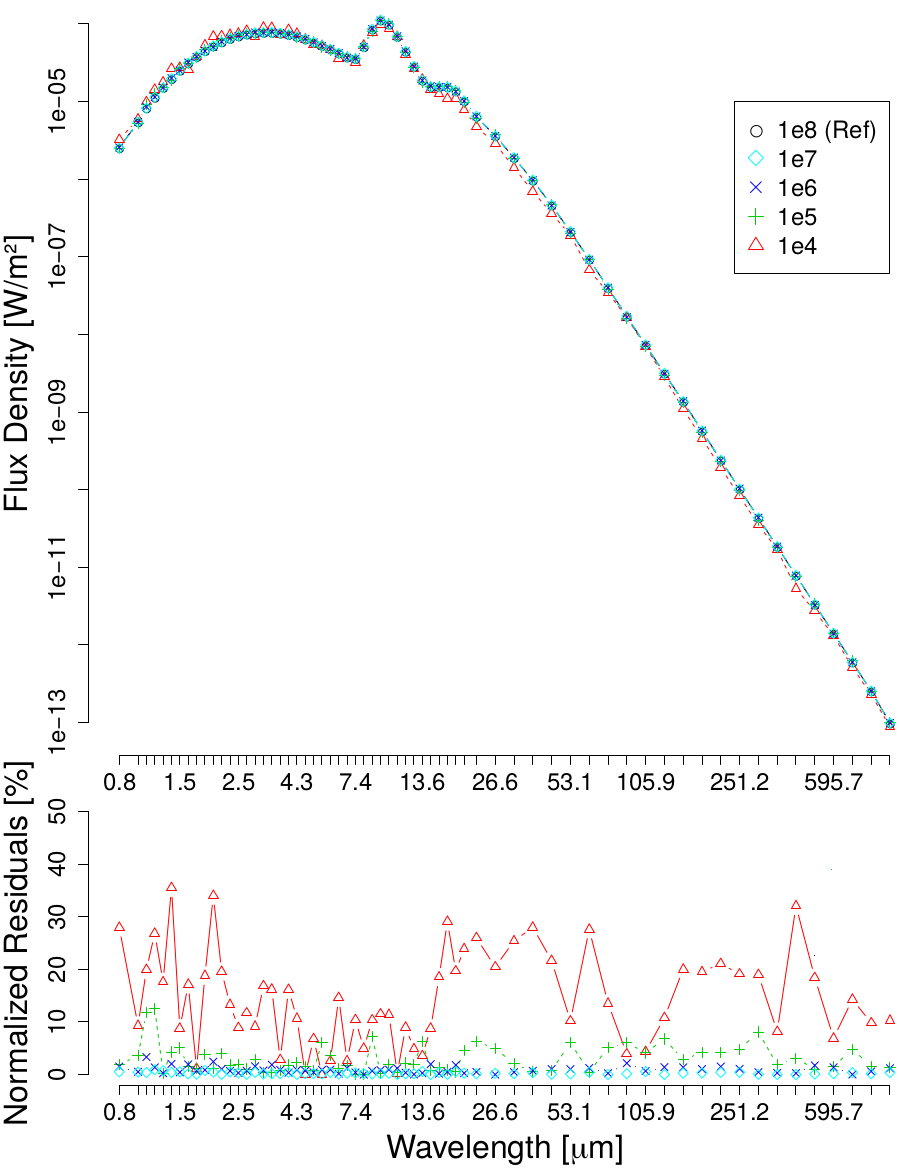}
        \caption{}
        \label{fig:SED1}
    \end{subfigure}
    \hspace{1mm}
    \begin{subfigure}[b]{.48\textwidth}
        \centering
        \includegraphics[width =\textwidth]{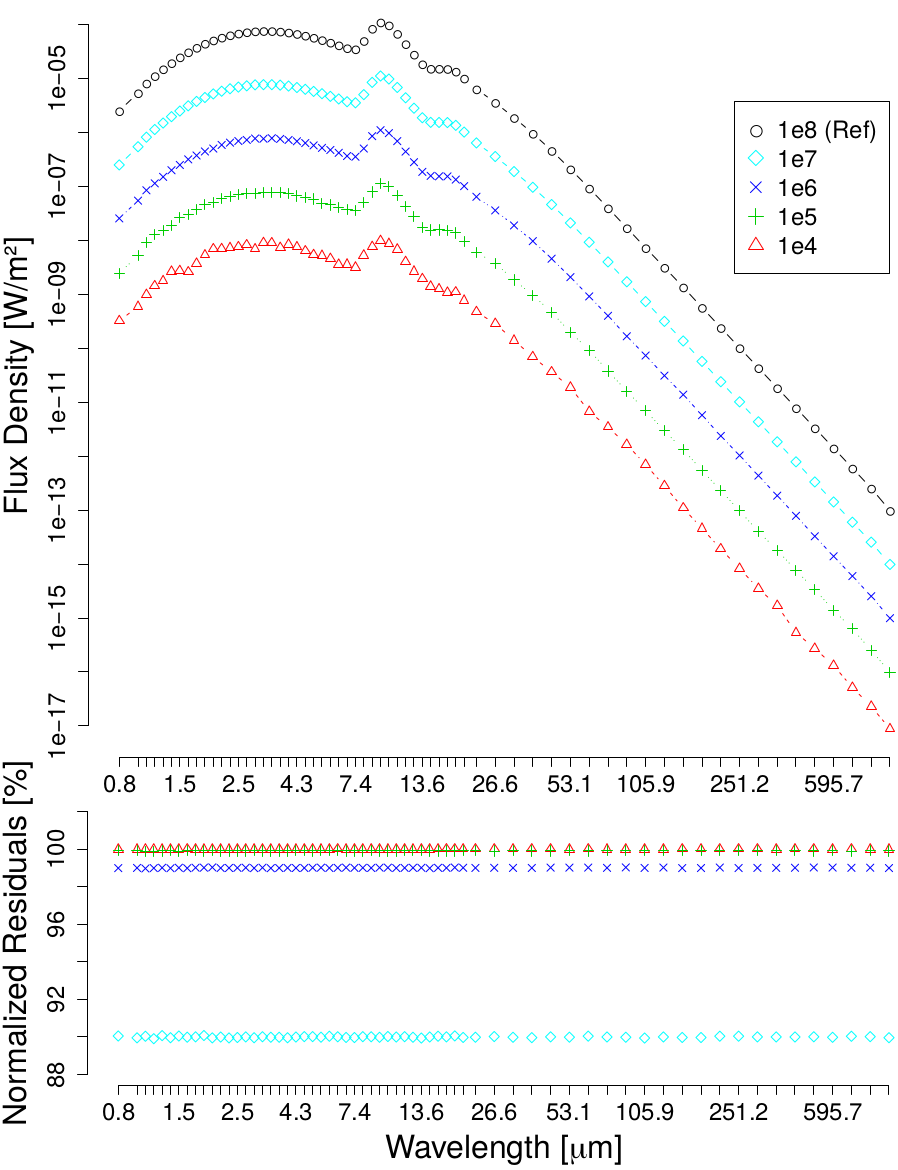}
        \caption{}
        \label{fig:SED2}
    \end{subfigure}

    \caption{Integrated SEDs (upper panel) and normalized residuals (lower panel) for each of the five SKIRT simulations in our dataset: a) due to normalization, realizations with different amounts of photons simulated display very similar integrated SEDs; b) integrated SEDs of realizations in AESet after de-normalization (total flux is here proportional to $N_p$). The labels indicate the value of $N_p$ for each realization.}
    \label{fig:SEDs_reg}
\end{figure}

Since the goal of our methodology is to reconstruct the reference simulation using LPN realizations as input, the values of each cube were multiplied by the ratio between the amount of photons simulated for the LPN input, $N_p^{LPN}$ and the amount of photons simulated for the HPN reference ($N_p^{HPN} = 10^8$). In Fig. \ref{fig:SEDs_reg} we can see the impact of this de-normalization on the integrated SEDs\footnote{An integrated SED results from integrating all spatial information at each wavelength.\label{fn:iSED}} of the LPN realizations: LPN realizations have less flux when not including SKIRTs normalization.\par

The spaxels of all cubes within AESet were used to train the DVAE model. For this we split AESet into training set (5/6) and test set (1/6). Spaxels of different cubes but which share the same spatial coordinates were assigned to the same ensemble. This strategy aimed for the DVAE model to achieve a denoising capability and consists on having input spaxels that result from realizations with different $N_p$ always match the HPN references version on the output layer.\par
In SKIRT, choosing to simulate fewer photons results in an output with more zero flux density pixels which in turn means that more spaxels will be null at more wavelength bins. Even though this is different than noise, it is akin to missing data whose impact we aim to curb by implementing a DVAE architecture.\par
Before training the DVAE model we perform some spaxel selection and pre-processing tasks on AESet which is thoroughly described in Appendix \ref{app:A}.\par
Later, when performing preliminary tests on EmulART, we used the 4 LPN realizations within AESet ($N_p \in \{10^4, 10^5, 10^6, 10^7\}$) as input.  

\subsubsection{EVASet}\label{sub:sub2}
This subset is comprised of the 25 SKIRT simulation cubes that also model a spherical shell of silicates and graphites surrounding a bright anisotropic point source but have different combinations of $\phi$, $\tau_{9.7}$ and $N_p$ from those in AESet. To all realizations in EVASet we performed the same feature selection and flux de-normalization tasks described both in Section \ref{sub:sub1} and Appendix \ref{app:A}.\par
The 20 LPN realizations within EVASet($N_p \in \{10^4, 10^5, 10^6, 10^7\}$) were used as input for EmulART for a deeper assessment of the capabilities of our emulation pipeline. The remaining 5 HPN cubes (those with $N_p = 10^8$) were used as references to compute performance metrics.

\subsubsection*{}
A list detailing the parameters of the SKIRT simulations used in this work, as well as their split into AESet and EVASet, can be consulted in Appendix \ref{app:B}.

\subsection{Training the DVAE}\label{sec:train}
To determine which set of hyper-parameters for the DVAE suited our needs best we performed some tests, grid-searches and explored: amounts of features in the latent space; activation, loss and optimization functions; batch size\footnote{Number of samples that will be passed through the network at one time. An epoch is complete once all samples are passed through the network. Increasing the batch size accelerates the completion of each epoch but it may also degrade the quality of the model.}; bias constraint\footnote{Limit, between 0 and 1, for the weight of bias neurons.}; learning rate\footnote{Learning rate is a parameter that determines the step size to take, at each iteration, in the direction determined by the optimization function so as to reach the minimum of the loss function.}; and, patience\footnote{Number of epochs to wait before implementing a change or stopping the training procedure.} values. These exploratory tests were performed by training the models during 100, 500 and 2000 epochs, according to the need to differentiate performance between hyper-parameter sets.\par
\begin{figure}[ht]
\centering
    \includegraphics[width =\textwidth]{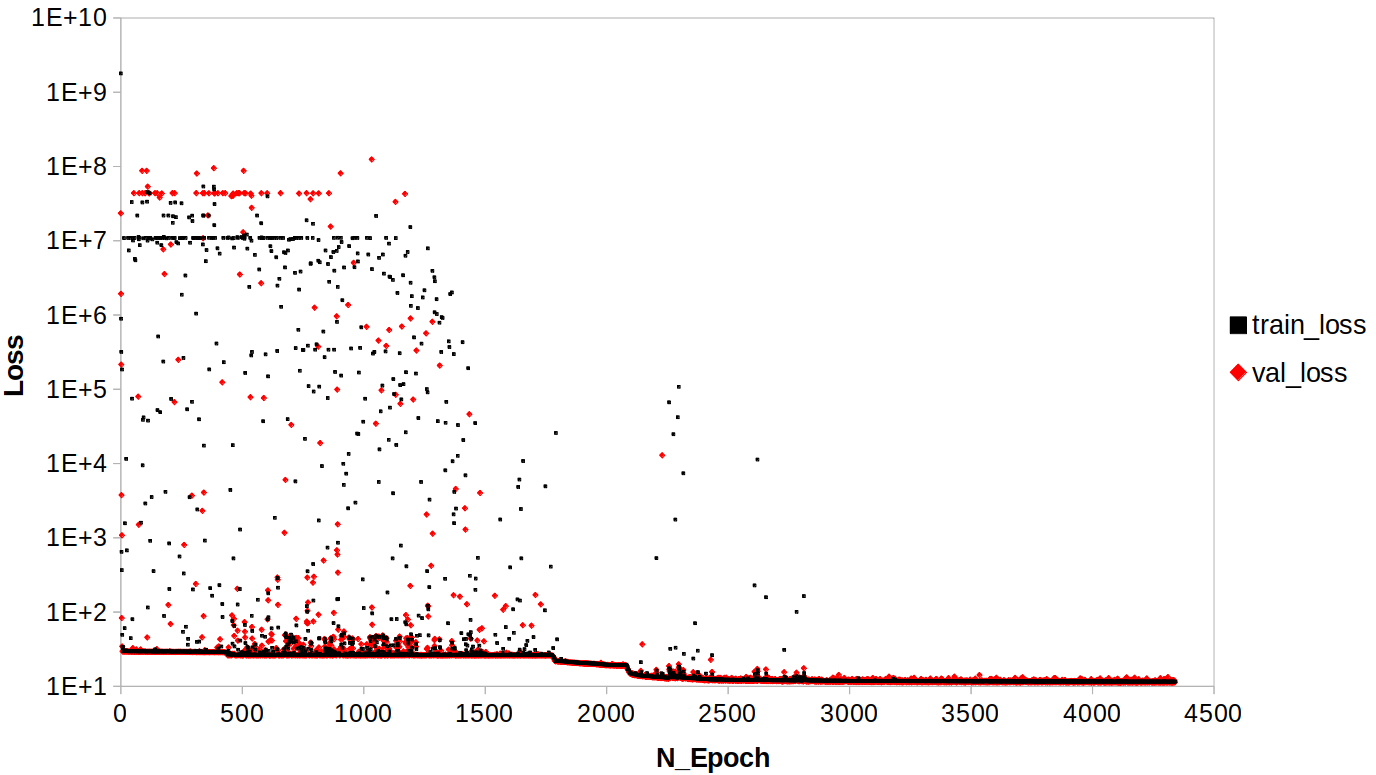}
    \caption{Loss, in log scale, as function of the number of epochs. The training loss is represented by $\blacksquare$ and the validation loss by red $\blacklozenge$. The plot shows that both metrics converge to comparable values.}
    \label{fig:losses}
\end{figure}

We measured performance by the percentage residuals of both the individually reconstructed spaxels, of the test subset of AESet, and of each of AESets cubes integrated SEDs. We selected the set of hyper-parameters listed below for being the most consistent across different tests. Fig. \ref{fig:losses} shows the validation loss closely following the training loss, indicating successful convergence of the model to our data.

\begin{itemize}
    \item Latent feature amount: 8
    \item Activation function: SELU \citep{selu} and sigmoid (for the output layer only)
    \item Loss function: weighted sum of the Kullback-Leibler divergence and mean percentual error\footnote{A custom variation of the one presented in Keras \href{documentation}{https://tensorflow.rstudio.com/guides/keras/making_new_layers_and_models_via_subclassing.html#putting-it-all-together-an-end-to-end-example} as the original version of this loss function makes use of the mean squared error instead.}
    \item Optimization function: Adam\footnote{A stochastic gradient descent method based on adaptive estimation of first-order and second-order moments.}
    \item Batch size: 32
    \item Bias constraint: 0.95
    \item Train - Validation split: 4/5, 1/5 
    \item Maximum $N_{Epoch}$: 4,500
    \item Patience 1\footnote{Number of epochs to wait, with no significant improvement in validation loss, before reducing the learning rate.}: 500 epochs
    \item Patience 2\footnote{Number of epochs to wait, with no significant improvement in validation loss, before stopping the training process.}: 3,000 epochs
    \item Initial learning Rate (LR): 0.001
    \item Learning rate decrease\footnote{Ratio between the new and old learning rate.}: 0.25
\end{itemize}

After training the weights were saved to files which are loaded into the pipeline (see Section \ref{sec:pipeDVAE}).\par 
Appendix \ref{app:C} shows the relationship between the compressed (or latent) features of the test set, as well as the Pearson's correlation coefficients (PCC) between those features. Based on the analysis of the correlations of the latent features we decided to stop further compression of the spectral dimensions of the data.\par

\subsection{DVAE Emulation Pipeline}\label{sec:pipeDVAE}
\begin{figure}[ht]
    \centering
    \includegraphics[width =\textwidth]{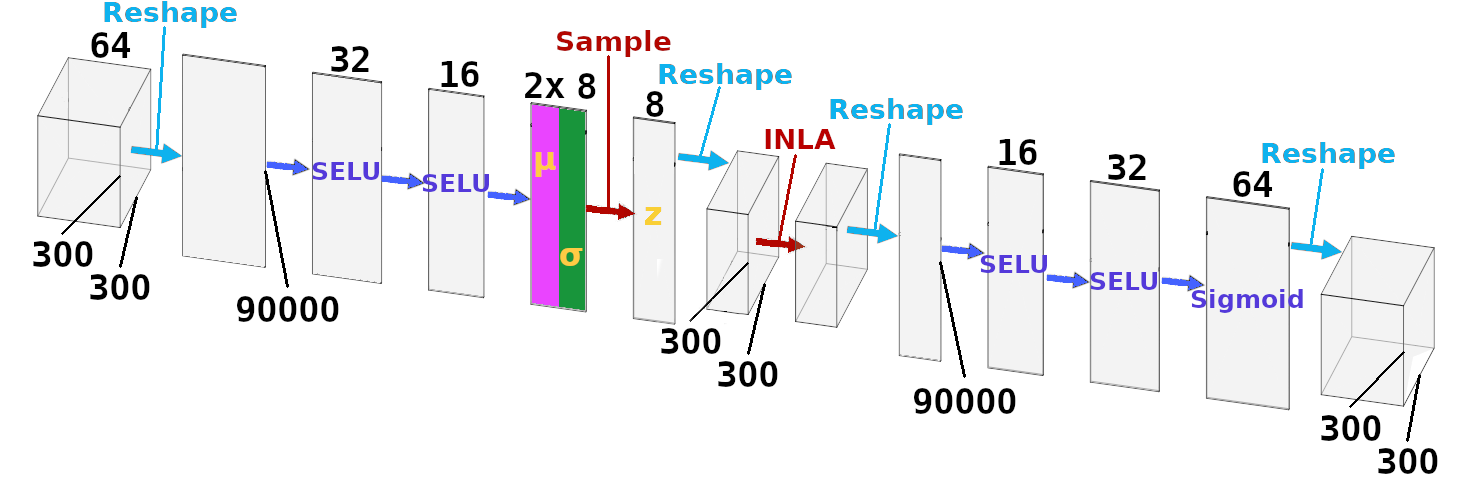}
    \caption{Scheme with the most relevant operations of the pipeline. Data shaping is represented in cyan; feature combination is in dark blue; and statistical operations are in red. Feature dimensionality is given along the axis of the arrows, at the top of each layer. $\bfmu$ is the vector holding the mean value of the latent features distribution, while $\bfsigma$ is the vector holding their variance; $\bfz$ is the vector built by randomly drawing values from the latent features distributions $G(\bfmu,\bfsigma)$.}
    \label{fig:pipe}
\end{figure}

Within EmulART the feature space is first compressed with the variational encoder; the latent space is then sampled and the resulting latent features spatial maps are reconstructed with INLA; finally the reconstructed wavelength (original feature space) maps are recovered with the decoder. Additionally, to conform the data to each of the different stages of the pipeline, we perform some operations described below. Fig. \ref{fig:pipe} presents a scheme\footnote{This illustration was drawn using  \href{NN SVG}{http://alexlenail.me/NN-SVG/index.html}.} of the emulation pipeline while a flowchart including all relevant data pre- and post-processing operations, integrated within the emulation pipeline, can be found in Appendix \ref{app:D}.\par

Prior to being parsed by the encoder network module, the data is initially pre-processed as described in Appendix \ref{app:A}. After the data goes through the encoding and variational sampling stages (see Fig. \ref{fig:pipe}), the resulting latent features maps, $Z_f(x,y)$\footnote{Where $f \in \{1,2,3,4,5,6,7,8\}$ denotes feature.}, are then transformed according to the following steps\footnote{The reader is reminded that the present, unchanged, dataset, being simulated flux density values, has an inferior limit of 0.} for each feature map, $Z_f(x,y)$:
\begin{enumerate}
    \item Determine the minimum, $m^0_f$, and maximum, $M^0_f$, values of the map,
    \begin{center}
        $m^0_f = $ min$(Z_f(x,y)),\;\;M^0_f = $ max$(Z_f(x,y))$\,\,;
    \end{center}
    \item Determine the value range, $R_f$, of the map,
    \begin{center}
        $R_f = M^0_f - m^0_f$\,\,;
    \end{center}
    \item Offset the value range of the map so that the new minimum is 0,
    \begin{center}
        $Z'_f(x,y) = Z_f(x,y) - m^0_f$\,\,;
    \end{center}
    \item For the offset map, $Z'_f(x,y)$, determine the minimum positive value, $m^1_f$, and divide it by $R^2_f$ to obtain the new minimum , $m^2_f$,
    \begin{center}
        $m^1_f =$  min$(Z'_f(x,y)):\;\; Z'_f(x,y)>0$\,\,,\\
        $m^2_f = m^1_f / R^2_f$\,\,;
    \end{center}
    \item Obtain the final map, $Z''_f(x,y)$, by offsetting the value range by $m^2_f$,
    \begin{center}
        $Z''_f(x,y) = Z'_f(x,y) + m^2_f$\,\,.
    \end{center}
\end{enumerate}

These transformations, found by trial and error, are useful in conveying the data to INLA in a value range where its performance is both consistent (across different inference task in this scientific domain), less prone to run-time errors and more accurate. It should be noted that the validity and the improvement on performance granted by this interval transformation has only been empirically verified for our particular case, and it may well be improved upon.\par
After these transformations the latent features maps are reconstructed by INLA. Then they are transformed back to the original value range and are parsed by the decoder network module.

\subsection{PCA Emulation Pipelines}\label{sec:pipePCA}

\begin{figure}[ht]
    \centering
    \includegraphics[width =.75\textwidth]{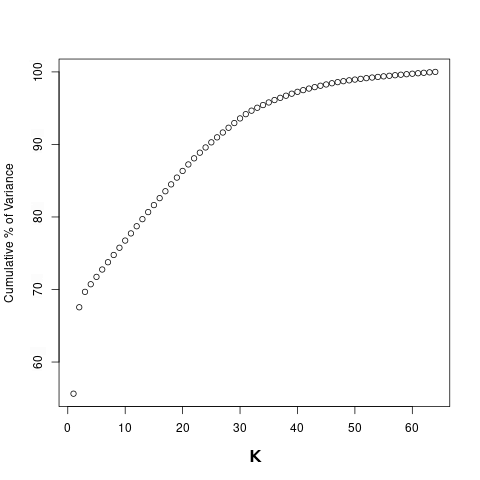}
    \caption{Plot of cumulative percentage of variance explained as a function of the number of principal components, K) used to encode the data. Increasing the amount of PCs used naturally increases the percentage of variance explained but it also risks over-fitting the model. In this case the elbow can be found at around K = 30.}
    \label{fig:elbow}
\end{figure}

For the PCA emulation pipeline a new PCA model of the input feature space is constructed every time (unlike with the EmulART which DVAE model was trained on a subset of AESet, as described in Section \ref{sec:train}). Once that model is constructed two approaches are followed: the first being the usage of the elbow method (see Fig. \ref{fig:elbow}) to find the threshold number of components, K, that would explain the data without over-fitting it, using INLA to spatially reconstruct those components maps, and then return to them to the original feature space. In the second approach, 8 principal components are used (the same number as the latent features in the DVAE emulation pipeline).\par

After some preliminary tests, the data range transformations described in the previous section, before and after spatial reconstruction, were also removed as they failed to lead to execution time or reconstruction accuracy improvement. Moreover, spaxel de-normalization (see Section \ref{sub:sub1}) was removed as it lead to the underestimation of the integrated SEDs of the reference model (possible reasons for this are presented in Section \ref{sec:pcares}).\par
A scheme presenting the most relevant operations for both implementations of the PCA emulation pipeline can be found in Appendix \ref{app:D}.\par

\section{Results and Discussion}\label{sec:results}
In this section we present and discuss some of the results from testing EmulART on AESet, these are compared against the results obtained with the PCA emulation implementations, and EVASet. Our goal in first using AESet was to evaluate the performance of the pipeline as a whole, mostly because the decoder network was not trained on latent and INLA reconstructed spaxels. The realizations from EVASet were then used to better gauge its performance.\par
We created the emulations using different LPN inputs ($N_p \in \{10^4, 10^5, 10^6, 10^7\}$). Because INLA performs faster on sparse maps, for each of those LPN realizations an emulation was performed by sampling different percentages of each latent feature map. These sampling percentages resulted from sampling 1 pixel in each bin of $2\times2$, $3\times3$ and $5\times5$ pixels, corresponding respectively to 25\%, 11\% and 4\% of the spatial data. With the intent of reducing the influence of null spaxels in the spatial inference, 90\% of the null spaxels were rejected from each map sampling pool.\par
Our analysis of the results consisted on inspecting how well EmulART reproduces the spectral and spatial features of the reference simulations, as well as the total computational time it took for the emulation to be completed.\par
To evaluate the spectral reconstruction we looked at the normalized residuals (see Eq. \ref{eq:normres}) between the integrated SEDs$^{\ref{fn:iSED}}$ of our emulations and of the HPN reference. We also inspected the spatial maps of the compressed features looking for spatial distributions compatible with physical properties of the simulated model. The spatial reconstruction was also evaluated by the median and MAD of the normalized residuals of our emulations as well as of their LPN inputs \footnote{When calculating residuals for individual spaxels their flux density was previously de-normalized as described in Section \ref{sub:sub1}. This is however not the case when calculating residuals for the SEDs that result from the spatial integration of the pixels at each wavelength.} at each wavelength. For the statistical analysis of the residuals, reference pixels with value 0 were not considered since this metric diverges, so the TIR for all emulations and simulations was calculated as well.\par

\subsection{AESet Predictions}\label{sec:aeres}
In this section we present and discuss the results obtained emulating the HPN reference of AESet using the different LPN realizations within it.\par
\begin{figure}[ht]
\centering
    \begin{subfigure}[b]{.45\textwidth}
        \centering
        \includegraphics[width =\textwidth]{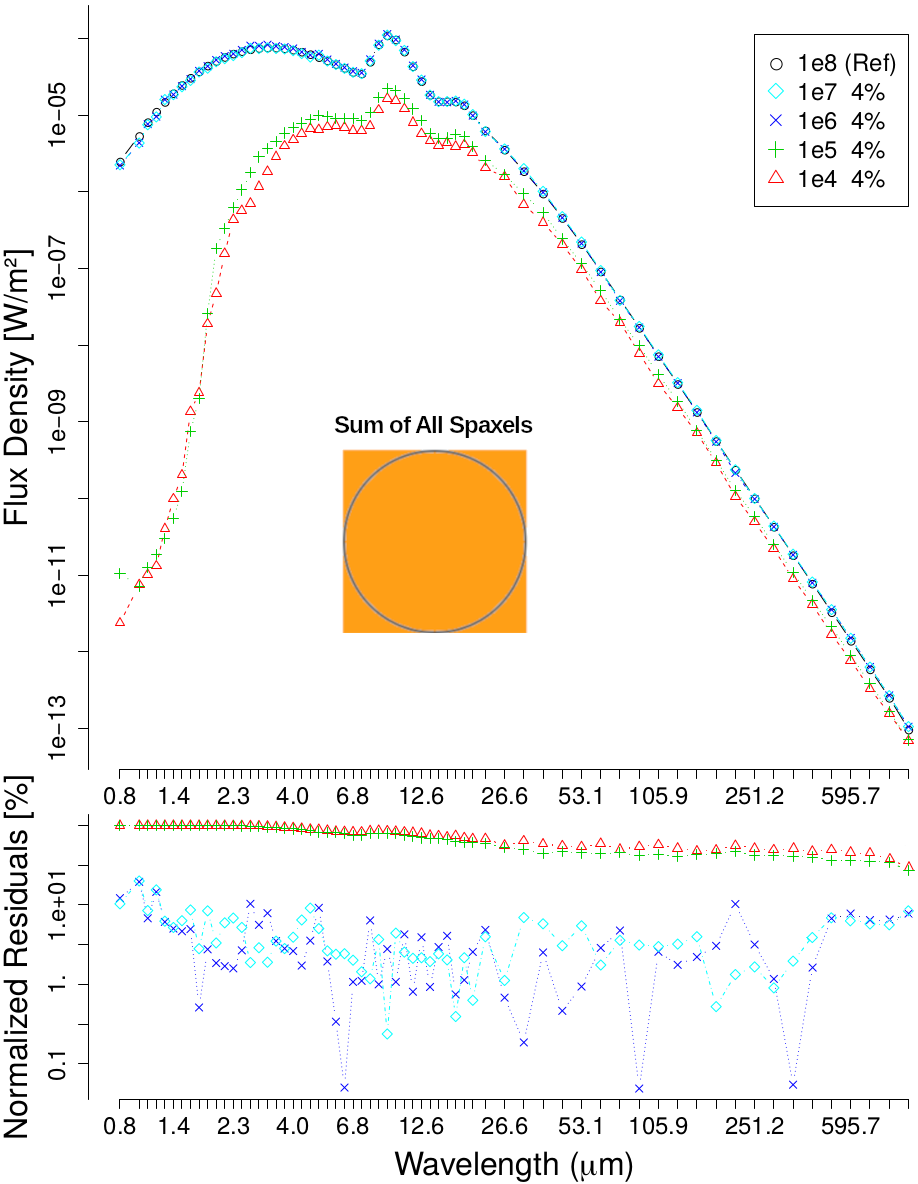}
        \caption{}
        \label{fig:int4}
        \vspace{3mm}
    \end{subfigure}
    \hspace{5mm}
    \begin{subfigure}[b]{.45\textwidth}
        \centering
        \includegraphics[width =\textwidth]{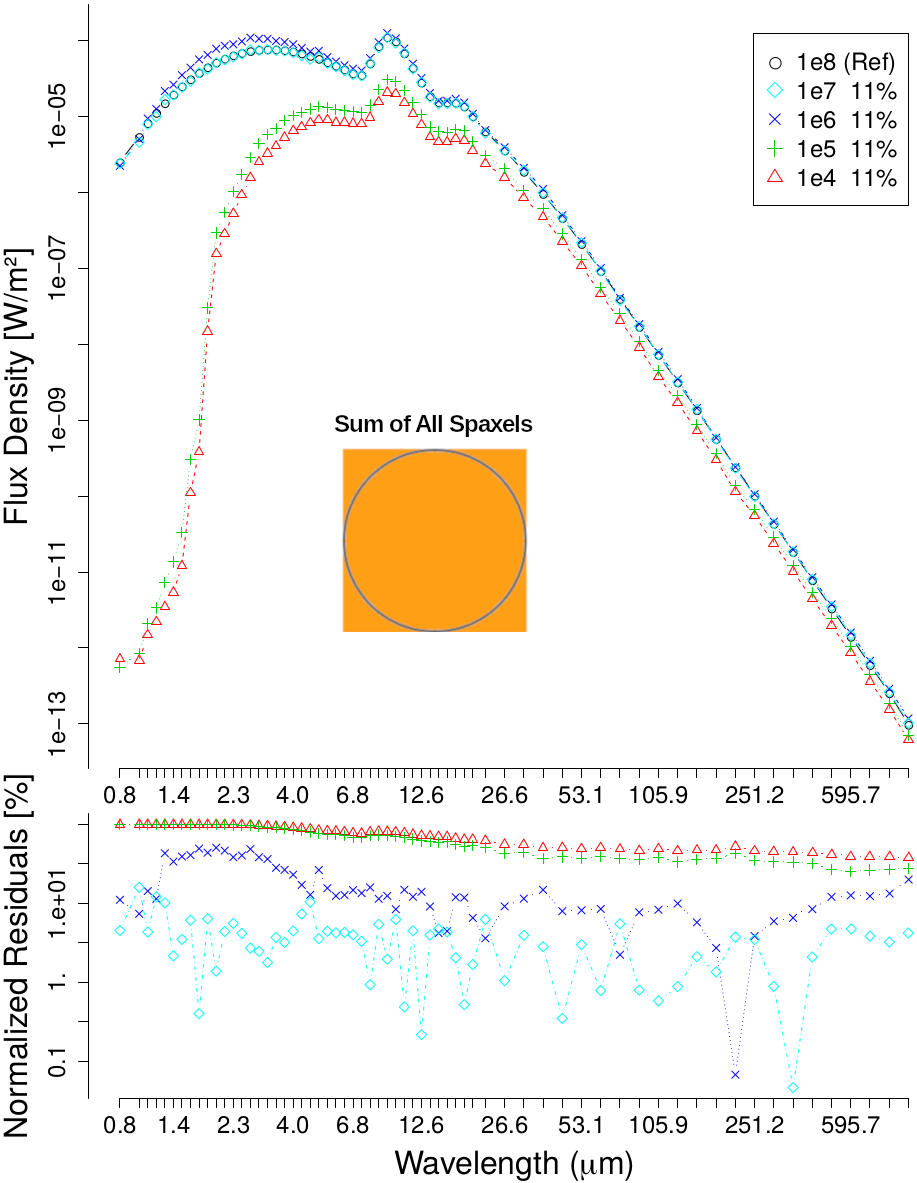}
        \caption{}
        \label{fig:int11}
        \vspace{3mm}
    \end{subfigure}
    \begin{subfigure}[b]{.45\textwidth}
        \centering
        \includegraphics[width =\textwidth]{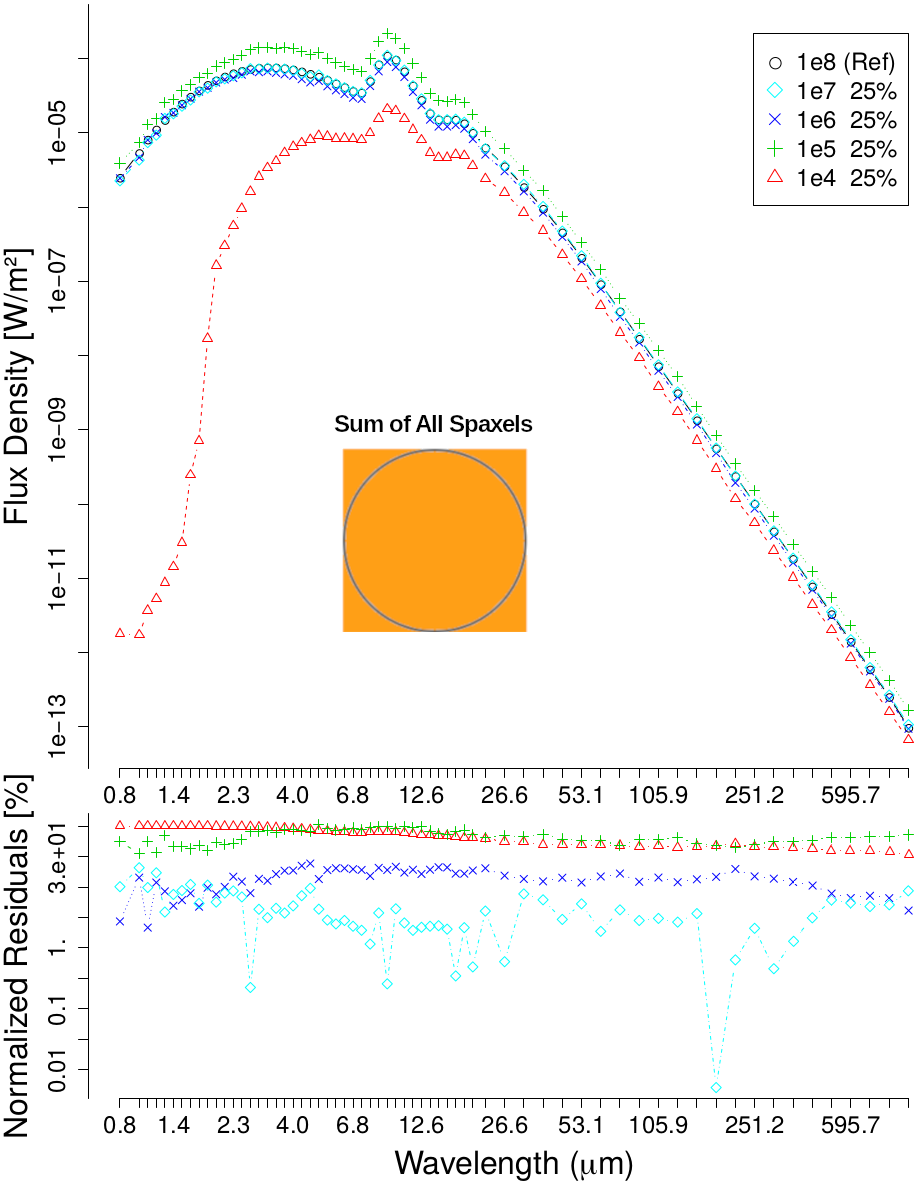}
        \caption{}
        \label{fig:int25}
    \end{subfigure}
    \caption{Emulations integrated SEDs resulting from spatial inference using 4\% (\ref{fig:int4}), 11\% (\ref{fig:int11}) and 25\% (\ref{fig:int25}) samples of the spatial information, and the respective normalized residuals, for the case of a dust shell with $\tau_{9.7} = 0.05$ and $\phi = 0^\circ$. The HPN reference, is represented in black ($\circ$), the emulation based on the $N_p = 10^4$ realization is in red ($\triangle$), on the $N_p = 10^5$ in green (+), on the $N_p = 10^6$ in blue ($\times$) and on the $N_p = 10^7$ in cyan ($\diamond$).}
    \label{fig:intSED}
\end{figure}

 
The upper panels of Fig. \ref{fig:intSED} show that the emulations integrated SEDs reproduce the shape of the references: a slow rise in the 1 $\mu$m to 8 $\mu$m range, the two emission bumps in the 8 $\mu$m to 20 $\mu$m range, and the steep decline towards longer wavelengths. Moreover, Tab. \ref{tab:intres} displays the median and MAD of the residuals of the integrated SEDs for the LPN input realizations before and after being de-normalized (as we describe in Section \ref{sub:sub1}), as well as those of the different emulations obtained from them. It is clear that using $N_p \geq 10^6$ realizations as input yields emulations integrated SEDs that closely (median residuals smaller than 15\%) follow the references throughout the whole wavelength range, independently (within this subset) of the sampling percentage chosen for the spatial inference task.\par

\begin{table}[ht]
\centering
\begin{tabular}{cccccccc}
\hline
\multirow{4}{*}{\textbf{\begin{tabular}[c]{@{}c@{}}LPN\\ Input\end{tabular}}} & \multicolumn{2}{c}{\multirow{3}{*}{\textbf{\begin{tabular}[c]{@{}c@{}}Residuals of \\ Integrated\\ Input (\%)\end{tabular}}}} & \multicolumn{2}{c}{\multirow{3}{*}{\textbf{\begin{tabular}[c]{@{}c@{}}Residuals of\\ Integrated\\ De-normalized Input (\%)\end{tabular}}}} & \multirow{4}{*}{\textbf{Sample}} & \multicolumn{2}{c}{\multirow{3}{*}{\textbf{\begin{tabular}[c]{@{}c@{}}Residuals of\\ Integrated\\ Emul. (\%)\end{tabular}}}} \\
   & \multicolumn{2}{c}{}   & \multicolumn{2}{c}{}  &       & \multicolumn{2}{c}{}  \\
   & \multicolumn{2}{c}{}   & \multicolumn{2}{c}{}  &       & \multicolumn{2}{c}{}  \\ \cline{2-5} \cline{7-8}
   & \textbf{M} & \textbf{MAD} & \textbf{M} & \textbf{MAD}  &       & \textbf{M} & \textbf{MAD}  \\ \cline{1-8}
   \multirow{5}{*}{1e4}     & \multirow{5}{*}{13.4} & \multirow{5}{*}{7.4} & \multirow{5}{*}{99.99} & \multirow{5}{*}{0.00}    & 4\%   & 82    & 18    \\ \cline{6-8}
   &        &       &       &       & 11\%  & 78    & 19      \\ \cline{6-8}
   &        &       &       &       & 25\%  & 77    & 20      \\ \cline{1-8}
   \multirow{5}{*}{1e5}     & \multirow{5}{*}{1.9}  & \multirow{5}{*}{2.0} & \multirow{5}{*}{99.90} & \multirow{5}{*}{0.00}    & 4\%   & 75    & 21    \\ \cline{6-8}
   &        &       &       &       & 11\%  & 68    & 23    \\ \cline{6-8}
   &        &       &       &       & 25\%  & 67    & 16    \\ \cline{1-8}
   \multirow{5}{*}{1e6}     & \multirow{5}{*}{0.81} & \multirow{5}{*}{0.51} & \multirow{5}{*}{99.00}& \multirow{5}{*}{0.01}    & 4\%   & 2.8   & 2.5   \\ \cline{6-8}
   &        &       &       &       & 11\%  & 12    & 10    \\ \cline{6-8}
   &        &       &       &       & 25\%  & 14.3  & 4.3   \\ \cline{1-8}
   \multirow{5}{*}{1e7}     & \multirow{5}{*}{0.19} & \multirow{5}{*}{0.17} & \multirow{5}{*}{90.00}& \multirow{5}{*}{0.03}    & 4\%   & 3.3   & 2.4   \\ \cline{6-8}
   &        &       &       &       & 11\%  & 3.6   & 2.0   \\ \cline{6-8}
   &        &       &       &       & 25\%  & 3.8   & 2.7   \\ \hline
\end{tabular}
\caption{Comparison of statistics for the residuals of the integrated SEDs, for the case of a dust shell with $\tau_{9.7} = 0.05$ and $\phi = 0^\circ$, for the different LPN realizations (columns 2 and 3), for those same realizations but after they have been de-normalized (columns 4 and 5), as described in Section \ref{sub:sub1}, and for the resulting emulations (columns 7 and 8), while using different sampling amounts (column 6) for the spatial reconstruction.}
\label{tab:intres}
\end{table}

From the residuals of the emulations integrated SEDs, shown in the lower panels of Fig. \ref{fig:intSED}, we conclude that: shorter wavelengths yield higher residuals; more input data for the spatial reconstruction does yield a better emulation but at the cost of an increased run time\footnote{INLAs reconstruction is as influenced by the amount of data it takes as input as by how that data is spatially distributed. Though on average having a larger uniform sample will be better than having a smaller uniform sample, it is possible that a particular smaller sample exists with a distribution that better captures the information of the field and that yields a better reconstruction. In the present case, we avoid this variability by using a regular sampling grid.}, as can be confirmed in Tab. \ref{tab:tab_emuls}; and, that the usage of the $N_p = 10^6$ realization as input greatly improves the quality of the emulation in relation to the two lowest photon number alternatives.\par
Looking at Tabs. \ref{tab:tab_sources} and \ref{tab:tab_emuls} we can also compare the overall performance of the pipeline at estimating information that was not available in the LPN input realizations. The amounts of differently classified spaxels in both LPN inputs and respective emulations show, together with median of the normalized residuals, that EmulART successfully estimates information missing from the input.\par

\begin{table}[ht]
\centering
\begin{tabular}{clcccccccc}
\hline
\multicolumn{2}{c}{\multirow{6}{*}{\textbf{\begin{tabular}[c]{@{}c@{}}LPN\\ Input\\ Sample\end{tabular}}}} & \multirow{6}{*}{\textbf{\begin{tabular}[c]{@{}c@{}}Total\\ Emul.\\ Time\\ (min)\end{tabular}}} & \multicolumn{2}{c}{\multirow{3}{*}{\textbf{\begin{tabular}[c]{@{}c@{}}Normalized\\ Residuals of\\ Emul. (\%)\end{tabular}}}} & \multicolumn{5}{c}{\multirow{3}{*}{\textbf{Information of Emulation}}} \\
\multicolumn{2}{c}{} &       & \multicolumn{2}{c}{} & \multicolumn{5}{c}{} \\
\multicolumn{2}{c}{} &       & \multicolumn{2}{c}{} & \multicolumn{5}{c}{} \\ \cline{4-10}
\multicolumn{2}{c}{} &       & \multirow{3}{*}{\textbf{M}}  & \multirow{3}{*}{\textbf{MAD}} & \multirow{3}{*}{\textbf{\begin{tabular}[c]{@{}c@{}}TIR\\ (\%)\end{tabular}}} & \multirow{3}{*}{\textbf{\begin{tabular}[c]{@{}c@{}}Null\\ Spxs\\ ($N$)\end{tabular}}} & \multirow{3}{*}{\textbf{\begin{tabular}[c]{@{}c@{}}Empty\\ Spxs\\ ($\Delta N$)\end{tabular}}} & \multirow{3}{*}{\textbf{\begin{tabular}[c]{@{}c@{}}Full\\ Spxs\\ ($N$)\end{tabular}}} & \multirow{3}{*}{\textbf{\begin{tabular}[c]{@{}c@{}}Partial\\ Spxs\\ ($N$)\end{tabular}}} \\
\multicolumn{2}{c}{} &       &       &       &       &       &       &       &       \\
\multicolumn{2}{c}{} &       &       &       &       &       &       &       &       \\ \cline{1-10}
\multicolumn{2}{c}{$10^4$ 4\%} & 7.2 & 77    & 33    & \multirow{5}{*}{98.3} & \multirow{5}{*}{20,712} & \multirow{5}{*}{2,856} & \multirow{5}{*}{69,288} & \multirow{5}{*}{0} \\ \cline{1-5}
\multicolumn{2}{c}{$10^4$ 11\%}& 33  & 76    & 35    &       &       &       &       &       \\ \cline{1-5}
\multicolumn{2}{c}{$10^4$ 25\%}& 20  & 76    & 35    &       &       &       &       &       \\ \cline{1-10}
\multicolumn{2}{c}{$10^5$ 4\%} & 6.9 & 76    & 36    & \multirow{5}{*}{102.3}& \multirow{5}{*}{17,858} & \multirow{5}{*}{2} & \multirow{5}{*}{72,142} & \multirow{5}{*}{0} \\ \cline{1-5}
\multicolumn{2}{c}{$10^5$ 11\%}& 22  & 75    & 37    &       &       &       &       &       \\ \cline{1-5}
\multicolumn{2}{c}{$10^5$ 25\%}& 19  & 196   & 272   &       &       &       &       &       \\ \cline{1-10}
\multicolumn{2}{c}{$10^6$ 4\%} & 8.9 & 42    & 45    & \multirow{5}{*}{103.6}& \multirow{5}{*}{16,969} & \multirow{5}{*}{-887} & \multirow{5}{*}{73,031} & \multirow{5}{*}{0} \\ \cline{1-5}
\multicolumn{2}{c}{$10^6$ 11\%}& 25  & 40    & 45    &       &       &       &       &       \\ \cline{1-5}
\multicolumn{2}{c}{$10^6$ 25\%}& 26  & 48    & 45    &       &       &       &       &       \\ \cline{1-10}
\multicolumn{2}{c}{$10^7$ 4\%} & 7.6 & 31    & 32    & \multirow{5}{*}{103.7}& \multirow{5}{*}{16,866} & \multirow{5}{*}{-990} & \multirow{5}{*}{73,134} & \multirow{5}{*}{0} \\ \cline{1-5}
\multicolumn{2}{c}{$10^7$ 11\%}& 31  & 32    & 33    &       &       &       &       &       \\ \cline{1-5}
\multicolumn{2}{c}{$10^7$ 25\%}& 19  & 26    & 26    &       &       &       &       &       \\ \hline
\end{tabular}
\caption{Statistics regarding how the different emulations, of the case of a dust shell with $\tau_{9.7} = 0.05$ and $\phi = 0^\circ$, compared to the HPN reference. The median normalized residuals and TIR display great improvement against all LPN realizations (see Tab. \ref{tab:tab_sources}). The emulations that used $N_p = 10^6$ and $N_p = 10^7$ realizations as input display a negative amount of empty spaxels, indicating that beyond reconstructing all empty spaxels it also inferred some of the references nulls.}
\label{tab:tab_emuls}
\end{table}

\begin{figure}[ht]
    \centering
    \begin{subfigure}[b]{.3\textwidth}
        \centering
        \includegraphics[width =\textwidth]{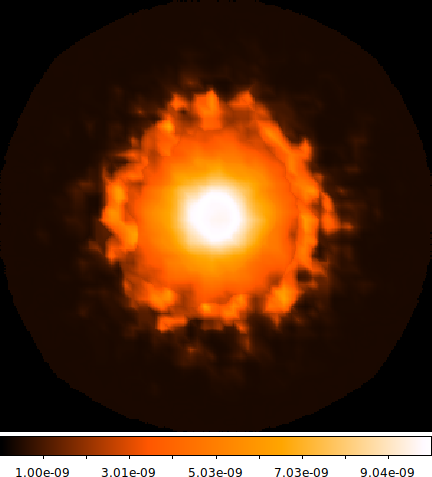}
        \caption{}
        \label{fig:lpn6}
    \end{subfigure}
    \begin{subfigure}[b]{.3\textwidth}
        \centering
        \includegraphics[width =\textwidth]{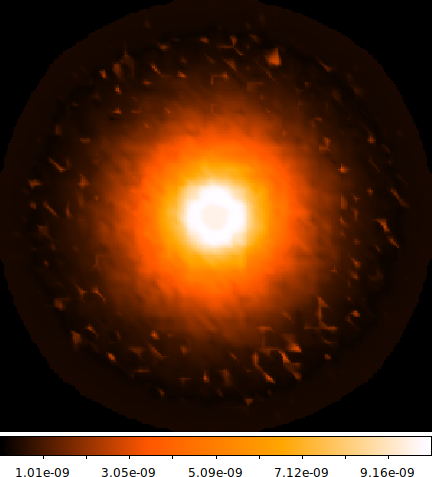}
        \caption{}
        \label{fig:lpn7}
    \end{subfigure}
        \begin{subfigure}[b]{.3\textwidth}
        \centering
        \includegraphics[width =\textwidth]{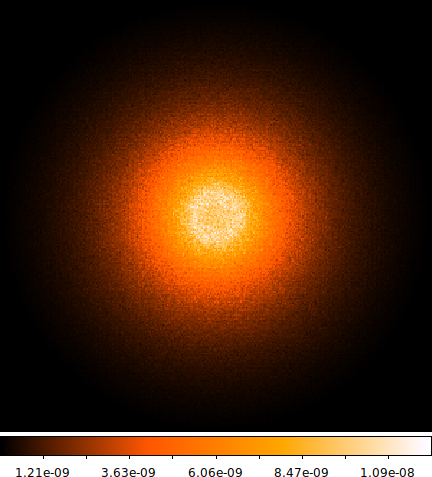}
        \caption{}
        \label{fig:hpn}
    \end{subfigure}
    \caption{Spatial maps of emulations, of the case of a dust shell with $\tau_{9.7} = 0.05$ and $\phi = 0^\circ$, at wavelength 9.28 $\mu$m, based on the input of 4\% of $N_p = 10^6$ LPN realization spatial information (Fig. \ref{fig:lpn6}), 4\% of $N_p = 10^7$ LPN realization spatial information (Fig. \ref{fig:lpn7}) and the HPN reference (Fig. \ref{fig:hpn}). Color indicates flux density in W/m$^2$.}
    \label{fig:emul}
\end{figure}

Fig. \ref{fig:emul} shows the comparison, at wavelength 9.28 $\mu$m, between the emulations resulting from the LPN inputs (see Fig. \ref{fig:obj}) and the HPN reference. Once again we can see that with the $N_p = 10^6$ realization the emulations start to display resemblances to the HPN reference not only in the range of flux density values but also in the morphology that emerges from their distribution.\par

From Tabs. \ref{tab:intres} and \ref{tab:tab_emuls}, as well as from Figs. \ref{fig:intSED} and \ref{fig:emul}, it would be natural to conclude that using the $N_p = 10^7$ realization as input would bring the most benefit in terms of the amount of information inferred by the emulation as well as its accuracy. Moreover, as can be seen in the second column of Tab. \ref{tab:tab_emuls}, the run time of the emulations is more dependent on the amount of information sampled for the spatial reconstruction than on the $N_p$ of the LPN input. Nevertheless, considering how SKIRTs run time for a model scales with the simulated $N_p$, the choice of LPN input to use with EmulART should weight the time it takes to produce that LPN input as well as the quality of the emulation we expect from it.\par

\subsubsection{PCA Pipeline Predictions}\label{sec:pcares}

The results of testing the PCA pipelines on AESet were processed in the same way as EmulARTs (statistical indicators regarding residuals of the emulation and of its spatial integration were calculated as well as the TIR; the number of different types of spaxels and execution time were measured).\par
The execution times\footnote{These do not include the time it took to train the PCA models. Training time for each PCA model was below 10 seconds, while training the DVAE took around 18 hours ($\sim 6500\times$ more). This was expected due to the methods themselves as well as the differences between training procedures. The DVAE model was trained with 6$\times$ the amount of data as each PCA model, with the purpose of being able to generalize across simulations resulting from different photon packet numbers; while each PCA model was trained on LPN simulation that was used as input for emulation, to compared against the results of EmulART.} for the pipeline implementing an 8 PC model were indiscernible from the execution times of EmulART on the same input, while the execution time for the pipeline implementing a PCA model with number of components, K, determined by the elbow method was in general much higher since for all except one of the LPN inputs K was larger than 30 (the time per component map was very similar depending mostly on the amount of spatial information sampled).\par
\begin{figure}[ht]
    \centering
    \begin{subfigure}[b]{.3\textwidth}
        \centering
        \includegraphics[width =\textwidth]{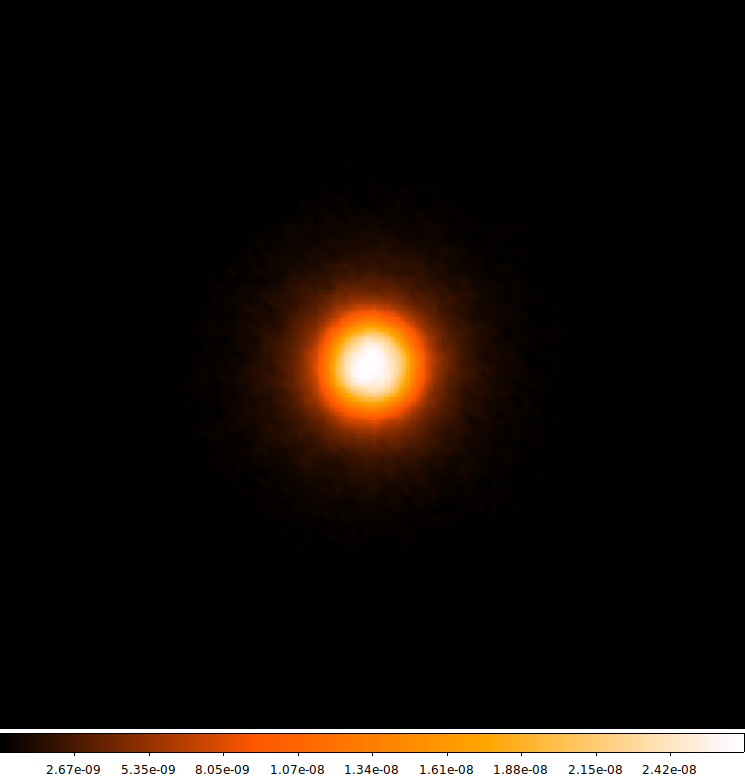}
        \caption{}
        \label{fig:pca10}
        \vspace{3mm}
    \end{subfigure}
    \begin{subfigure}[b]{.3\textwidth}
        \centering
        \includegraphics[width =\textwidth]{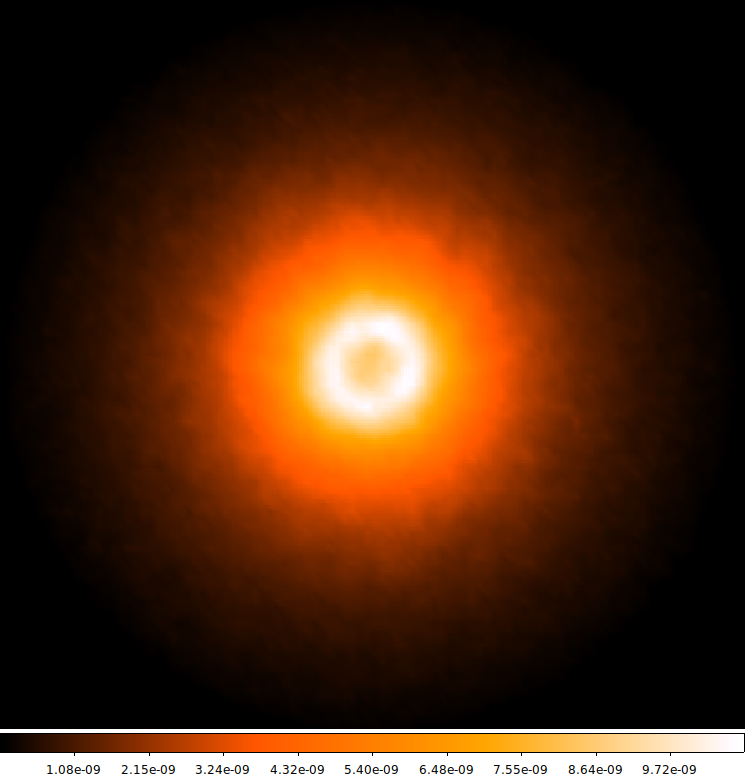}
        \caption{}
        \label{fig:pca31}
        \vspace{3mm}
    \end{subfigure}
        \begin{subfigure}[b]{.3\textwidth}
        \centering
        \includegraphics[width =\textwidth]{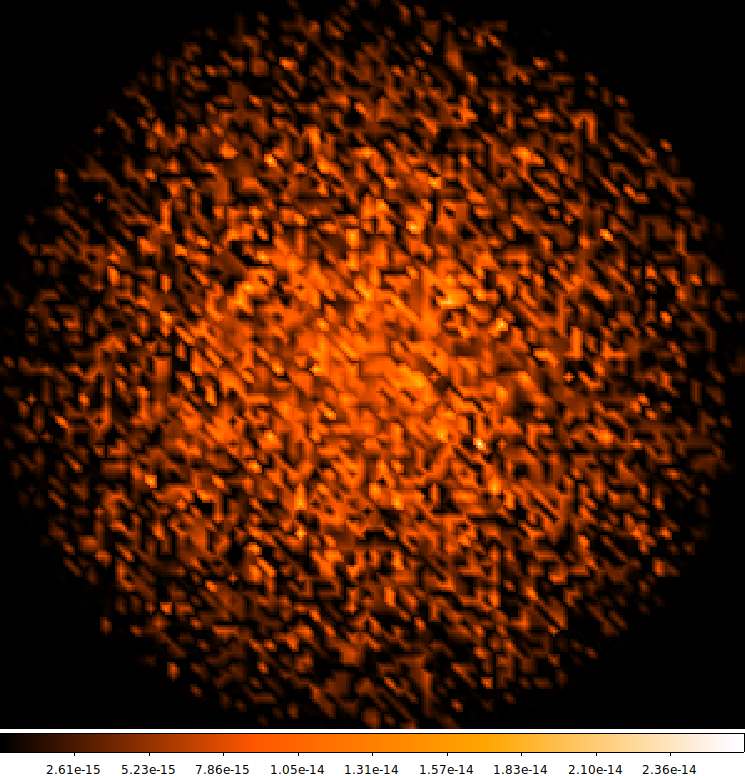}
        \caption{}
        \label{fig:pca55}
        \vspace{3mm}
    \end{subfigure}
        \begin{subfigure}[b]{.3\textwidth}
        \centering
        \includegraphics[width =\textwidth]{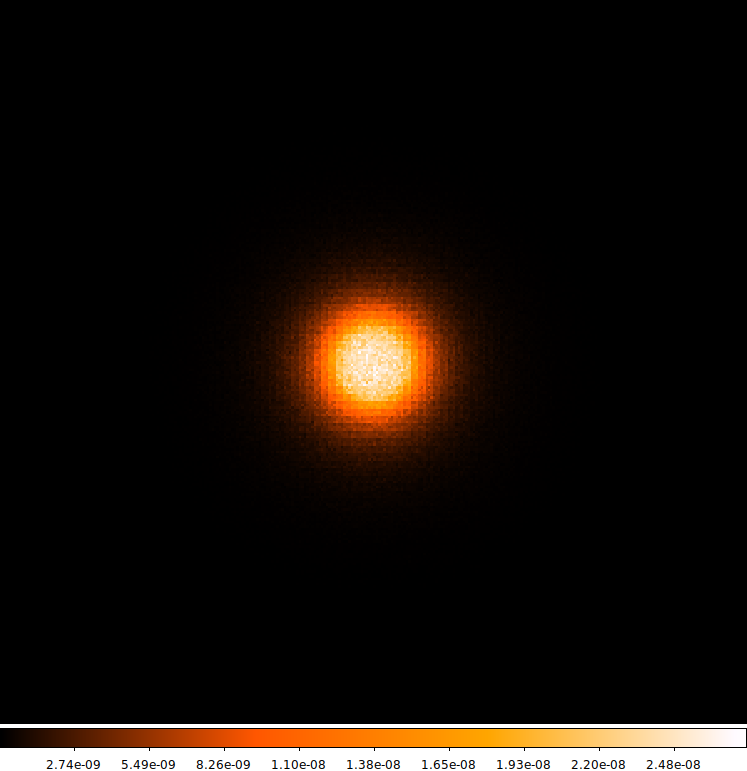}
        \caption{}
        \label{fig:ref10}
    \end{subfigure}
    \begin{subfigure}[b]{.3\textwidth}
        \centering
        \includegraphics[width =\textwidth]{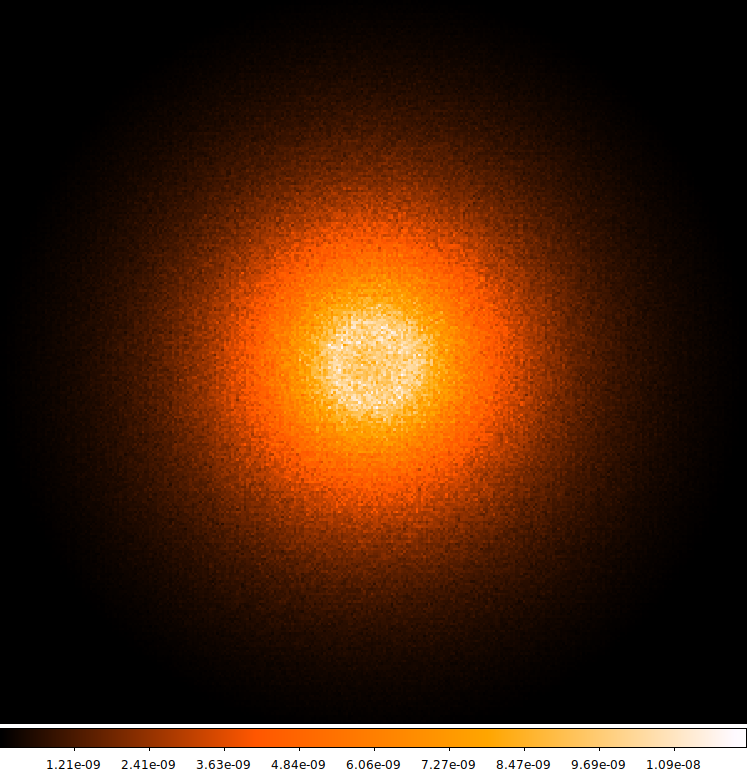}
        \caption{}
        \label{fig:ref31}
    \end{subfigure}
        \begin{subfigure}[b]{.3\textwidth}
        \centering
        \includegraphics[width =\textwidth]{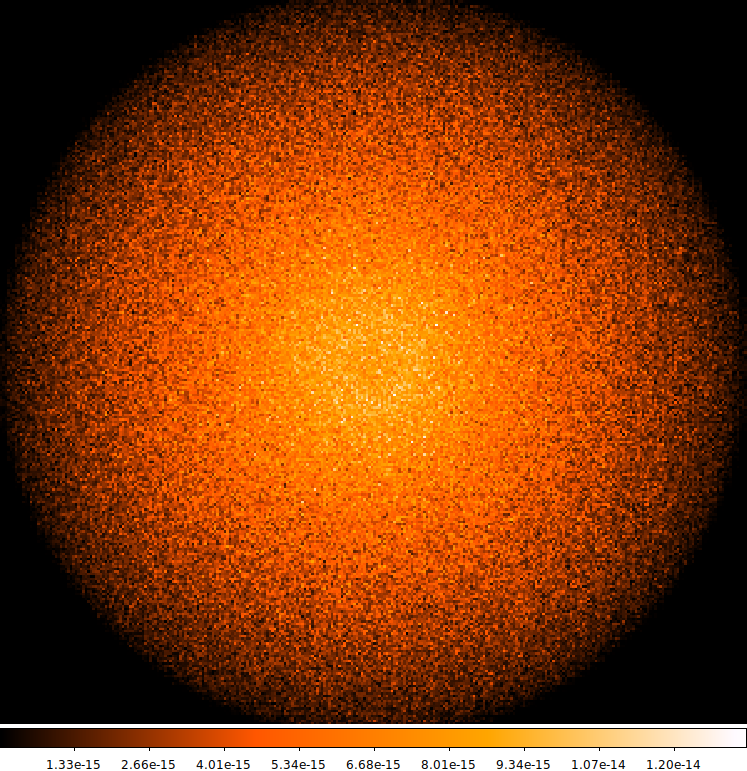}
        \caption{}
        \label{fig:ref55}
    \end{subfigure}
    \caption{Spatial maps of emulations (Top), performed with a pipeline including a PCA model with K = 30 components based on the input of 11\% of the spatial information of the $N_p = 10^7$ LPN realization; and, reference simulation (Bottom) of the case of a dust shell with $\tau_{9.7} = 0.05$ and $\phi = 0^\circ$, at wavelengths 1.85 $\mu$m  (Figs. \ref{fig:pca10} and \ref{fig:ref10}), 9.28 $\mu$m (Figs. \ref{fig:pca31} and \ref{fig:ref31}) and 211.35 $\mu$m (Figs. \ref{fig:pca55} and \ref{fig:ref55}). Color indicates flux density in W/m$^2$.}
    \label{fig:emulpca}
\end{figure}

Tabs. \ref{tab:tab_emuls_pca1} and \ref{tab:tab_emuls_pca2} present the most significant indicators to be compared to EmulARTs. At first sight the performance of both PCA models on the emulation pipeline seems to be similar to, and in some cases even better than that of EmulARTs, showing statistically similar residuals, and presenting residuals regarding the spatially integrated SEDs ~2 to 3 times lower. They are however unable to consistently recover complete spaxel information which then leads to poor spatial reconstructions at longer wavelengths, even when using K $>$ 8 (see Fig. \ref{fig:emulpca}), as can be inferred from the TIR values (as well as the number of full and partial spaxels).\par

\begin{table}[ht]
\centering
\begin{tabular}{cclcccccccc}
\hline
\multirow{6}{*}{\textbf{\begin{tabular}[c]{@{}c@{}}Input\end{tabular}}} &
\multirow{6}{*}{\textbf{\begin{tabular}[c]{@{}c@{}}K\end{tabular}}} &
\multirow{3}{*}{\textbf{\begin{tabular}[c]{@{}c@{}}Normalized\\ Residuals of\\ Emul. (\%)\end{tabular}}} &
\multicolumn{2}{c}{\multirow{3}{*}{\textbf{\begin{tabular}[c]{@{}c@{}}Residuals of\\ Integrated\\ Emul. (\%)\end{tabular}}}} & \multicolumn{5}{c}{\multirow{3}{*}{\textbf{Information of Emulation}}} \\
    &       & \multicolumn{2}{c}{} & \multicolumn{5}{c}{}   \\
    &       &       & \multicolumn{2}{c}{} & \multicolumn{5}{c}{}   \\ \cline{3-10} 
    &       & \multirow{3}{*}{\textbf{\begin{tabular}[c]{@{}c@{}}M\end{tabular}}}   &   \multirow{3}{*}{\textbf{M}} & \multirow{3}{*}{\textbf{MAD}} &   \multirow{3}{*}{\textbf{\begin{tabular}[c]{@{}c@{}}TIR\\ (\%)\end{tabular}}} & \multirow{3}{*}{\textbf{\begin{tabular}[c]{@{}c@{}}Null\\ Spxs\\ ($N$)\end{tabular}}} & \multirow{3}{*}{\textbf{\begin{tabular}[c]{@{}c@{}}Empty\\ Spxs\\ ($\Delta N$)\end{tabular}}} & \multirow{3}{*}{\textbf{\begin{tabular}[c]{@{}c@{}}Full\\ Spxs\\ ($N$)\end{tabular}}} & \multirow{3}{*}{\textbf{\begin{tabular}[c]{@{}c@{}}Partial\\ Spxs\\ ($N$)\end{tabular}}} \\
    &       &       &       &       &       &       &       &       &       \\
    &       &       &       &       &       &       &       &       &       \\ \cline{1-10}
    $0^\circ$, $10^4$   & 55    & 100   & 63    & 54    & 53    & 20712 & 2856  & 0     & 69288 \\ \cline{1-10}
    $0^\circ$, $10^5$   & 64    & 99    & 13    & 13    & 70    & 17858 & 2     & 10589 & 61553 \\ \cline{1-10}
    $0^\circ$, $10^6$   & 48    & 99    & 3     & 3     & 77    & 17051 & -805  & 2458  & 70491 \\  \cline{1-10}
    $0^\circ$, $10^7$   & 31    & 52    & 1     & 2     & 86    & 16866 & -990  & 3820  & 69314 \\ \cline{1-10}
    $90^\circ$, $10^4$  & 8     & 97    & 25    & 44    & 66    & 20497 & 2641  & 0     & 69503 \\ \cline{1-10}
    $90^\circ$, $10^5$  & 55    & 100   & 18    & 19    & 60    & 18002 & 146   & 0     & 71998 \\ \cline{1-10}
    $90^\circ$, $10^6$  & 48    & 98    & 3     & 4     & 75    & 16872 & -984  & 1322  & 71806 \\ \cline{1-10}
    $90^\circ$, $10^7$  & 40    & 48    & 1     & 2     & 90    & 16833 & -1023 & 9346  & 63821 \\ \hline
\end{tabular}
\caption{Statistics regarding PCA pipeline emulations, using K PCs (where K was determined by the elbow method) of the case of a dust shell with $\tau_{9.7} = 0.05$, $\phi \in \{0^\circ, 90^\circ\}$ and spatial sampling of 11\%.}
\label{tab:tab_emuls_pca1}
\end{table}

\begin{table}[ht]
\centering
\begin{tabular}{clcccccccc}
\hline
\multirow{6}{*}{\textbf{\begin{tabular}[c]{@{}c@{}}Input\end{tabular}}} &
\multirow{3}{*}{\textbf{\begin{tabular}[c]{@{}c@{}}Normalized\\ Residuals of\\ Emul. (\%)\end{tabular}}} &
\multicolumn{2}{c}{\multirow{3}{*}{\textbf{\begin{tabular}[c]{@{}c@{}}Residuals of\\ Integrated\\ Emul. (\%)\end{tabular}}}}    & \multicolumn{5}{c}{\multirow{3}{*}{\textbf{Information of Emulation}}} \\
    & \multicolumn{2}{c}{} & \multicolumn{5}{c}{} \\
    &       & \multicolumn{2}{c}{} & \multicolumn{5}{c}{} \\ \cline{2-9} 
    & \multirow{3}{*}{\textbf{\begin{tabular}[c]{@{}c@{}}M\end{tabular}}} & \multirow{3}{*}{\textbf{M}} & \multirow{3}{*}{\textbf{MAD}}& \multirow{3}{*}{\textbf{\begin{tabular}[c]{@{}c@{}}TIR\\ (\%)\end{tabular}}} & \multirow{3}{*}{\textbf{\begin{tabular}[c]{@{}c@{}}Null\\ Spxs\\ ($N$)\end{tabular}}} & \multirow{3}{*}{\textbf{\begin{tabular}[c]{@{}c@{}}Empty\\ Spxs\\ ($\Delta N$)\end{tabular}}} & \multirow{3}{*}{\textbf{\begin{tabular}[c]{@{}c@{}}Full\\ Spxs\\ ($N$)\end{tabular}}} & \multirow{3}{*}{\textbf{\begin{tabular}[c]{@{}c@{}}Partial\\ Spxs\\ ($N$)\end{tabular}}} \\
    &       &       &       &       &       &       &       &       \\
    &       &       &       &       &       &       &       &       \\ \cline{1-9}
    $0^\circ$, $10^4$   & 96    & 24    & 52    & 73     & 20712    & 2856  & 0     & 69288 \\ \cline{1-9}
    $0^\circ$, $10^5$   & 100   & 22    & 15    & 75     & 17858    & 2     & 3127  & 69015 \\ \cline{1-9}
    $0^\circ$, $10^6$   & 64    & 4     & 6     & 84     & 17051    & -805  & 2352  & 70597 \\ \cline{1-9}
    $0^\circ$, $10^7$   & 33    & 1     & 2     & 88     & 16866    & -990  & 3122  & 70012 \\ \cline{1-9}
    $90^\circ$, $10^4$  & 97    & 25    & 44    & 66     & 20497    & 2641  & 0     & 69503 \\ \cline{1-9}
    $90^\circ$, $10^5$  & 100   & 20    & 13    & 70     & 18002    & 146   & 151   & 71847 \\ \cline{1-9}
    $90^\circ$, $10^6$  & 55    & 2     & 5     & 85     & 16872    & -984  & 2884  & 70244 \\ \cline{1-9}
    $90^\circ$, $10^7$  & 30    & 1     & 2     & 91     & 16833    & -1023 & 6845  & 66322 \\ \hline
\end{tabular}
\caption{Statistics regarding PCA pipeline emulations, using 8 PCs, of the case of a dust shell with $\tau_{9.7} = 0.05$, $\phi \in \{0^\circ, 90^\circ\}$ and spatial sampling of 11\%.}
\label{tab:tab_emuls_pca2}
\end{table}

In the present application we thus find the implementation of a DVAE model for spectral compression to be justified. Unlike the PCA models, it not only captures non-linear relationships between the spectral features but also achieves comparable results. Despite some loss regarding the reconstruction of integrated SED profile, when compared to PCA models, it achieves equal/higher compression rate, lower/equal execution time and most importantly the spatial structure can successfully be reconstructed by the remaining parts of the pipeline.\par
As such, the test results of EmulART on EVASet are compared against its results on AESet. Testing results obtained with PCA emulation pipelines on EVASet did not offer a perspective different from the one above and as such will not be discussed further here (a successful implementation of PCA in an emulation pipeline is described in \cite{majda}).\par


\subsection{EVASet Predictions}\label{sec:evares}
In this section we present and discuss the results obtained by emulating the HPN references of EVASet using the respective LPN realizations within it. The DVAE model was not trained on any data within this set, which allows us to evaluate whether it manages to accurately predict HPN-like spaxels from LPN spaxels that originate from simulations with different $\tau_{9.7}$ and $\phi$ values.\par
First we tested EmulART on realizations with $\tau_{9.7} = 0.05$ and $\phi = 90^\circ$; we then applied the pipeline to different LPN realizations with $\tau_{9.7} \in \{0.1, 1.0\}$ and $\phi \in \{0^\circ, 90^\circ\}$.\par
\begin{figure}[ht]
\centering
    \begin{subfigure}[b]{.45\textwidth}
        \centering
        \includegraphics[width =\textwidth]{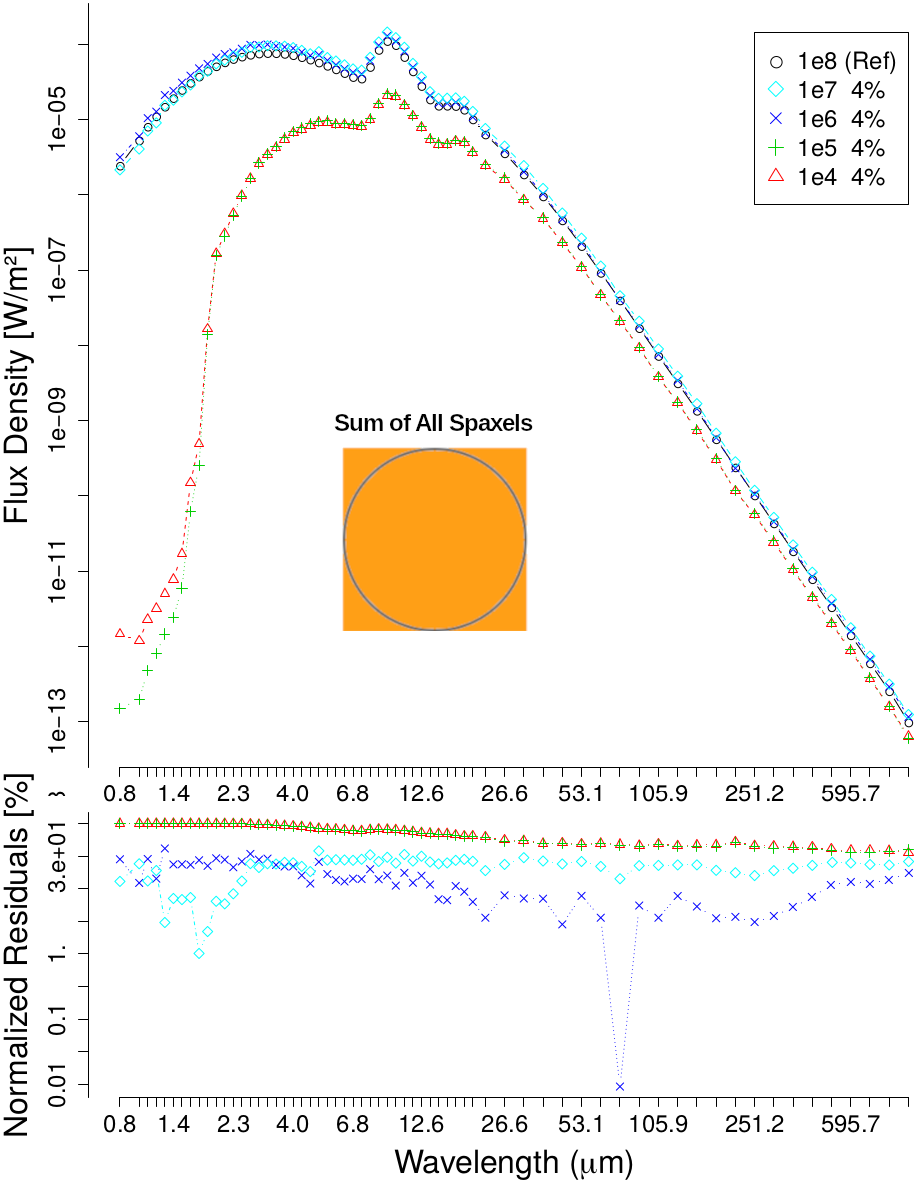}
        \caption{}
        \label{fig:int4ed}
        \vspace{3mm}
    \end{subfigure}
    \hspace{5mm}
    \begin{subfigure}[b]{.45\textwidth}
        \centering
        \includegraphics[width =\textwidth]{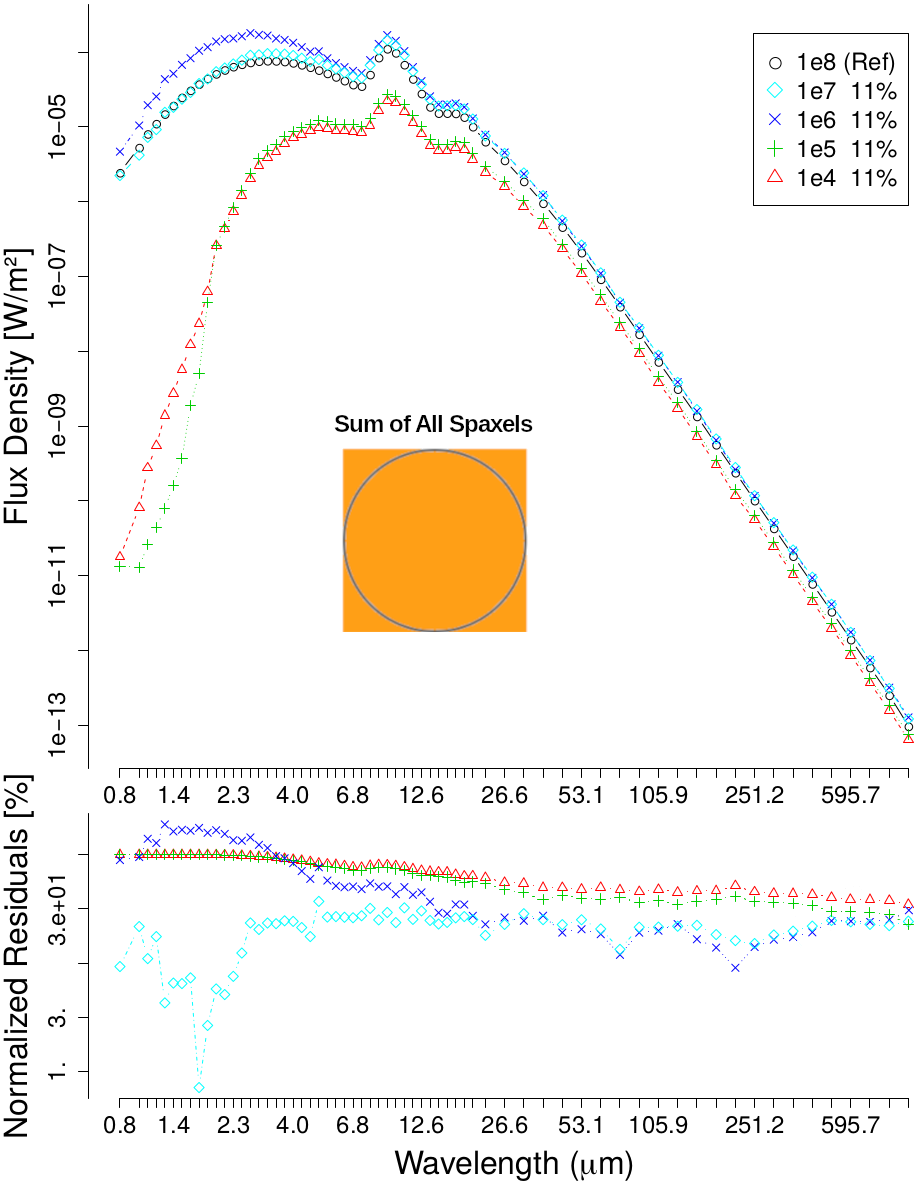}
        \caption{}
        \label{fig:int11ed}
        \vspace{3mm}
    \end{subfigure}
    \begin{subfigure}[b]{.45\textwidth}
        \centering
        \includegraphics[width =\textwidth]{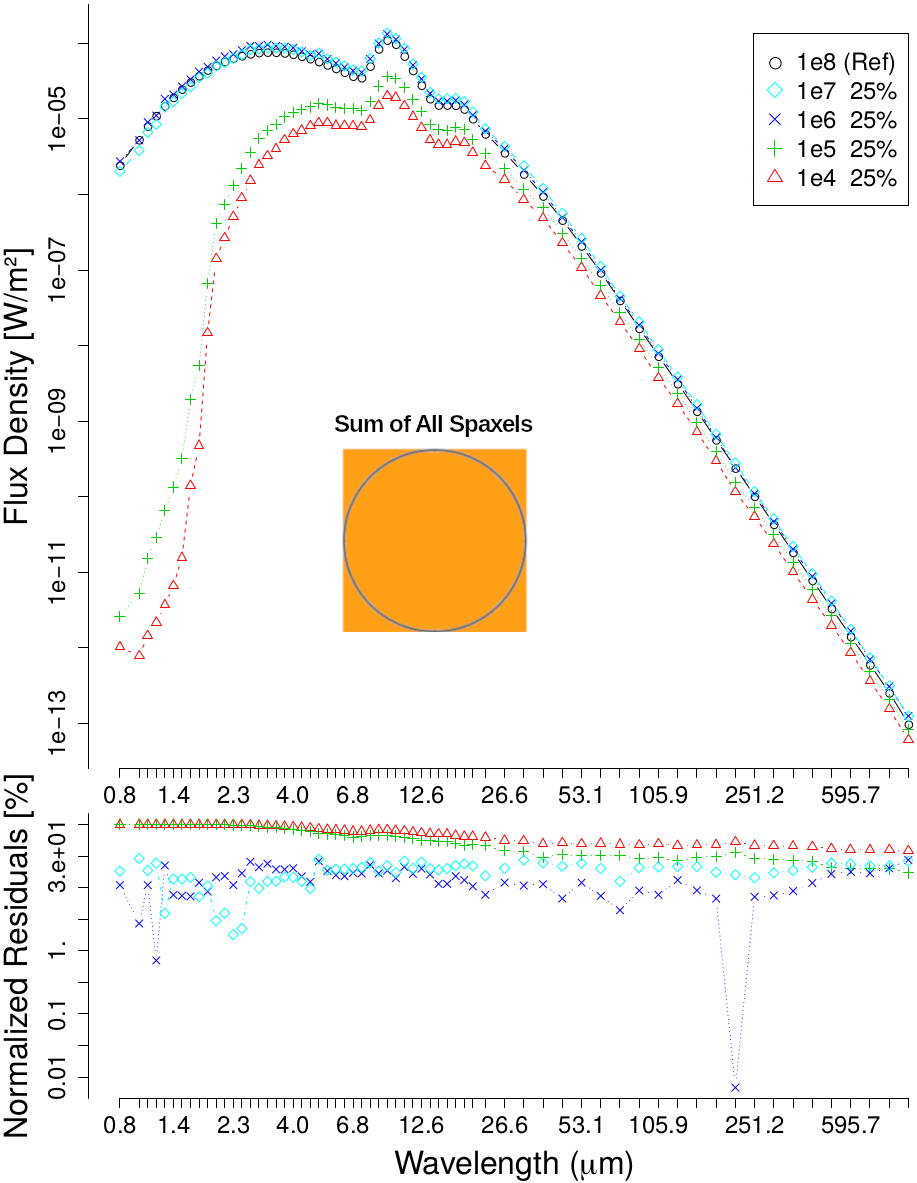}
        \caption{}
        \label{fig:int25ed}
    \end{subfigure}
    \caption{Emulations integrated SEDs, resulting from spatial inference using 4\% (\ref{fig:int4ed}), 11\% (\ref{fig:int11ed}) and 25\% (\ref{fig:int25ed}) samples of the spatial information, and the respective normalized residuals, for the case of a dust shell with $\tau_{9.7} = 0.05$ and $\phi = 90^\circ$. The HPN reference is represented in black ($\circ$), the emulation based on the $N_p = 10^4$ realization is in red ($\triangle$), on the $N_p = 10^5$ in green (+), on the $N_p = 10^6$ in blue ($\times$) and on the $N_p = 10^7$ in cyan ($\diamond$) .}
    \label{fig:intSEDed005}
\end{figure}

Similarly to the $\tau_{9.7} = 0.05$ and $\phi = 0^\circ$ emulations, the edge-on, $\phi = 90^\circ$, emulations appear to preserve well the spectral information, reproducing the slow rise in the 1 $\mu$m to 8 $\mu$m range, the two emission bumps in the 8 $\mu$m to 20 $\mu$m range, and the steep decline towards longer wavelengths, as can be seen in Fig. \ref{fig:intSEDed005}. Tab. \ref{tab:intres2} shows that 4\% sampling of the spatial information of the $N_p = 10^6$ realization is enough to get median integrated residuals below 15\%. We note however the abnormal performance of the emulations that took as input the $N_p = 10^7$ realizations, displaying higher median integrated residuals than the ones that used different samplings of the $N_p = 10^6$ LPN. This may indicate that one or more of the spatial data manipulation modules, or their interface, should be improved upon.\par

\begin{table}[ht]
\centering
\begin{tabular}{cccccccc}
\hline
\multirow{4}{*}{\textbf{\begin{tabular}[c]{@{}c@{}}LPN\\ Input\end{tabular}}} & \multicolumn{2}{c}{\multirow{3}{*}{\textbf{\begin{tabular}[c]{@{}c@{}}Residuals of \\ Integrated\\ Input (\%)\end{tabular}}}} & \multicolumn{2}{c}{\multirow{3}{*}{\textbf{\begin{tabular}[c]{@{}c@{}}Residuals of\\ Integrated\\ De-normalized Input (\%)\end{tabular}}}} & \multirow{4}{*}{\textbf{Sample}} & \multicolumn{2}{c}{\multirow{3}{*}{\textbf{\begin{tabular}[c]{@{}c@{}}Residuals of\\ Integrated\\ Emul. (\%)\end{tabular}}}} \\
    & \multicolumn{2}{c}{} & \multicolumn{2}{c}{} &     & \multicolumn{2}{c}{} \\
    & \multicolumn{2}{c}{} & \multicolumn{2}{c}{} &     & \multicolumn{2}{c}{} \\ \cline{2-5} \cline{7-8} 
    & \textbf{M} & \textbf{MAD} & \textbf{M} & \textbf{MAD} &       & \textbf{M} & \textbf{MAD} \\ \cline{1-8}
\multirow{5}{*}{1e4} & \multirow{5}{*}{13.4} & \multirow{5}{*}{7.4} & \multirow{5}{*}{99.99} & \multirow{5}{*}{0.00}& 4\%  & 77    & 19    \\ \cline{6-8}
    &       &       &       &       & 11\%  & 16    & 19    \\ \cline{6-8} 
    &       &       &       &       & 25\%  & 78    & 19    \\ \cline{1-8}
\multirow{5}{*}{1e5} & \multirow{5}{*}{1.9} & \multirow{5}{*}{2.0} & \multirow{5}{*}{99.90} & \multirow{5}{*}{0.00}& 4\%  & 76    & 20    \\ \cline{6-8} 
    &       &       &       &       & 11\%  & 71    & 23    \\ \cline{6-8} 
    &       &       &       &       & 25\%  & 63    & 25    \\ \cline{1-8}
\multirow{5}{*}{1e6}& \multirow{5}{*}{0.83} & \multirow{5}{*}{0.51} & \multirow{5}{*}{99.00} & \multirow{5}{*}{0.01} & 4\% & 13.2  & 7.4   \\ \cline{6-8}
    &       &       &       &       & 11\%  & 46    & 41    \\ \cline{6-8} 
    &       &       &       &       & 25\%  & 13.1  & 5.0   \\ \cline{1-8}
\multirow{5}{*}{1e7} & \multirow{5}{*}{0.19} & \multirow{5}{*}{0.17} & \multirow{5}{*}{90.00} & \multirow{5}{*}{0.03}& 4\%  & 23.7  & 5.7   \\ \cline{6-8}
    &       &       &       &       & 11\%  & 22.7  & 5.7   \\ \cline{6-8}
    &       &       &       &       & 25\%  & 19.6  & 4.9   \\ \hline
\end{tabular}
\caption{Comparison of statistics for the residuals of the spatial integration SEDs, for the case of a dust shell with $\tau_{9.7} = 0.05$ and $\phi = 90^\circ$, for the different LPN realizations (columns 2 and 3), for those same realizations but after they have been de-normalized (columns 4 and 5), as described in Section \ref{sub:sub1}, and for the resulting emulations (columns 7 and 8).}
\label{tab:intres2}
\end{table}

\begin{figure}[ht]
\centering
    \begin{subfigure}[b]{.45\textwidth}
        \centering
        \includegraphics[width =\textwidth]{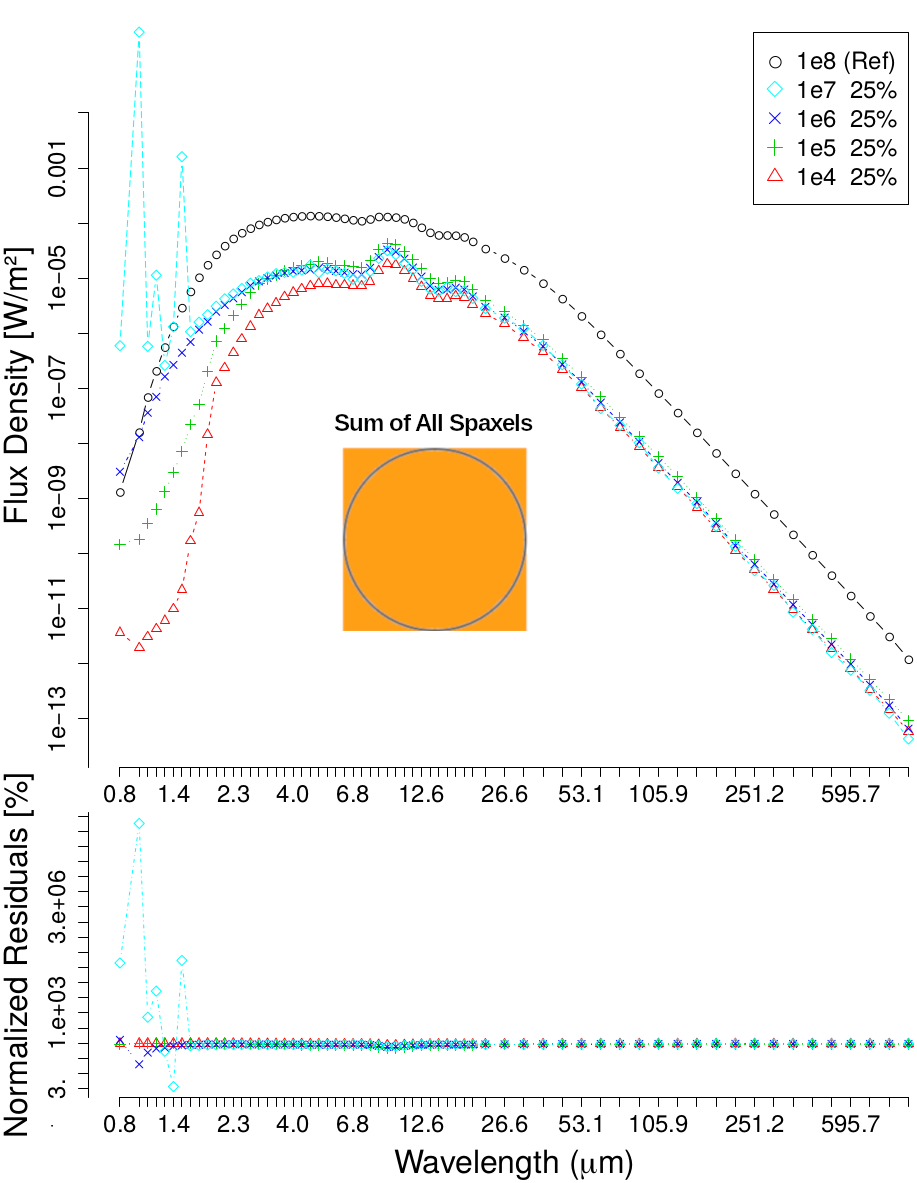}
        \caption{$\phi = 90^\circ$}
        \label{fig:10ed25}
    \end{subfigure}
    \hspace{5mm}
    \begin{subfigure}[b]{.45\textwidth}
        \centering
        \includegraphics[width =\textwidth]{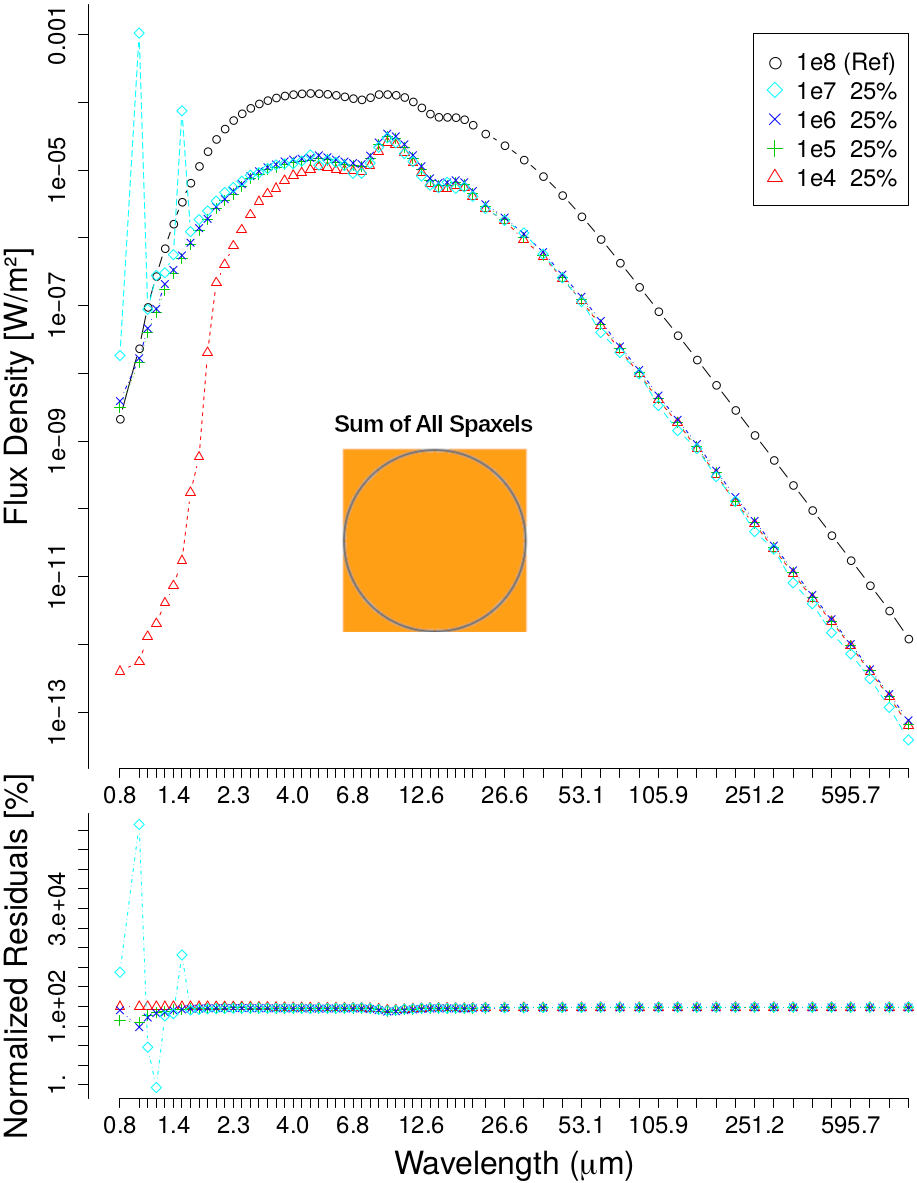}
        \caption{$\phi = 0^\circ$}
        \label{fig:10fa25}
    \end{subfigure}
    \caption{Emulations integrated SEDs, resulting from spatial inference using 25\% of spatial information, and the respective normalized residuals, for the case of a dust shell with $\tau_{9.7} = 1.0$, $\phi = 90^\circ$ (Fig. \ref{fig:10ed25}) and $\phi = 0^\circ$ (Fig. \ref{fig:10fa25}). The HPN reference is represented in black ($\circ$), the emulation based on the $N_p = 10^4$ realization is in red ($\triangle$), on the $N_p = 10^5$ in green (+), on the $N_p = 10^6$ in blue ($\times$) and on the $N_p = 10^7$ in cyan ($\diamond$) .}
    \label{fig:intSED10}
\end{figure}

As for the emulations using as input LPN realizations of simulations with $\tau_{9.7} \in \{0.1, 1.0\}$, at both tilt angles, we observe that both the shape and flux density value range of the integrated SED degrades as $\tau_{9.7}$ increases. As shown in Fig. \ref{fig:intSED10}, the emulations integrated SEDs fail to reproduce the shape of the HPN references, reproducing instead the shape that characterized the realizations present in AESet, a clear sign of over-fitting of the DVAE model. This can be solved by expanding the training set of our DVAE architecture to include spaxels originating from simulations with different optical depths.\par

\begin{figure}[ht]
\centering
    \begin{subfigure}[b]{\textwidth}
    \includegraphics[width =\textwidth]{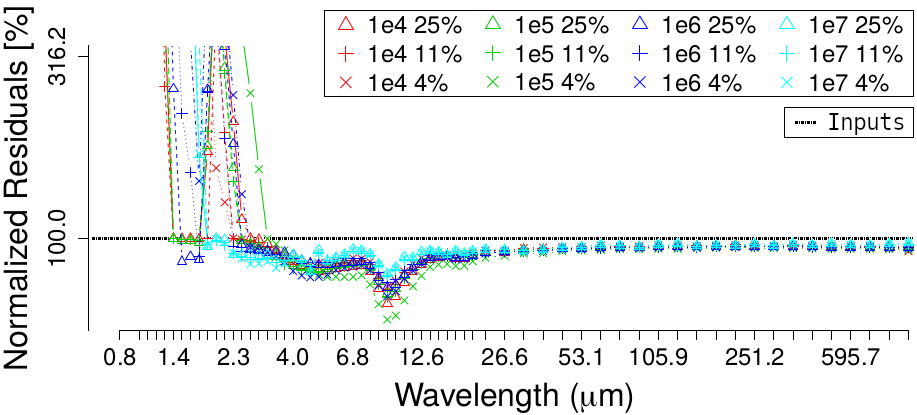}
    \caption{$\phi = 0^\circ$}
    \label{fig:sp_res_face10}
    \vspace{3mm}
    \end{subfigure}
    \begin{subfigure}[b]{\textwidth}
    \includegraphics[width =\textwidth]{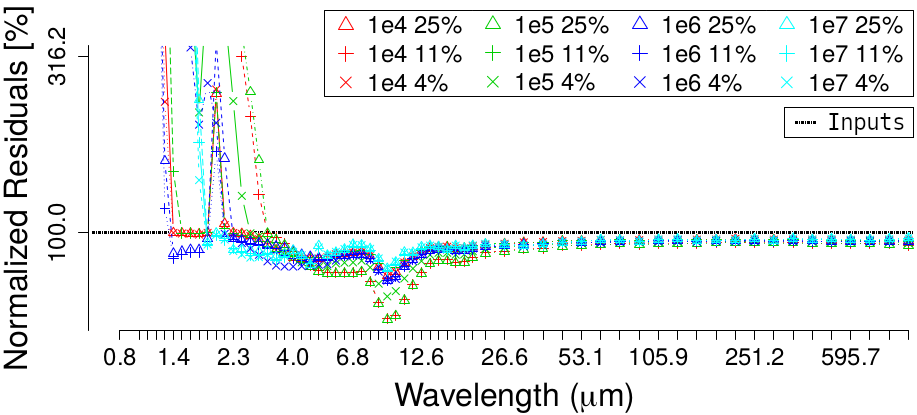}
    \caption{$\phi = 90^\circ$}
    \label{fig:sp_res_edge10}
    \end{subfigure}
\caption{Median of the normalized residuals, at each wavelengths spatial map, for every emulation obtained with $N_p = 10^4$ (red), $N_p = 10^5$ (green), $N_p = 10^6$ (blue) and $N_p = 10^7$ (cyan) realizations, for the case of a dust shell with $\tau_{9.7}=1.0$, with $\phi = 0^\circ$ (Fig. \ref{fig:sp_res_face10}) and $\phi = 90^\circ$ (Fig. \ref{fig:sp_res_edge10}). Emulations whose spatial inference was performed using 25\% of data are represented by ($\triangle$), 11\% by (+) and 4\% by ($\times$). The interrupted black line marks the same metric for the LPN inputs.}
\label{fig:sp_res_10}
\end{figure}

\begin{figure}[ht]
\centering
    \begin{subfigure}[b]{\textwidth}
        \centering
        \includegraphics[width =\textwidth]{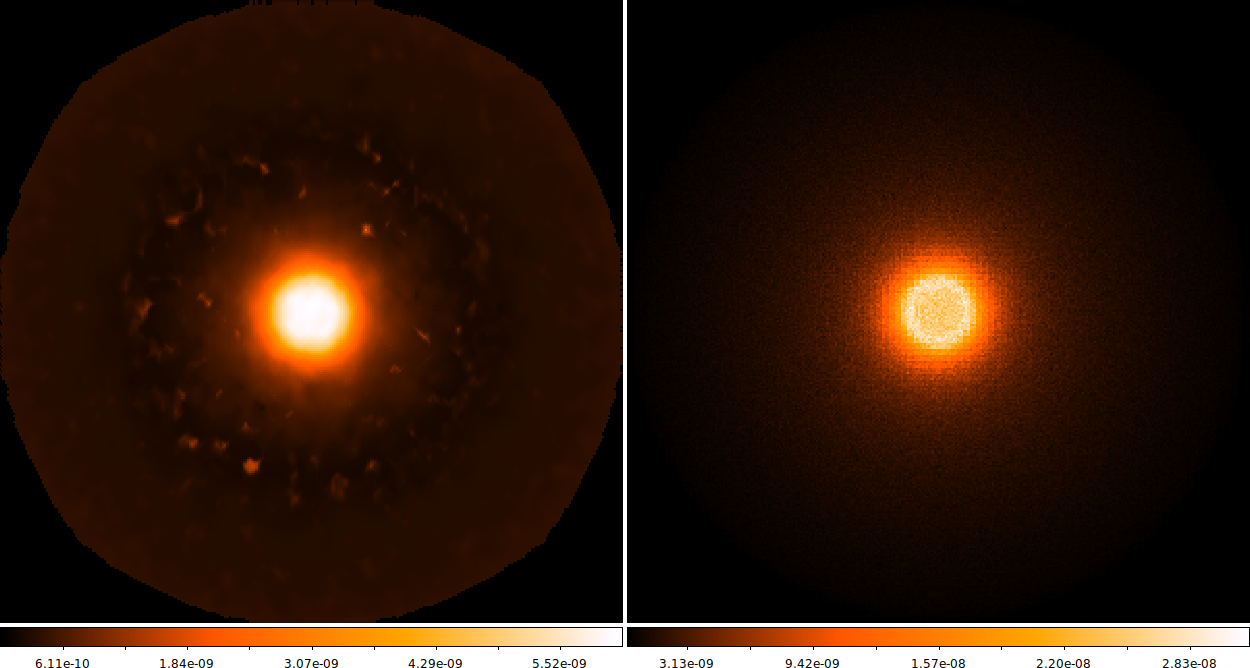}
        \caption{$\phi = 0^\circ$}
        \label{fig:face10}
    \vspace{3mm}
    \end{subfigure}
    \begin{subfigure}[b]{\textwidth}
        \centering
        \includegraphics[width =\textwidth]{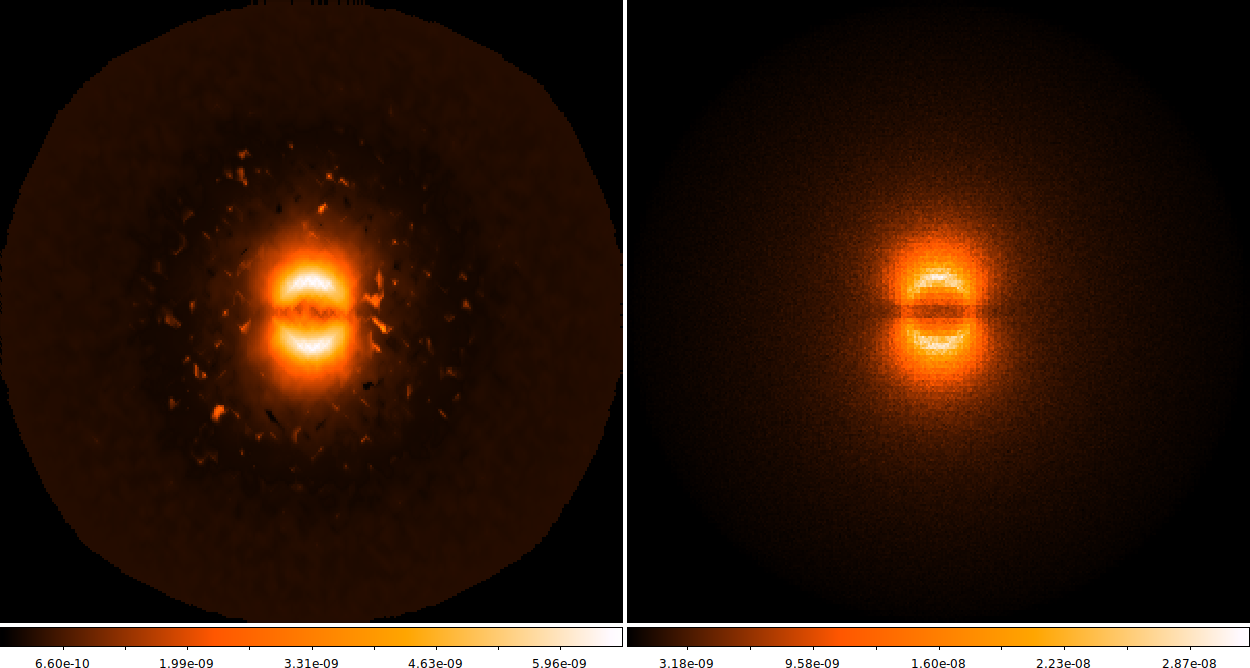}
        \caption{$\phi = 90^\circ$}
        \label{fig:edge10}
    \end{subfigure}
\caption{Emulation produced using as input a 25\% sample of the $N_p = 10^6$ realization (left panel) and HPN reference (right panel) at wavelength 9.28 $\mu$m, for the case of a dust shell with $\tau_{9.7} = 1.0$, $\phi = 0^\circ$ (Fig. \ref{fig:face10}) and $\phi = 90^\circ$ (Fig. \ref{fig:edge10}).}
\label{fig:928_10}
\end{figure}

For $\tau_{9.7} = 1.0$, with both $\phi$ cases, we observe the influence of the first wavelengths (see Fig. \ref{fig:sp_res_10}) in the overall residuals\footnote{For more details consult the data products available at repository.}. Fig. \ref{fig:928_10} shows that though the overall morphology of the spatial distribution is well recovered the value range for the emulations flux density value range is drastically underestimated while the contrast between the central and peripheral regions is considerably higher than what the HPN references display.\par

\subsection*{}
These results appear to show that our pipeline is capable of recovering $40\%$ to $60\%$ of the emergent spatial information of HPN MCRT models from LPN realizations, taking as input as little as 0.04\% of the information that would be present in the HPN model, all while preserving $85\%$ to $95\%$ of the spectral information.\par
Furthermore the results also show a clear bias in the performance of the DVAE model as a compressor and decompressor of spectral information, with the performance degrading substantially as the LPN inputs models depart from the optical depth, $\tau_{9.7}$, value present within the training set.\par
Further details of the results we obtained with AESet and EVASet are discussed in Appendix \ref{app:E}.\par

\section{Summary}\label{sec:conc}
We report the development of a pipeline that implements in conjunction a denoising variational autoencoder (or alternatively PCA), and a reconstruction method based on approximate Bayesian inference, INLA, with the purpose of emulating high information-like MCRT simulations using LPN MCRT models, created with SKIRT, as input. With this approach we aim for the hastily expansion of libraries of synthetic models against which to compare future observations. By producing positive preliminary indicators we show that such a framework is worth pursuing further, with multiple alleys to explore.\par
Conditions for systematically measuring the computational cost are necessary to properly evaluate the merit of this approach. However, in this work our aim was to qualitatively assess the potential of this method to be applied to MCRT simulations images. In this pilot study we chose a very simple model of a centrally-illuminated spherical dust shell which is computationally inexpensive, whose reference simulations took around two hours to be computed. Thus, in our particular examples we reduced the computational time by approximately 6$\times$. Nevertheless, the computational cost of SKIRT simulations scales with the amount of photon packets simulated, the spatial resolution of the grid and the actual geometrical complexity of the model. As for our emulation pipeline, its computational cost is only impacted noticeably when increasing the size and density of the spatial grid to be processed by INLA. This leads us to believe that a generalized version of this pipeline may expedite, by up to 50$\times$, the study of dust environments through this kind of radiative transfer models.\par
Further exploration of the proposed DVAE architecture is being undertaken, via expansion and diversification of the training set to improve the prediction of the spectral features.\par
Other approaches to be tested include the use of dropout (cutting the connections, mid training, whose weights are below a given threshold) and incremental learning (training a model with new data with the starting point of the network being the weights obtained in a previous training session). The first serves as feature selection tool, removing those of little importance, which also prevents the model from over-fitting the training data; the second would be of great use to quickly adapt an already trained model to new data, which is important in the context of emulating simulations.\par
To improve the reconstruction of the spatial features, using non-uniform sampling methods based on the information spatial density may help improve the reconstruction of the latent features that result from the compression of models simulated with insufficient number of photon ($N_p < 10^6$, in the particular case of our study). Alternative pre-INLA data pre-processing, such as data value range manipulations and sampling grids, may also be worth exploring.\par

\backmatter

\bmhead{Acknowledgments}\label{sec:ack}
Computations were performed at the cluster “Baltasar-Sete-S\'ois” and supported by the H2020 ERC Consolidator Grant ``Matter and strong field gravity: New frontiers in Einstein’s theory" grant agreement no. MaGRaTh-646597.\par
J. R.-S. is funded by Funda\c{c}\~ao para a Ci\^encia e a Tecnologia (\href{FCT}{https://www.fct.pt/}) (PD/BD/150487/2019), via the International Doctorate Network in Particle Physics, Astrophysics and Cosmology (\href{IDPASC}{https://idpasc.lip.pt/}). The work here described is integrated in the CRISP project, which is also funded by the FCT (PTDC/FIS-AST/31546/2017). M.S. and M.S. acknowledge support by the Ministry of Education, Science and Technological Development of the Republic of Serbia through the contract no. 451-03-9/2021-14/200002 and by the Science Fund of the Republic of Serbia, PROMIS 6060916, BOWIE. \par

\section*{Declarations}

The authors declare no conflict of interest. The founding sponsors had no role in the design of the study; in the collection, analyses, or interpretation of data; in the writing of the manuscript, and in the decision to publish the results.

\begin{appendices}

\section{Data Pre-processing}\label{app:A}

\subsection{Identifying and Sampling Null-Spaxels}\label{app:A1}
The SKIRT simulations of the spherical simulated models result in a peripheral region filled with spaxels with value 0 at all wavelength bins, these will be henceforth be referred to as null-spaxels. We perform two routines to identify and filter out null-spaxels, to curb their potential impact in both training the DVAE model and when (within the emulation procedure) reconstructing the spatial information of the latent feature maps:

\begin{enumerate}
    \item Each feature (wavelength) of each spaxel is checked, if all features of a spaxel have value 0, then that spaxel is temporarily tagged as null-spaxel;
    \item An iterative process then checks the immediate neighborhood of each previously tagged spaxel (a 5pix$\times$5pix grid centered at the spaxel being tested) for other tagged spaxels, if a minimum of 4 neighbors share the null-spaxel tag, then that spaxel is tagged as a true null-spaxel, otherwise its temporary tag is removed.
\end{enumerate}

\begin{figure}[ht]
    \centering
    \includegraphics[width =\textwidth]{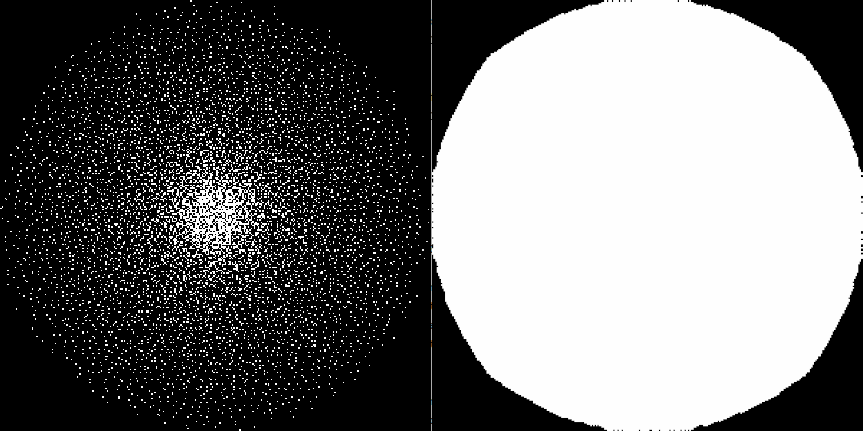}
    \caption{On the left, in black, the spaxels temporarily tagged as null-spaxels by routine 1; on the right, also in black, the final selection of null-spaxels that resulted from applying routine 2 with a search grid of 5pix$\times$5pix around each temporary null-spaxel. The result on the right shows that for this particular case of a spherical shell of dust surrounding an isotropic emitter we only find null-spaxels on the outskirts of the shell, as expected.}
    \label{fig:zeros}
\end{figure}

The second routine serves as a way to discriminate spaxels that have zero information at given wavelengths, possibly from the spaxel being the result of a low photon count simulation, from actual null-spaxels - spaxels resulting from the lack of photons traveling from the source to the observer through that direction, neither by direct emission nor by scattering nor by absorption and re-emission by dust. Fig. \ref{fig:zeros} illustrates the results of routine 1 and 2.\par

After identifying the null-spaxels we sample them to incorporate the filtered dataset.\par
For the DVAE training, complete removal of these null-spaxels could be done but we decided to reduce their incidence rate instead. Since there is no guarantee that their latent representation (where the spatial reconstruction will take place) will also be a null-vector we want the DVAE model to learn null-spaxels, too many of these will nevertheless be uninformative and may lead to unintended biases. For this reason we sample 10\% of them, reducing the null-spaxel prevalence in the training dataset from $\sim$25\% to $\sim$2.5\%. Empty spaxels (spaxels tagged as null for LPN realizations but non-null for the respective HPN reference) were completely removed from training.\par
For the emulation process, and more specifically for the spatial reconstruction of the latent feature maps, we sample 10\% of the null-spaxels selected by the spatial sampling grid, for INLA is faster with little data and that amount is sufficient for the null-region to be properly recovered\footnote{Different simulated objects with different distributions of regions with 0 emission may need a higher sampling of null-spaxels.}. After the emulation process, spaxels that were at this stage identified as being null-spaxels have their values changed to 0. 

\subsection{Reshaping Data Structures}\label{app:A2}
Once all spaxels are selected the data cubes are reshaped into a 2D structure, with a given pixel corresponding to a row with its spectral information on the columns (see Fig. \ref{fig:pipe}), to match the input format of the DVAE. The inverse reshaping, from a 2D to a 3D structure, is performed after the DVAE output layer.\par
Within the emulation pipeline the data is reshaped two more times. First, after sampling the latent feature distributions (to obtain the $Z$ arrays, as seen in Fig. \ref{fig:pipe}) the resulting sample is reshaped from a 2D to a 3D structure to obtain latent feature spatial maps (the 8 1D arrays become 8 2D maps) whose spatial information can then be reconstructed with INLA. The second reshape is performed on INLAs reconstructed results to conform that data to the decoder accepted input format, once again going from a 3D to a 2D structure.\par

\newpage

\section{Dataset}\label{app:B}
Tab. \ref{tab:cubes} presents the list of SKIRT simulations used in this work. Every realization models a spherical dust, composed by silicates and graphites, surrounding a bright point source with anisotropic emission, differing in parameters such as the tilt angle, $\phi$, the optical depth, $\tau_{9.7}$, and the amount of photon packets simulated, $N_p$.\par

\begin{table}[ht]
\centering
\begin{tabular}{ccccc}
\hline
\textbf{$N$} & \textbf{$\tau_{9.7}$} & \textbf{$\phi$ (deg)} & \textbf{$N_p$} & \textbf{Subset} \\ \hline
1       & 0.05      & 0     & $10^4$    & AESet \\
2       & 0.05      & 0     & $10^5$    & AESet \\
3       & 0.05      & 0     & $10^6$    & AESet \\
4       & 0.05      & 0     & $10^7$    & AESet \\
5*      & 0.05      & 0     & $10^8$    & AESet \\ \hline
6       & 0.05      & 90    & $10^4$    & EVASet\\
7       & 0.05      & 90    & $10^5$    & EVASet\\
8       & 0.05      & 90    & $10^6$    & EVASet\\
9       & 0.05      & 90    & $10^7$    & EVASet\\
10*     & 0.05      & 90    & $10^8$    & EVASet\\ \hline
11      & 0.1       & 0     & $10^4$    & EVASet\\
12      & 0.1       & 0     & $10^5$    & EVASet\\
13      & 0.1       & 0     & $10^6$    & EVASet\\
14      & 0.1       & 0     & $10^7$    & EVASet\\
15*     & 0.1       & 0     & $10^8$    & EVASet\\ \hline
16      & 0.1       & 90    & $10^4$    & EVASet\\
17      & 0.1       & 90    & $10^5$    & EVASet\\
18      & 0.1       & 90    & $10^6$    & EVASet\\
19      & 0.1       & 90    & $10^7$    & EVASet\\
20*     & 0.1       & 90    & $10^8$    & EVASet\\ \hline
21      & 1.0       & 0     & $10^4$    & EVASet\\
22      & 1.0       & 0     & $10^5$    & EVASet\\
23      & 1.0       & 0     & $10^6$    & EVASet\\
24      & 1.0       & 0     & $10^7$    & EVASet\\
25*     & 1.0       & 0     & $10^8$    & EVASet\\ \hline
26      & 1.0       & 90    & $10^4$    & EVASet\\
27      & 1.0       & 90    & $10^5$    & EVASet\\
28      & 1.0       & 90    & $10^6$    & EVASet\\
29      & 1.0       & 90    & $10^7$    & EVASet\\
30*     & 1.0       & 90    & $10^8$    & EVASet\\ \hline     
\end{tabular}
\caption{Parametrical description of the SKIRT realizations used in this work, along with the distribution of these realizations by subset. Realizations marked with (*) were not parsed as input to EmulART while testing the pipeline, being used merely as the HPN reference for the purpose of computing performance metrics.}
\label{tab:cubes}
\end{table}

\newpage
\setcounter{figure}{0}
\section{Latent Features}\label{app:C}

In this section we briefly discuss the relationship between each of the compressed features, as well as how they may relate to some physical attributes of the simulated models.\par
The PCC measures how strongly linearly correlated different parameters are, and is defined as the ratio between the covariance of the parameters and the product of their standard deviations (see Eq. \ref{eq:pcc}):\par

\begin{equation}\label{eq:pcc}
    \rho_{X,Y} = \frac{cov(X,Y)}{\sigma_X\sigma_Y}\,\, .
\end{equation}

\begin{figure}[ht]
    \centering
    \includegraphics[height =.85\textheight]{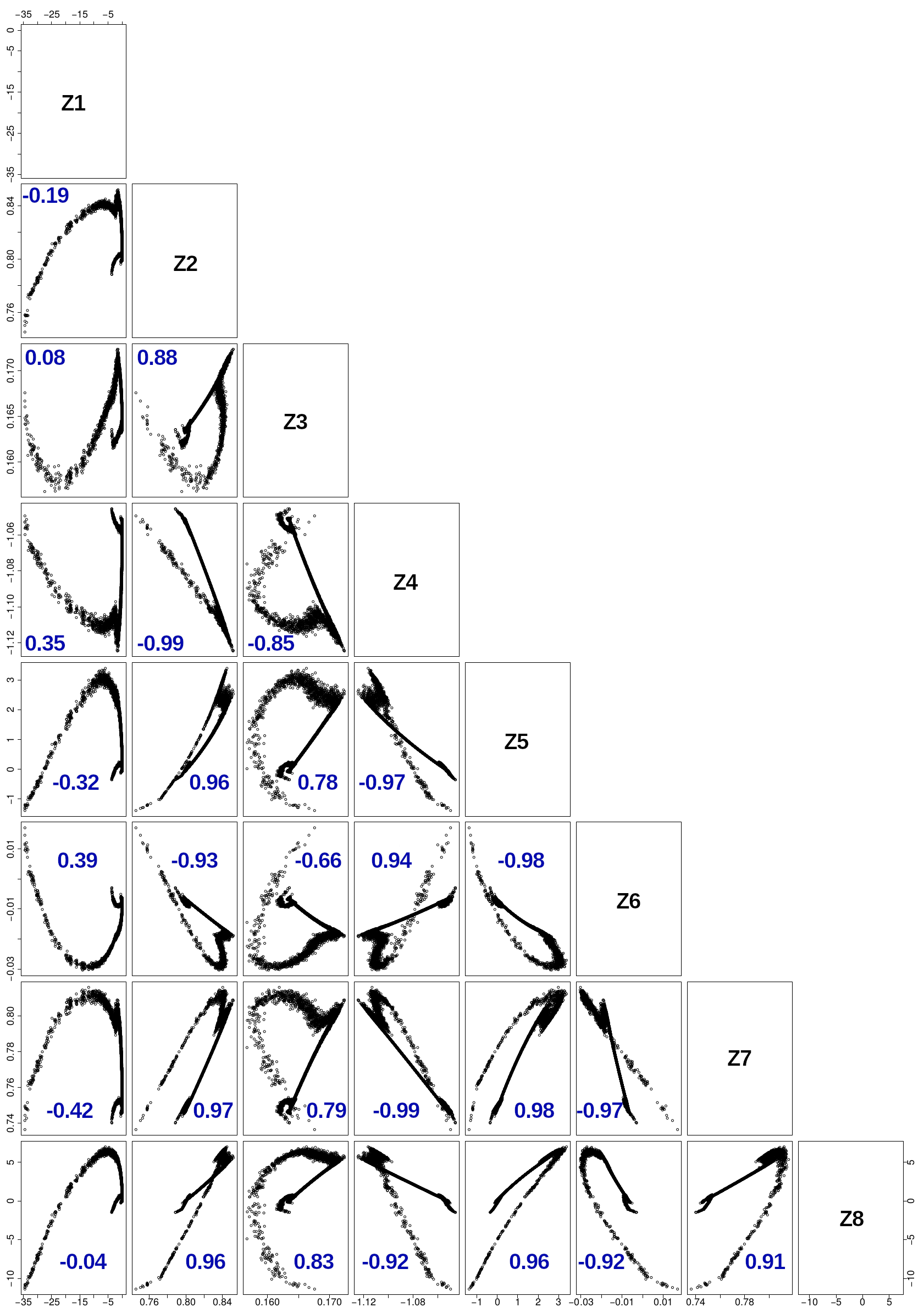}
    \caption{Corner plot of the latent features, $Z_f$ of the test set. Pearson's correlation coefficients are indicated in navy blue.}
    \label{fig:corner}
\end{figure}

Fig. \ref{fig:corner} illustrates the relationships between our datasets latent features along with their respective PCC. The PCC indicates that only the first feature shows little correlation with the others, ranging from -0.42 to 0.39, while the remaining features display high correlations (and anti-correlations) between themselves, most either above 0.9 or below -0.8. This would indicate that the information in those features presents a high level of redundancy.\par
The point spread in the panels of Fig. \ref{fig:corner} shows, however, something unexpected given the PCC values. There are clearly distinct populations, some presenting linear behavior, while others show highly non-linear relationships (see, for example, Z3 vs Z5) and even dual linear populations (see Z6 vs Z7).\par
The overall large Pearson correlation values thus indicate that the population density within the linear spreads is higher. But even if the compressed features show at best linear correlations (and anti-correlations) for a greater fraction of our data, we decided not to ignore the non-linear behavior being expressed for the remaining data. In future implementations we will tackle this by both expanding the dataset, including models with different dust distributions and compositions and employing different training strategies (this can be achieved by changing loss function terms and adding different training stages).\par

\begin{figure}[ht]
\centering
    \begin{subfigure}[b]{.30\textwidth}
        \centering
        \includegraphics[width =\textwidth]{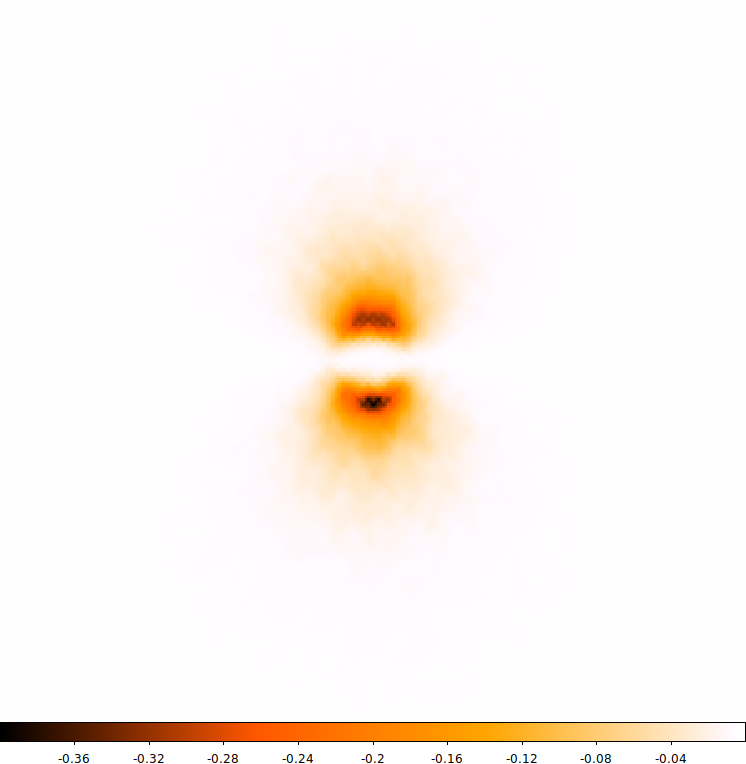}
        \subcaption{Z1}
        \label{fig:e5f1}
    \end{subfigure}
    \begin{subfigure}[b]{.30\textwidth}
        \centering
        \includegraphics[width =\textwidth]{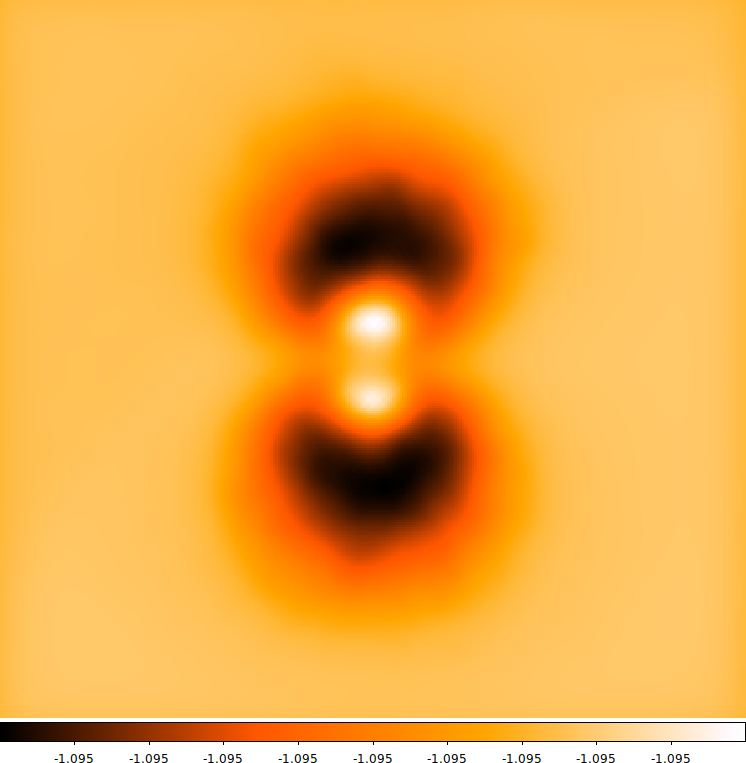}
        \subcaption{Z2}
        \label{fig:e5f2}
    \end{subfigure}
    \begin{subfigure}[b]{.30\textwidth}
        \centering
        \includegraphics[width =\textwidth]{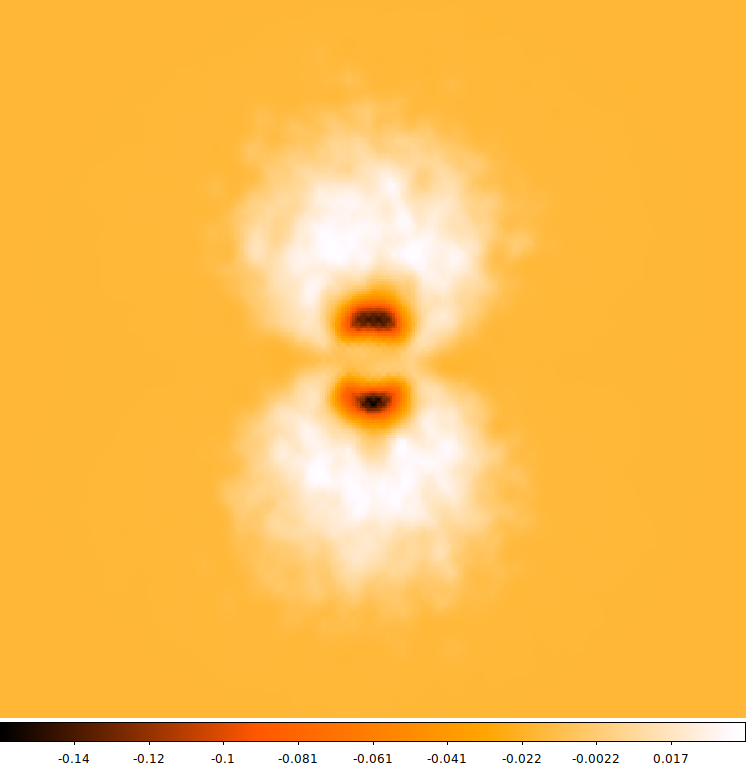}
        \subcaption{Z8}
        \label{fig:e5f8}
    \end{subfigure}
    \caption{Spatial reconstruction of 3 latent feature maps of the LPN case $N_p = 10^6$, $\tau_{9.7}=0.05$ and $\phi = 90^\circ$, using 25\% of the available data.}
    \label{fig:ed5lat}
\end{figure}

Fig. \ref{fig:ed5lat} shows the spatial reconstruction of 3 latent feature maps\footnote{Though our pipeline yields 8 latent features for every spaxel, we chose to only show these 3 maps since the remaining latent feature maps either were very similar to these or no other information could be extracted from them.} obtained from the $N_p = 10^6$ LPN realization of the case of a dust shell with $\tau_{9.7}=0.05$ and $\phi = 90^\circ$.\par
In the map of latent feature Z1 (Fig. \ref{fig:e5f1}) we can relate the spatial distribution to the hot dust emission, at wavelengths 1-3$\mu$m, of the region close to the sublimation zone around the central source. Conversely, in the map of latent feature Z2 (Fig. \ref{fig:e5f2}), the larger structure further away form the center, which corresponds to the colder dust emission at longer wavelengths is outlined.\par

As mentioned in Section \ref{sec:dataset} the model used along this work is that of a bright point like source with anisotropic emission \citep{marko}, where the emission is defined by Eq. \ref{eq:geo_sig}, following \cite{netzer}. Looking at the map of latent feature Z8 (Fig. \ref{fig:e5f8}) the geometric signature of the model is clearly identifiable.\par 

\newpage
\setcounter{figure}{0}
\setcounter{table}{0}
\section{Emulation Flowcharts}\label{app:D}

This section presents flowcharts for different emulation pipelines used in this work. In these, circles and ellipses represent data cubes at different stages of the pipeline and round cornered squares present lists of indexes. Grey rectangle boxes are operations using the data, purple diamond boxes indicate set operations ($\cap$ for intersection and $\cup$ for union) on indexes, green boxes indicate operations related to dimensionality transformation models. $\blacklozenge$ refers to the de-normalization procedure mentioned in Section \ref{sub:sub1} and $\dagger$ refers to operations described in Section \ref{sec:pipeDVAE}.

\begin{figure}[ht]
    \centering
    \includegraphics[height = .85\textheight]{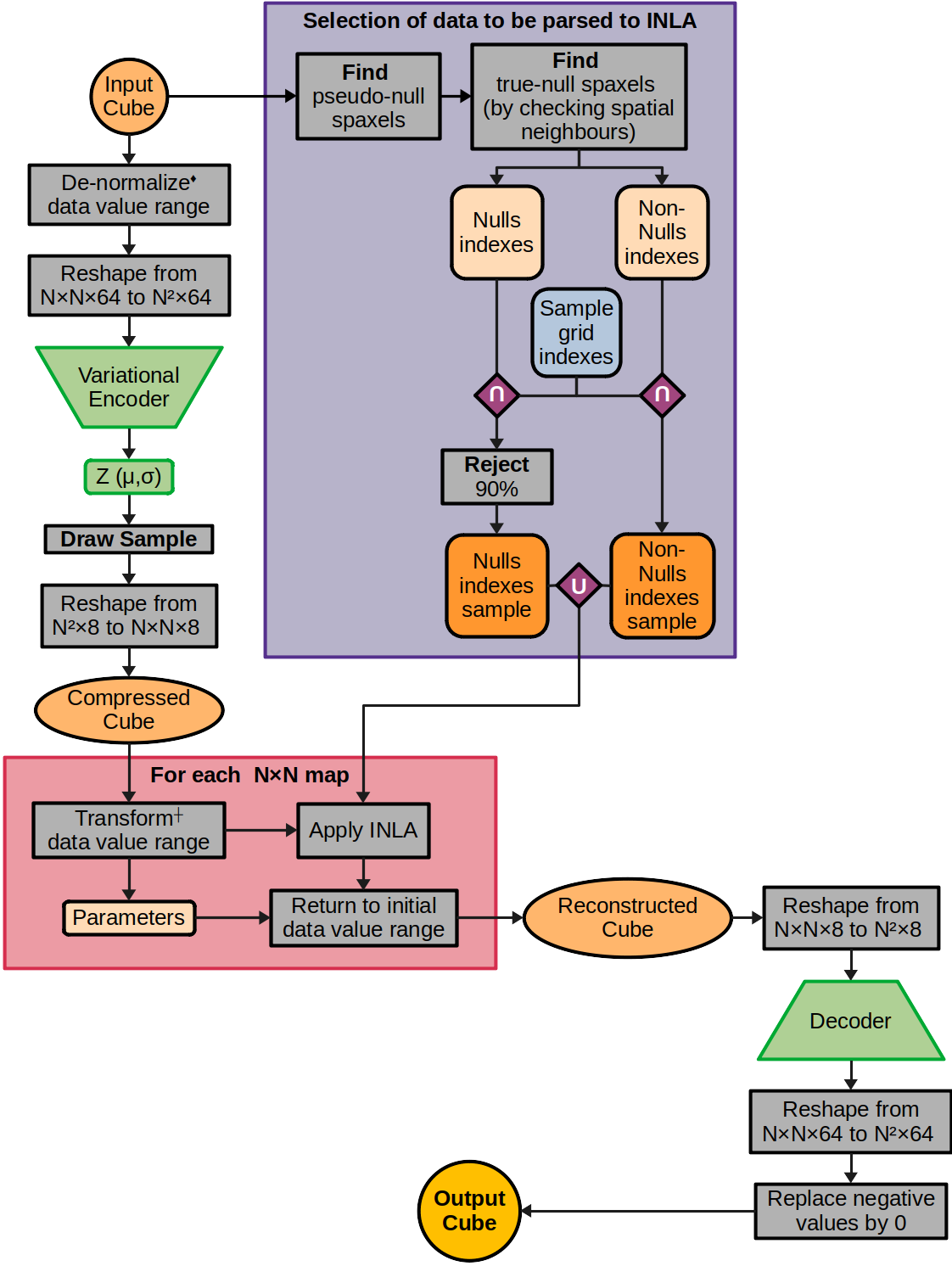}
    \caption{Flowchart for EmulART.}
    \label{fig:scheme}
\end{figure}

\begin{figure}[ht]
    \centering
    \includegraphics[height = .85\textheight]{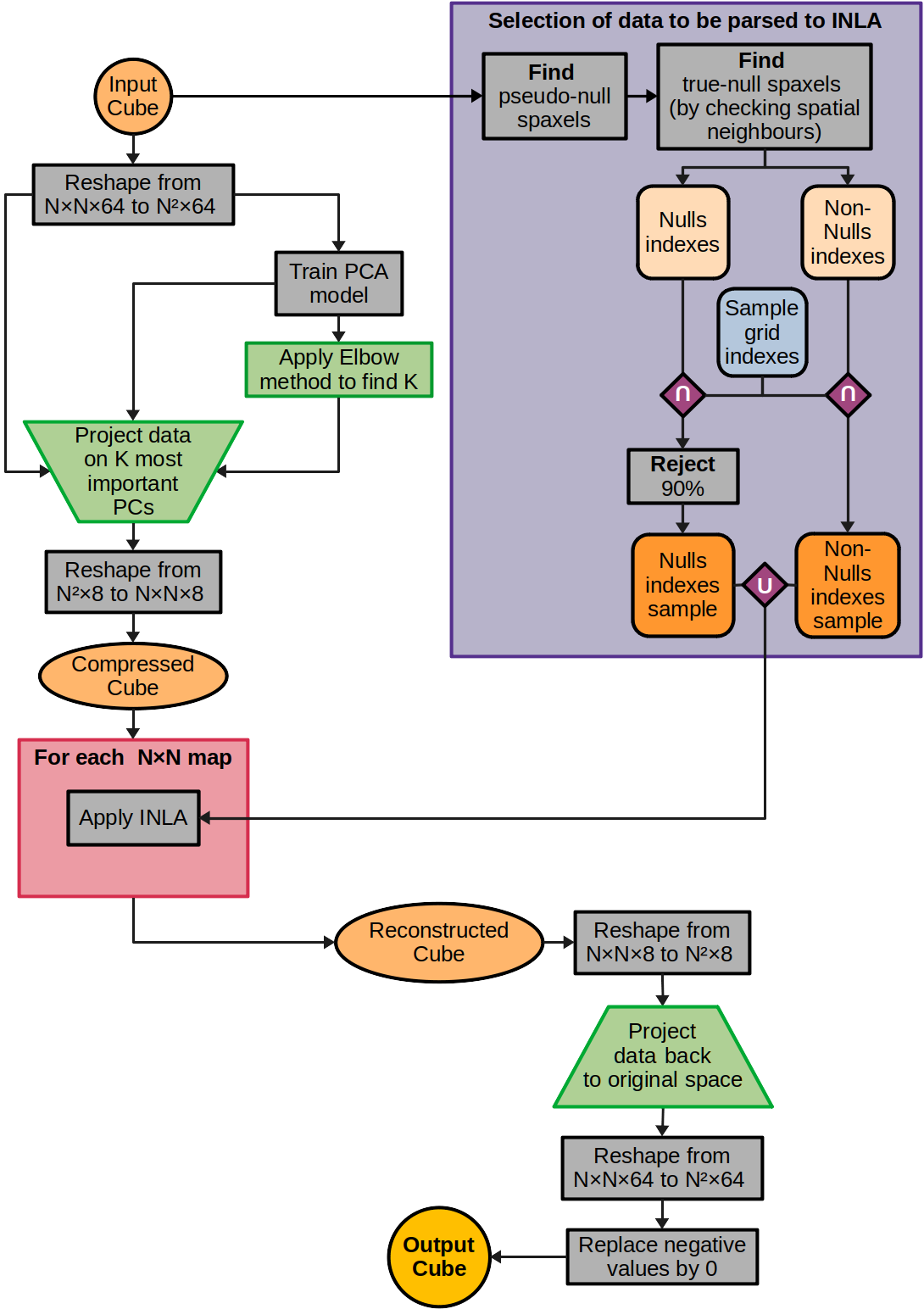}
    \caption{Flowchart for the emulation pipeline that employs the PCA model with the number of components, K, determined by the elbow method. For the emulation pipeline that employs the PCA model with a number of components equal to the number of latent features of the DVAE model in EmulART (K = 8) the Elbow method block is skipped.}
    \label{fig:schemePCA1}
\end{figure}

\newpage
\setcounter{figure}{0}
\setcounter{table}{0}
\section{Result Analysis}\label{app:E}

This section presents a detailed view of the results obtained with both AESet and EVASet. As mentioned in Section \ref{sec:results}, for every $\tau_{9.7}$ and $\phi$ case the realization with $N_p = 10^8$ was considered as the HPN reference, or ``ground truth". The remaining LPN realizations and the emulations obtained from those were compared to the HPN reference in regards to their spectral and spatial features.\par
For the spectral analysis we computed the normalized residuals (Eq. \ref{eq:normres}) of the spatially integrated SEDs, compared the shape of those SEDs and inspected the spatial maps of the DVAE compressed features looking for distributions compatible with the physical properties of the HPN reference.\par
For the spatial analysis we computed the normalized residuals and measured statistical descriptors - median and mean absolute deviation (MAD) - for each wavelength bin spatial map, for both LPN inputs and resulting emulations. To measure the impact of INLA on the reconstruction of input spaxels we count how many full, partial and null spaxels are present in the HPN reference, LPN input and the resulting emulations, and compute the TIR of each.\par

\subsection{AESet Predictions}
\subsubsection{Spectral Predictions}
In this section we discuss how well the spectral information is preserved, after the LPN inputs of AESet are processed by EmulART, by looking into the emulations spatially integrated SEDs.\par
From both Fig. \ref{fig:intSED} and Tab. \ref{tab:intres} we determine the $N_p = 10^6$ realization, with spaxel sampling of 4\%, to be the most favorable input for our pipeline. This simulation is the result of 100$\times$ less photons simulated than the reference, and with the corresponding emulation using as input 4\% of its spaxels which results in 2500$\times$ less input information than the present in the HPN reference. This emulation presents median residuals of $\sim$85\% between 1 and 4$\mu$m, $\sim$33\% between 4$\mu$m and 1mm and $\sim$42\% over the whole spectral range, and 2.8\% when considering the SED that results from the spatial integration of each wavelength map (less than 2\% worse than the realization which was used as input). It also reconstructs over 71,000 spaxels presenting almost two times the amount of full spaxels of the reference. Appendix \ref{app:F} shows how the pipeline goes even further with the emulated individual spaxels presenting smoother profiles than the references.\par 

\subsubsection{Spatial Predictions}
This section looks at the reconstructions of sparse samples of pixels from latent feature maps and the spatial distributions that result after returning those to the original spectral space.\par
\begin{figure}[ht]
    \centering
    \includegraphics[width =\textwidth]{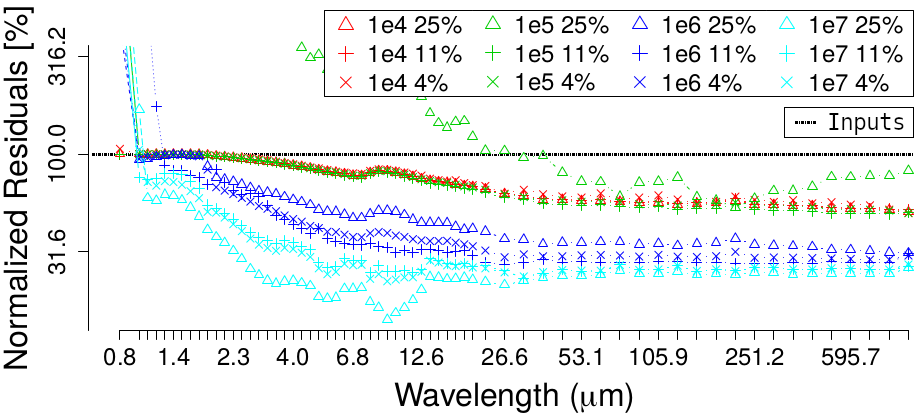}
    \caption{Median of the normalized residuals, at each wavelengths spatial map, for every emulation obtained with $N_p = 10^4$ (red), $N_p = 10^5$ (green), $N_p = 10^6$ (blue) and $N_p = 10^7$ (cyan) realizations, for the case of a dust shell with $\tau_{9.7}=0.05$ and $\phi = 0^\circ$. Emulations whose spatial inference was performed using as input 25\% of data are represented by ($\triangle$), 11\% by (+) and 4\% by ($\times$). The interrupted black line marks the same metric for the LPN realizations.}
    \label{fig:res_sp_wl}
\end{figure}

The impact of INLAs spatial reconstruction is evidenced by the emulations residuals having lower median values than all realizations but the reference, and mean deviations of the same order of magnitude. In the case of the emulations performed using the $N_p = 10^7$ realizations as input that is even clearer, with median of the spatial residuals at wavelengths greater than 5 $\mu$m going below 30\%. Furthermore, as Fig. \ref{fig:res_sp_wl} shows, there are two transitions in quality evidenced by the gaps between the median residuals. It is also shown that all lower photon count simulations have their residuals dominated by lack of information (the reader is reminded that a 100\% normalized residual means that pixel holds no information), and that the spatial inference at lower wavelengths results in large over-estimations. The absence of information which heavily influences the statistical descriptors of the simulations residuals is thus curbed. This is confirmed when comparing the TIR values, where we can see that the emulation based on the lowest $N_p$ realizations has a TIR of 98.3\% compared to the reference realization (see Tab. \ref{tab:intres}), while the $N_p = 10^7$ realization's is 65.2\% (see Tab. \ref{tab:tab_sources}).\par

\begin{figure}[ht]
    \centering
    \begin{subfigure}[b]{.45\textwidth}
        \centering
        \includegraphics[width =\textwidth]{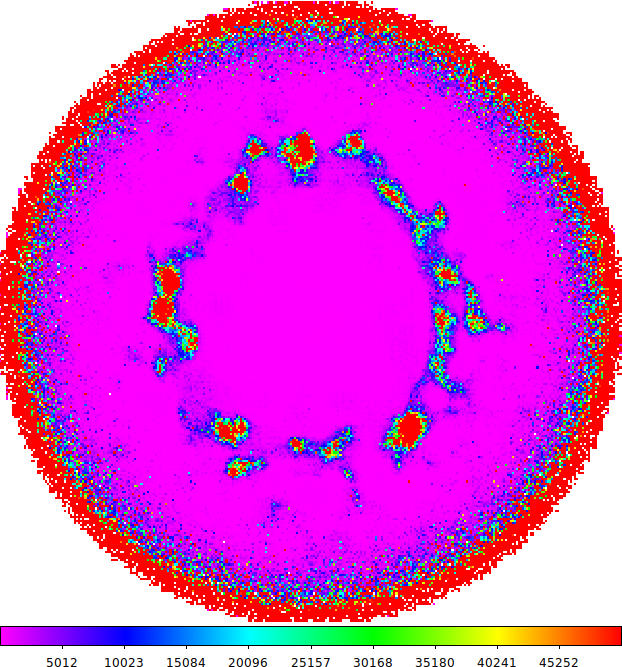}
        \caption{0.84 $\mu$m}
        \label{fig:1um1e6face005}
        \vspace{3mm}
    \end{subfigure}
    \hspace{5mm}
    \begin{subfigure}[b]{.45\textwidth}
        \centering
        \includegraphics[width =\textwidth]{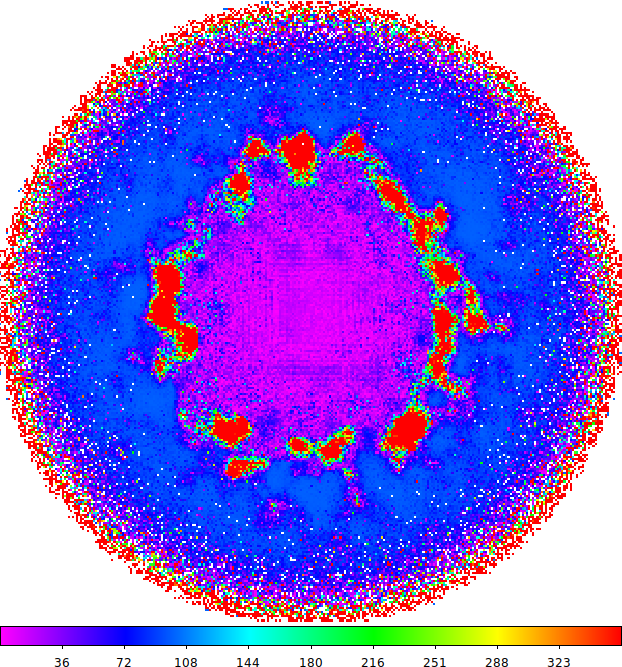}
        \caption{1.85 $\mu$m}
        \label{fig:2um1e6face005}
        \vspace{3mm}
    \end{subfigure}
    \centering
    \begin{subfigure}[b]{.45\textwidth}
        \centering
        \includegraphics[width =\textwidth]{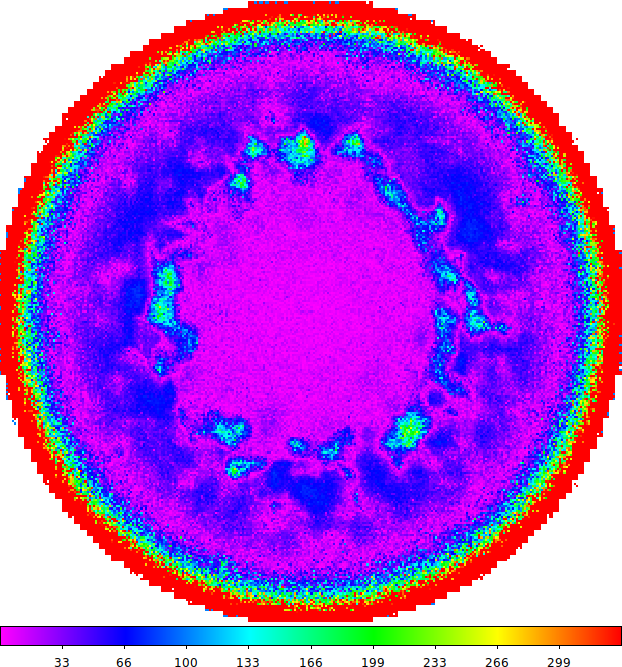}
        \caption{9.28 $\mu$m}
        \label{fig:9um1e6face005}
    \end{subfigure}
    \hspace{5mm}
    \begin{subfigure}[b]{.45\textwidth}
        \centering
        \includegraphics[width =\textwidth]{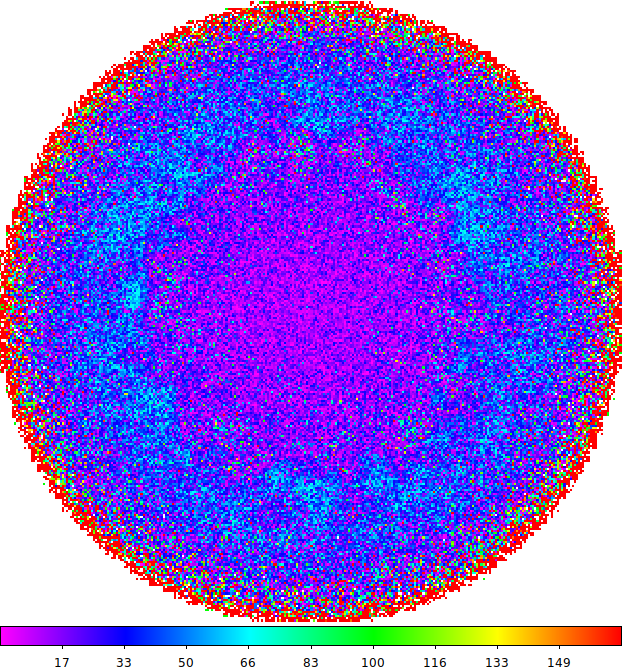}
        \caption{211.35 $\mu$m}
        \label{fig:211um1e6face005}
    \end{subfigure}
    \caption{Spatial distribution of residuals, as defined in Eq. \ref{eq:normres}, for the emulation produced using as input a 4\% sample of the $N_p = 10^6$ realization, for the case of a dust shell with $\tau_{9.7} = 0.05$ and $\phi = 0^\circ$, at wavelengths 0.84 $\mu$m (Fig. \ref{fig:1um1e6face005}), 1.85 $\mu$m (Fig. \ref{fig:2um1e6face005}), 9.28 $\mu$m (Fig. \ref{fig:9um1e6face005}) and 211.35 $\mu$m (Fig. \ref{fig:211um1e6face005}).}
    \label{fig:res_face005_1e6}
\end{figure}

Fig. \ref{fig:res_face005_1e6} shows the spatial distribution of the emulations residuals at some wavelengths, it is clear that:\par

\begin{itemize}
    \item The residuals are correlated with the distance towards the center of the object, with the highest residuals populating the outer edge and the lowest in the center. 
    \item Shorter wavelengths display higher residuals. 
\end{itemize} 

The above results were expected given that: we performed a uniform spatial sampling of an object central to a spherically symmetric dust distribution; more spatial information is available at the center of the simulated object than in the outskirts; and, at shorter wavelengths there is overall less information, as seen in the previous section, from which to build the spatial reconstruction.\par 
It is harder to assess how accurate INLAs inference of the outskirt regions is. Those regions, given the spatial distribution of the simulated model, tend to be less dense. Nevertheless, at most wavelengths and most of the spacial distribution of the object, the residuals place the emulated value at the same order of magnitude as the reference values. Moreover, it is in these regions that INLA recovers more information, filling in the gaps.\par

\subsection{EVASet Predictions}

\subsubsection{Spectral Predictions}
In this section we discuss how well the spectral information is preserved, after the LPN inputs of EVASet, with $\tau_{9.7} \neq 0.05$, are processed by EmulART, by once again looking into the emulations spatially integrated SEDs.\par
\begin{figure}[ht]
\centering
    \begin{subfigure}[b]{.45\textwidth}
        \centering
        \includegraphics[width =\textwidth]{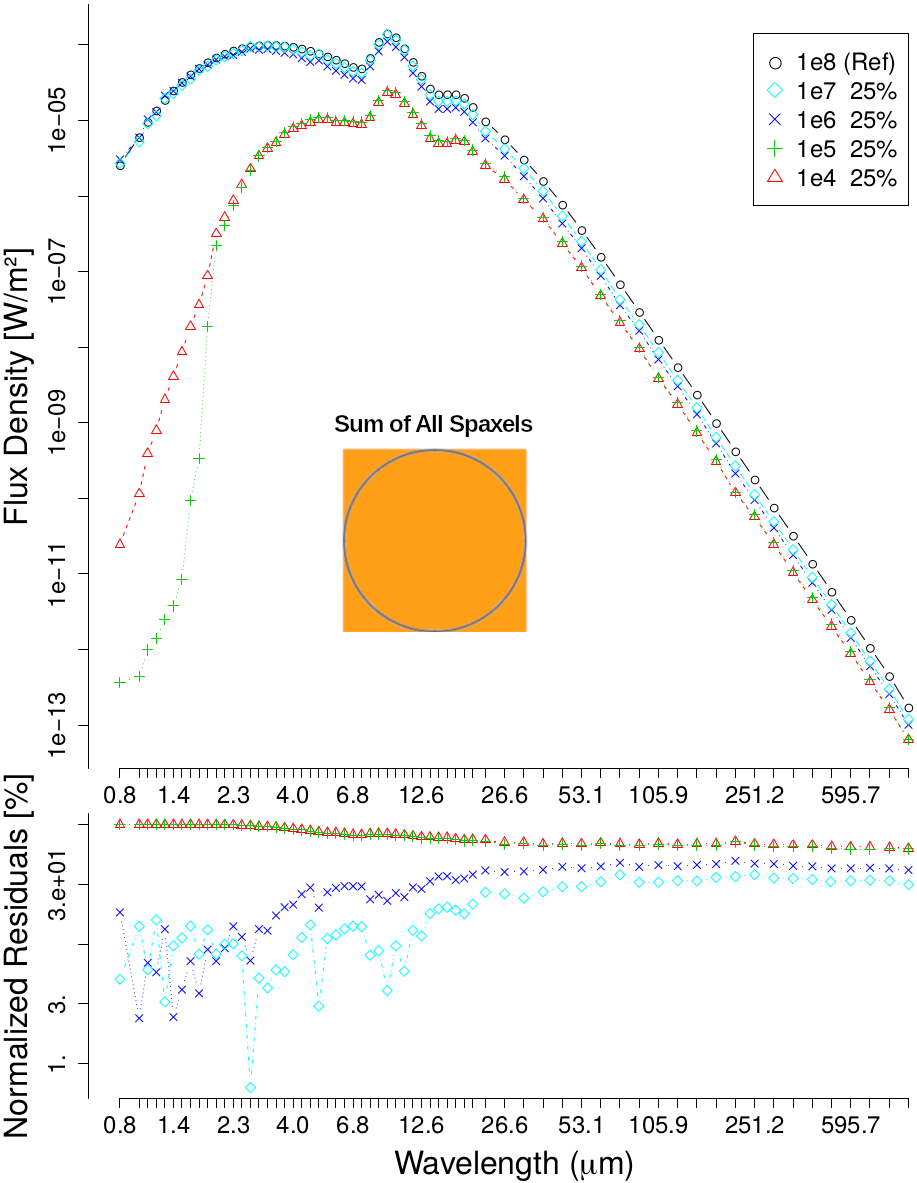}
        \caption{$\phi = 90^\circ$}
        \label{fig:01ed25}
    \end{subfigure}
    \hspace{5mm}
    \begin{subfigure}[b]{.45\textwidth}
        \centering
        \includegraphics[width =\textwidth]{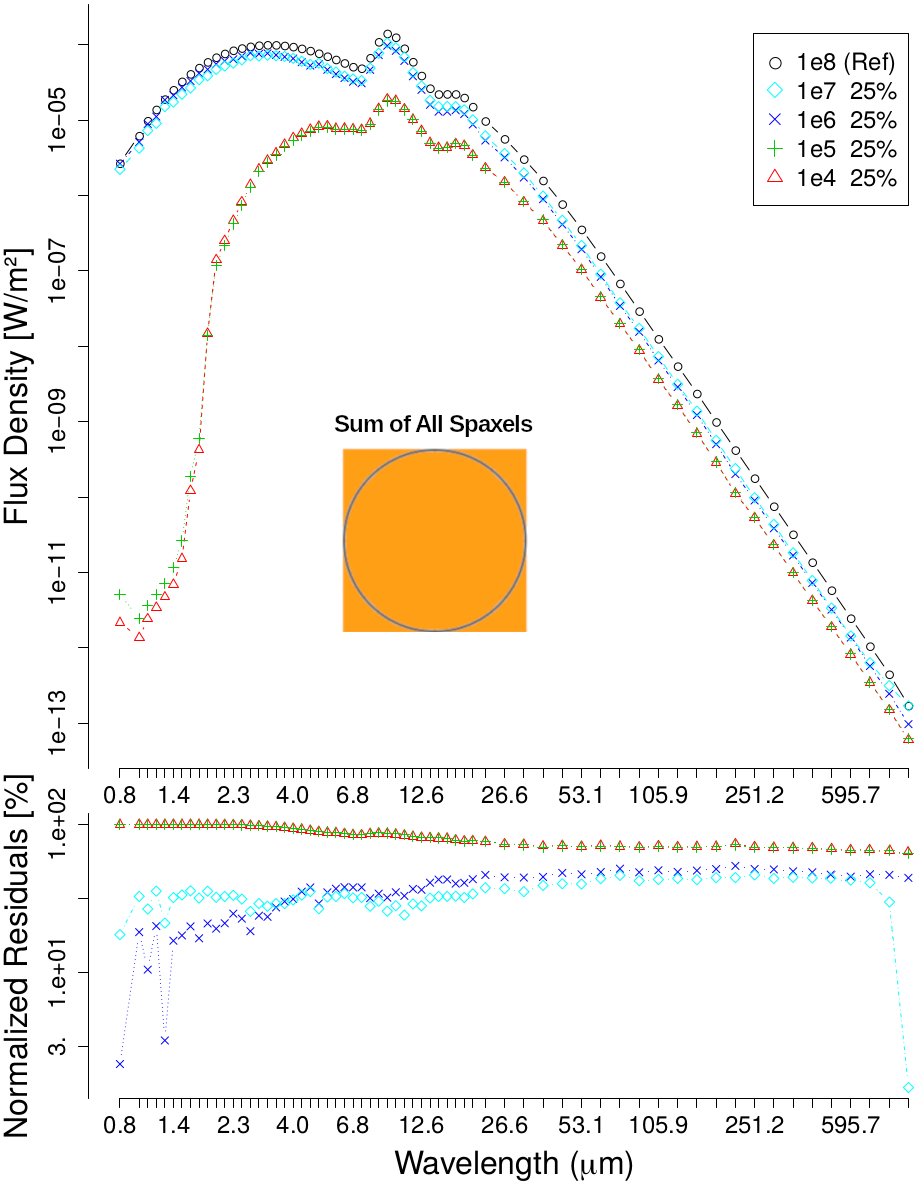}
        \caption{$\phi = 0^\circ$}
        \label{fig:01fa25}
    \end{subfigure}
    \caption{Emulations integrated SEDs, resulting from spatial inference using 25\% of spatial information, and the respective normalized residuals, for the cases of dust shells with $\tau_{9.7} = 0.1$, $\phi = 90^\circ$ (Fig. \ref{fig:01ed25}) and $\phi = 0^\circ$ (Fig. \ref{fig:01fa25}). The HPN reference is represented in black ($\circ$), the emulation based on the $N_p = 10^4$ realization is in red ($\triangle$), on the $N_p = 10^5$ in green (+), on the $N_p = 10^6$ in blue ($\times$) and on the $N_p = 10^7$ in cyan ($\diamond$) .}
    \label{fig:intSED01}
\end{figure}

Regarding the case with $\tau_{9.7} = 0.1$, Fig. \ref{fig:intSED01} and Tab. \ref{tab:intres3} show that, for both tilt angles, while the shape of the emulations integrated spectra still follows closely that of the references, the residuals are now over 2$\times$ higher than with $\tau_{9.7} = 0.05$. The results for the cases with $\tau_{9.7} = 1.0$ confirm this trend in degrading quality of the pipeline's spectral predictions. As seen in Fig. \ref{fig:intSED10}, the emulation integrated spectrum completely misses the shape of the reference spectrum, keeping the shape displayed for the cases with $\tau_{9.7} = 0.05$ and $\tau_{9.7}=0.1$, a sign of over-fitting which is confirmed when inspecting individual spaxels (see Appendix \ref{app:F}). It also underestimates the flux density and overestimates the relative magnitude of the two emission bumps in the 8 $\mu$m to 20 $\mu$m range. The median residual of the total flux density per wavelength, displayed in Tab. \ref{tab:intres3}, further confirms the poor quality of the spectral predictions.\par 

\begin{table}[ht]
\centering
\begin{tabular}{cccccccllll}
\cline{1-7}
\multicolumn{2}{c}{\multirow{2}{*}{\textbf{Case}}} & \multirow{5}{*}{\textbf{\begin{tabular}[c]{@{}c@{}}LPN\\ Input\\ Sample\end{tabular}}} & \multicolumn{2}{c}{\multirow{4}{*}{\textbf{\begin{tabular}[c]{@{}c@{}}Residuals of\\ Integrated\\ Emul. (\%)\end{tabular}}}} & \multicolumn{2}{c}{\multirow{4}{*}{\textbf{\begin{tabular}[c]{@{}c@{}}Normalized\\ Residuals of\\ Emul. (\%)\end{tabular}}}} & \multicolumn{1}{c}{\multirow{3}{*}{\textbf{}}} \\
\multicolumn{2}{c}{} &      & \multicolumn{2}{c}{} & \multicolumn{2}{c}{} & \multicolumn{1}{c}{} & \multicolumn{2}{l}{} \\ \cline{1-2}
\multirow{3}{*}{\textbf{$\tau_{9.7}$}} & \multirow{3}{*}{\textbf{$\phi$ (deg)}} &       & \multicolumn{2}{c}{} & \multicolumn{2}{c}{} & \multicolumn{1}{c}{} & \multicolumn{2}{l}{} \\ \cline{4-7}
    &       &       & \textbf{M} & \textbf{MAD} & \textbf{M} & \textbf{MAD} \\ \cline{1-7}
\multirow{14}{*}{0.1} & \multirow{6}{*}{0} & $10^6$ 11\%    & 279   & 256   & 561   & 805 \\ \cline{3-7}
    &       & $10^6$ 25\%   & 37    & 10    & 56    & 32 \\ \cline{3-7}
    &       & $10^7$ 11\%   & 33.2  & 5.9   & 46    & 31 \\ \cline{3-7}
    &       & $10^7$ 25\%   & 32.6  & 5.3   & 48    & 35 \\ \cline{2-7}
    & \multirow{6}{*}{90}   & $10^6$ 11\%   & 33    & 12    & 58    & 35 \\ \cline{3-7}
    &       & $10^6$ 25\%   & 30    & 13    & 54    & 34 \\ \cline{3-7}
    &       & $10^7$ 11\%   & 15    & 10    & 36    & 29 \\ \cline{3-7}
    &       & $10^7$ 25\%   & 14.1  & 9.9   & 35    & 28 \\ \cline{1-7}
\multirow{14}{*}{1.0} & \multirow{6}{*}{0}  & $10^6$ 11\%   & 89.8  & 5.2   & 92.1  & 9.8 \\ \cline{3-7}
    &       & $10^6$ 25\%   & 89.8  & 5.3   & 92    & 10 \\ \cline{3-7}
    &       & $10^7$ 11\%   & 91    & 2e+9  & 93.5  & 9.4 \\ \cline{3-7}
    &       & $10^7$ 25\%   & 91    & 7e+5  & 94.1  & 7.8 \\ \cline{2-7}
    & \multirow{6}{*}{90}   & $10^6$ 11\%   & 89.6  & 7.8   & 92.3  & 8.1 \\ \cline{3-7}
    &       & $10^6$ 25\%   & 90.6  & 5.8   & 92.6  & 7.7 \\ \cline{3-7}
    &       & $10^7$ 11\%   & 91    & 4e+8  & 93.0  & 9.1 \\ \cline{3-7}
    &       & $10^7$ 25\%   & 91    & 3e+7  & 93.5  & 8.1 \\ \cline{1-7}
\end{tabular}
\caption{Comparison of statistics for the residuals of the spatial integration SEDs (columns 4 and 5) and of the residuals overall wavelengths (columns 6 and 7), for the emulations that took as input LPN realizations of dust shells with $\tau_{9.7} \in \{0.1, 1.0\}$, $\phi \in \{0^\circ, 90^\circ\}$, and $N_p \in \{10^6, 10^7\}$.}
\label{tab:intres3}
\end{table}

\subsubsection{Spatial Predictions}
\begin{figure}[ht]
    \centering
    \includegraphics[width =\textwidth]{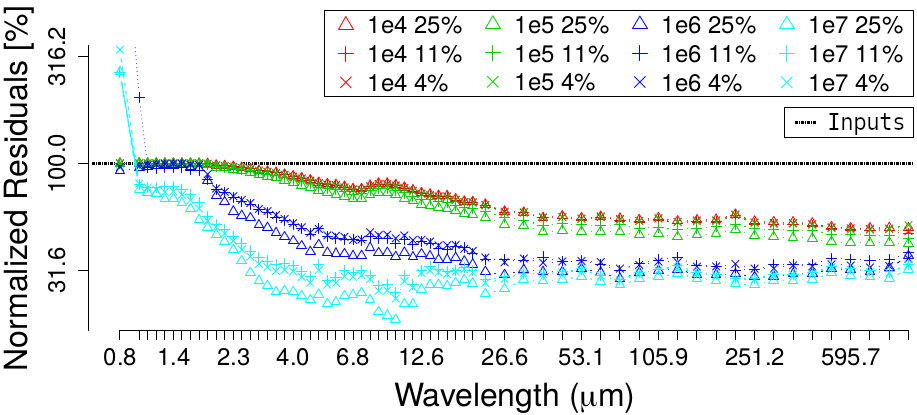}
    \caption{Median of the normalized residuals, at each wavelengths spatial map, for every emulation obtained with $N_p = 10^4$ (red), $N_p = 10^5$ (green), $N_p = 10^6$ (blue) and $N_p = 10^7$ (cyan) realizations, for the case of a dust shell with $\tau_{9.7}=0.05$ and $\phi = 90^\circ$. Emulations whose spatial inference was performed taking as input 25\% of the data are represented by ($\triangle$), 11\% by (+) and 4\% by ($\times$). The interrupted black line marks the same metric for the realizations used as input.}
    \label{fig:sp_res_edge005}
\end{figure}

Regarding spatial information, for emulations with LPN inputs of $\tau_{9.7}=0.05$ and $\phi = 90^\circ$ we observe the same behavior in the performance metrics as with the $\phi = 0^\circ$ case. Tab. \ref{tab:tab_emuls2} shows that the overall median residuals decrease with the increase of $N_p$. The median residuals per wavelength, displayed in Fig. \ref{fig:sp_res_edge005}, present a similar behavior to those of the $\phi = 0^\circ$ case. 

\begin{table}[ht]
\centering
\begin{tabular}{clcccccccc}
\hline
\multicolumn{2}{c}{\multirow{6}{*}{\textbf{\begin{tabular}[c]{@{}c@{}}LPN\\ Input\\ Sample\end{tabular}}}} & \multirow{6}{*}{\textbf{\begin{tabular}[c]{@{}c@{}}Total\\ Emul.\\ Time\\ (min)\end{tabular}}} & \multicolumn{2}{c}{\multirow{3}{*}{\textbf{\begin{tabular}[c]{@{}c@{}}Normalized\\ Residuals of\\ Emul. (\%)\end{tabular}}}} & \multicolumn{5}{c}{\multirow{3}{*}{\textbf{Information of Emulation}}} \\
\multicolumn{2}{c}{} &      & \multicolumn{2}{c}{} & \multicolumn{5}{c}{} \\
\multicolumn{2}{c}{} &      & \multicolumn{2}{c}{} & \multicolumn{5}{c}{} \\ \cline{4-10} 
\multicolumn{2}{c}{} &      & \multirow{3}{*}{\textbf{M}} & \multirow{3}{*}{\textbf{MAD}} & \multirow{3}{*}{\textbf{\begin{tabular}[c]{@{}c@{}}TIR\\ (\%)\end{tabular}}} & \multirow{3}{*}{\textbf{\begin{tabular}[c]{@{}c@{}}Null\\ Spxs\\ ($N$)\end{tabular}}} & \multirow{3}{*}{\textbf{\begin{tabular}[c]{@{}c@{}}Empty\\ Spxs\\ ($\Delta N$)\end{tabular}}} & \multirow{3}{*}{\textbf{\begin{tabular}[c]{@{}c@{}}Full\\ Spxs\\ ($N$)\end{tabular}}} & \multirow{3}{*}{\textbf{\begin{tabular}[c]{@{}c@{}}Partial\\ Spxs\\ ($N$)\end{tabular}}} \\
\multicolumn{2}{c}{} &      &       &       &       &       &       &       &       \\
\multicolumn{2}{c}{} &      &       &       &       &       &       &       &       \\ \cline{1-10}
\multicolumn{2}{c}{$10^4$ 4\%} & 7.2    & 74    & 36    & \multirow{5}{*}{98.1} & \multirow{5}{*}{20,497} & \multirow{5}{*}{2,641} & \multirow{5}{*}{69,503} & \multirow{5}{*}{0} \\ \cline{1-5}
\multicolumn{2}{c}{$10^4$ 11\%}& 22.2   & 74    & 36    &       &       &       &       &       \\ \cline{1-5}
\multicolumn{2}{c}{$10^4$ 25\%} & 16.3  & 75    & 36    &       &       &       &       &       \\ \cline{1-10}
\multicolumn{2}{c}{$10^5$ 4\%}  & 6.2   & 73    & 36    & \multirow{5}{*}{101.6} & \multirow{5}{*}{18,002} & \multirow{5}{*}{146} & \multirow{5}{*}{71,998} & \multirow{5}{*}{0} \\ \cline{1-5}
\multicolumn{2}{c}{$10^5$ 11\%} & 21.7  & 70    & 40    &       &       &       &       &       \\ \cline{1-5}
\multicolumn{2}{c}{$10^5$ 25\%} & 13.4  & 67    & 44    &       &       &       &       &       \\ \cline{1-10}
\multicolumn{2}{c}{$10^6$ 4\%}  & 7.2   & 47    & 43    & \multirow{5}{*}{103.2} & \multirow{5}{*}{16,872} & \multirow{5}{*}{-984} & \multirow{5}{*}{73,128} & \multirow{5}{*}{0} \\ \cline{1-5}
\multicolumn{2}{c}{$10^6$ 11\%} & 22.7  & 48    & 47    &       &       &       &       &       \\ \cline{1-5}
\multicolumn{2}{c}{$10^6$ 25\%} & 15.9  & 42    & 40    &       &       &       &       &       \\ \cline{1-10}
\multicolumn{2}{c}{$10^7$ 4\%} & 8.0    & 36    & 38    & \multirow{5}{*}{103.2} & \multirow{5}{*}{16,833} & \multirow{5}{*}{-1,023} & \multirow{5}{*}{73,167} & \multirow{5}{*}{0} \\ \cline{1-5}
\multicolumn{2}{c}{$10^7$ 11\%} & 19.1  & 36    & 39    &       &       &       &       &       \\ \cline{1-5}
\multicolumn{2}{c}{$10^7$ 25\%} & 15.2  & 32    & 34    &       &       &       &       &       \\ \hline
\end{tabular}
\caption{Statistics regarding how the different emulations, of the case of a dust shell with $\tau_{9.7} = 0.05$ and $\phi = 90^\circ$, compared to the HPN reference. The results are similar to those displayed in Tab.  \ref{tab:tab_emuls}.}
\label{tab:tab_emuls2}
\end{table}

\begin{figure}[ht]
\centering
    \begin{subfigure}[b]{\textwidth}
        \centering
        \includegraphics[width =\textwidth]{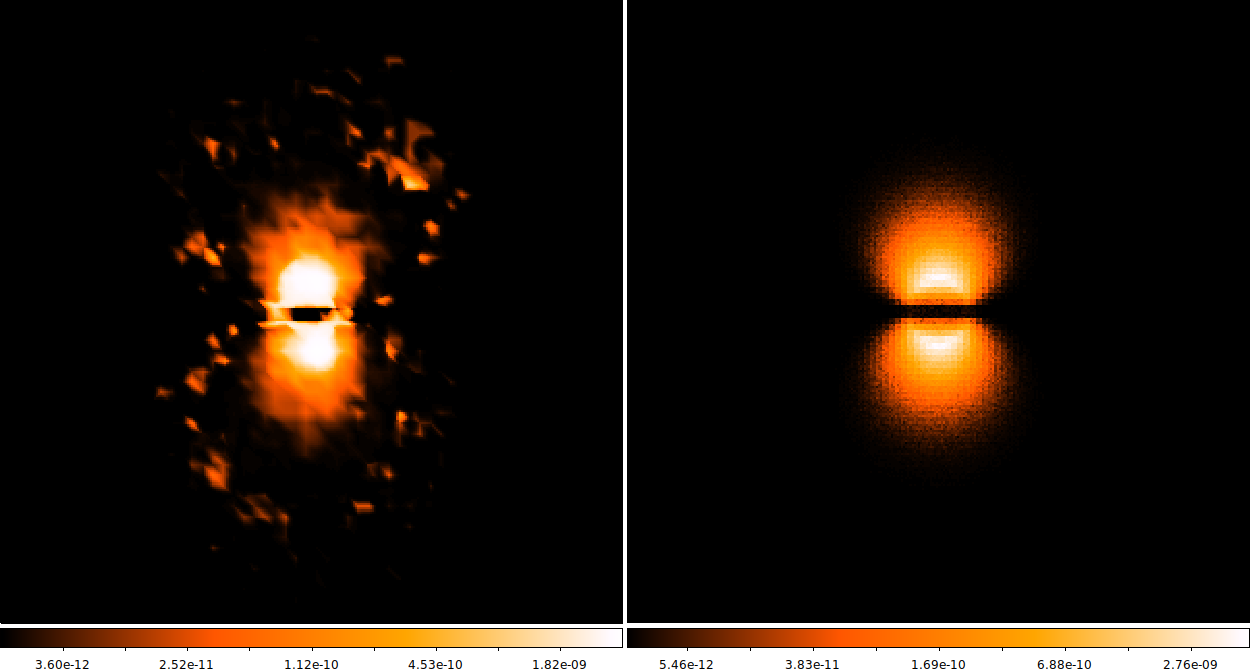}
        \caption{0.84 $\mu$m}
        \label{fig:edge1}
    \vspace{3mm}
    \end{subfigure}
    \begin{subfigure}[b]{\textwidth}
        \centering
        \includegraphics[width =\textwidth]{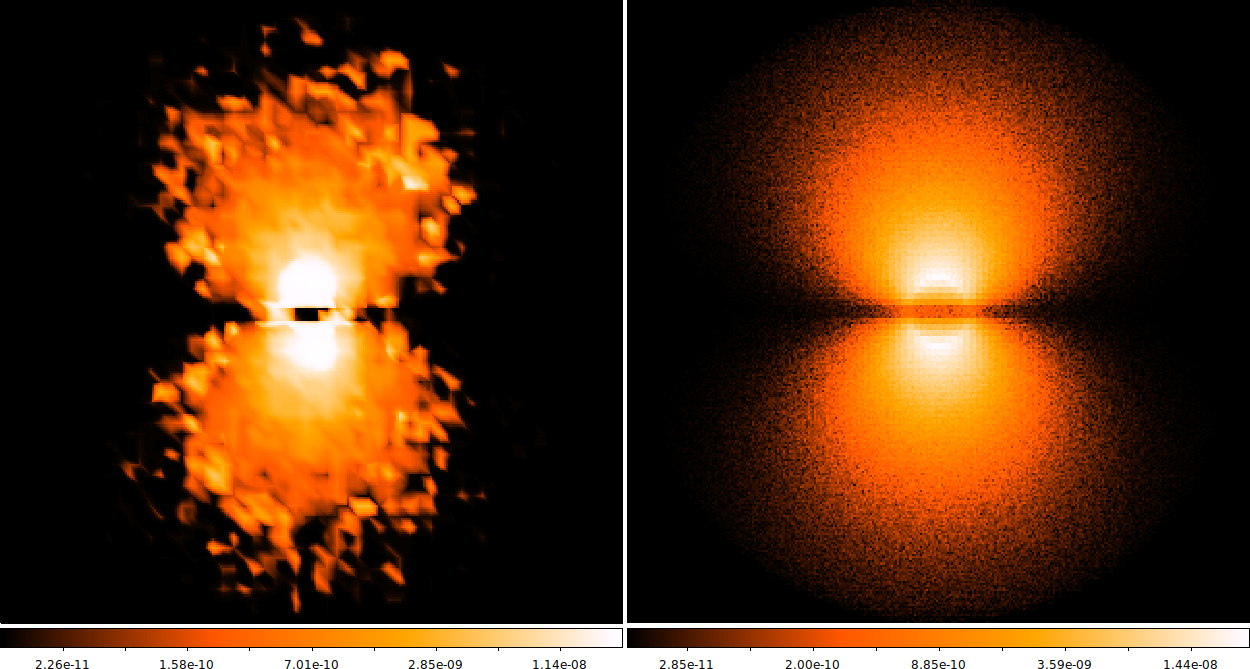}
        \caption{1.85 $\mu$m}
        \label{fig:edge2}
    \end{subfigure}
    \caption{Emulation produced using as input a 4\% sample of the $N_p = 10^6$ realization, left, and HPN reference, right, for the case of a dust shell with $\tau_{9.7} = 0.05$ and $\phi = 90^\circ$, at wavelengths 0.84 $\mu$m (Fig. \ref{fig:edge1}) and 1.85 $\mu$m (Fig. \ref{fig:edge2}).}
    \label{fig:4wledge1}
\end{figure}

\begin{figure}[ht]
\centering
    \begin{subfigure}[b]{\textwidth}
        \centering
        \includegraphics[width =\textwidth]{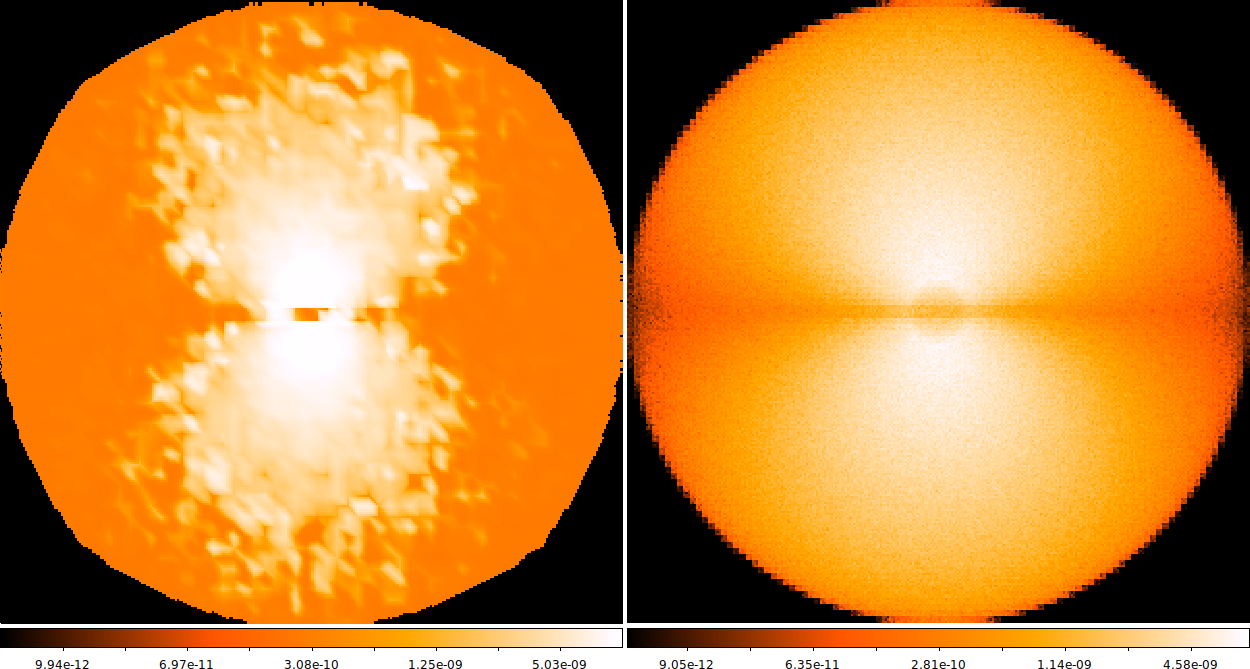}
        \caption{9.28 $\mu$m}
        \label{fig:edge3}
            \vspace{3mm}
    \end{subfigure}
    \begin{subfigure}[b]{\textwidth}
        \centering
        \includegraphics[width =\textwidth]{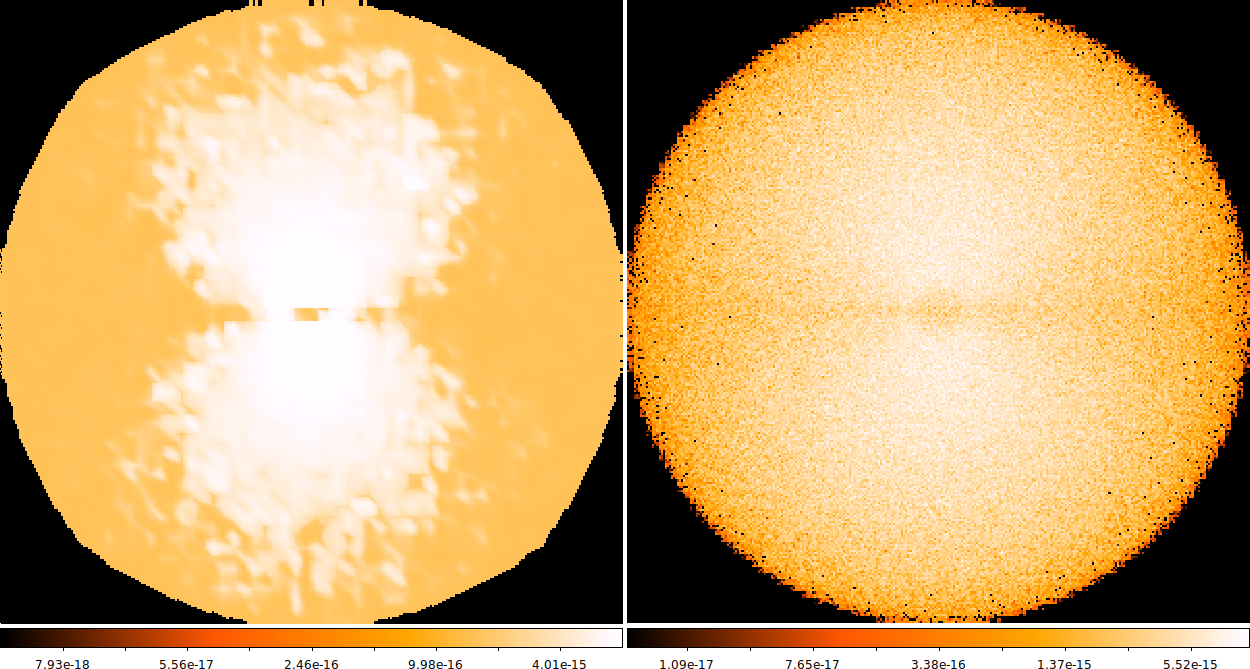}
        \caption{211.35 $\mu$m}
        \label{fig:edge4}
    \end{subfigure}
\caption{Emulation produced using as input a 4\% sample of the $N_p = 10^6$ realization, left, and HPN reference, right, for the case of a dust shell with $\tau_{9.7} = 0.05$ and $\phi = 90^\circ$, at wavelengths 9.28 $\mu$m (Fig. \ref{fig:edge3}) and 211.35 $\mu$m (Fig. \ref{fig:edge4}).}
\label{fig:4wledge2}
\end{figure}

\begin{figure}[ht]
    \centering
    \begin{subfigure}[b]{.45\textwidth}
        \centering
        \includegraphics[width =\textwidth]{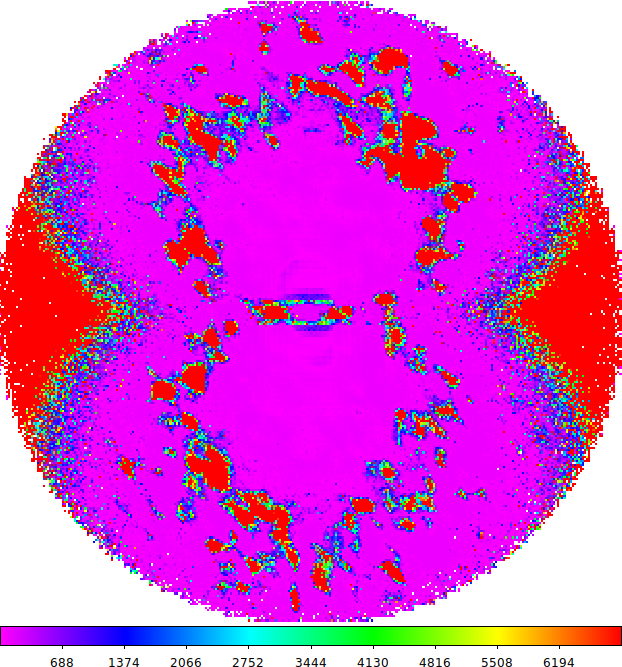}
        \caption{0.84 $\mu$m}
        \label{fig:1um1e6edge005}
        \vspace{3mm}
    \end{subfigure}
    \hspace{5mm}
    \begin{subfigure}[b]{.45\textwidth}
        \centering
        \includegraphics[width =\textwidth]{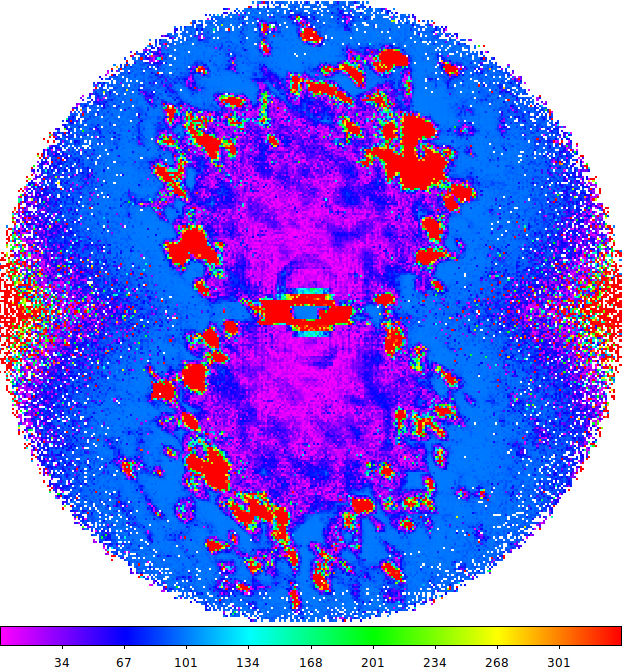}
        \caption{1.85 $\mu$m}
        \label{fig:2um1e6edge005}
        \vspace{3mm}
    \end{subfigure}
    \centering
    \begin{subfigure}[b]{.45\textwidth}
        \centering
        \includegraphics[width =\textwidth]{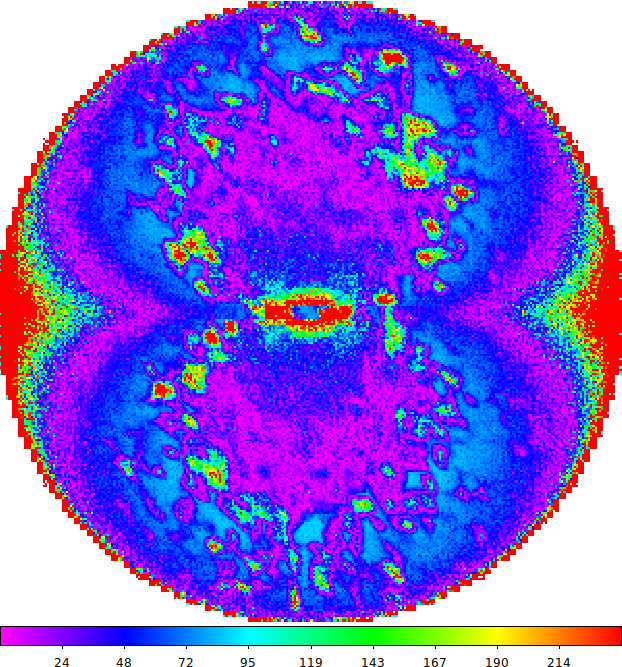}
        \caption{9.28 $\mu$m}
        \label{fig:9um1e6edge005}
    \end{subfigure}
    \hspace{5mm}
    \begin{subfigure}[b]{.45\textwidth}
        \centering
        \includegraphics[width =\textwidth]{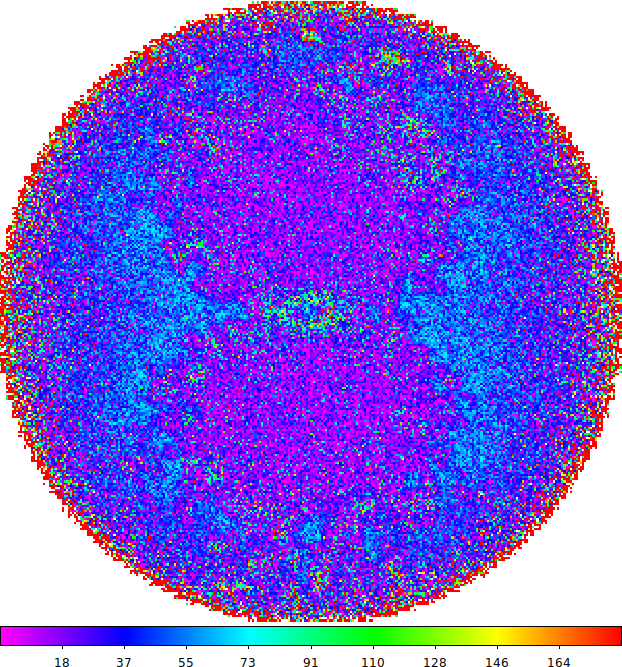}
        \caption{211.35 $\mu$m}
        \label{fig:211um1e6edge005}
    \end{subfigure}
    \caption{Spatial distribution of residuals, as defined in Eq. \ref{eq:normres}, for the emulation produced using as input a 4\% sample of the $N_p = 10^6$ realization, for the case of a dust shell with $\tau_{9.7} = 0.05$ and $\phi = 90^\circ$, at wavelengths 0.84 $\mu$m (Fig. \ref{fig:1um1e6edge005}), 1.85 $\mu$m (Fig. \ref{fig:2um1e6edge005}), 9.28 $\mu$m (Fig. \ref{fig:9um1e6edge005}) and 211.35 $\mu$m (Fig. \ref{fig:211um1e6edge005}).}
    \label{fig:res_edge005_1e6}
\end{figure}

\textbf{Fig. \ref{fig:4wledge1} shows that the emulation of models with excessive emission within the central regions in comparison to the reference, while in Fig. \ref{fig:4wledge2} the prediction for the peripheral region is more isotropic than the reference, this is even clearer when consulting the spatial distribution of residuals for these emulations in Fig. \ref{fig:res_edge005_1e6}.}\par

\begin{figure}[ht]
\centering
    \begin{subfigure}[b]{\textwidth}
        \centering
        \includegraphics[width =\textwidth]{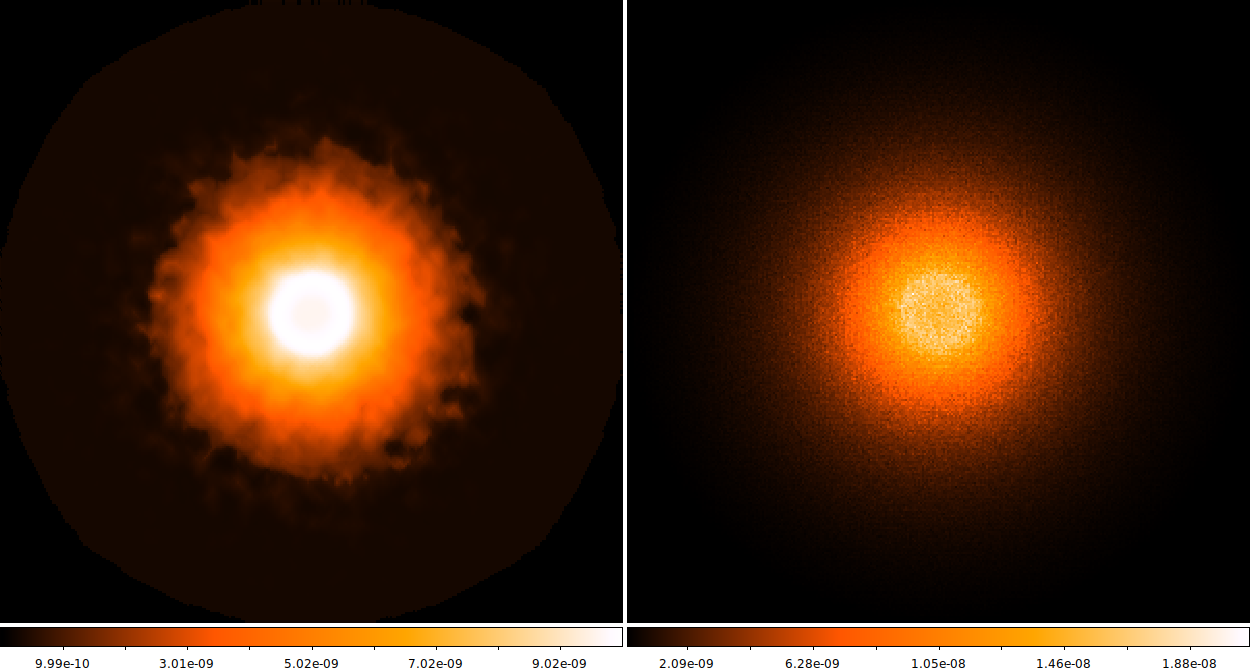}
        \caption{$\phi = 0^\circ$}
        \label{fig:face01}
    \vspace{3mm}
    \end{subfigure}
    \begin{subfigure}[b]{\textwidth}
        \centering
        \includegraphics[width =\textwidth]{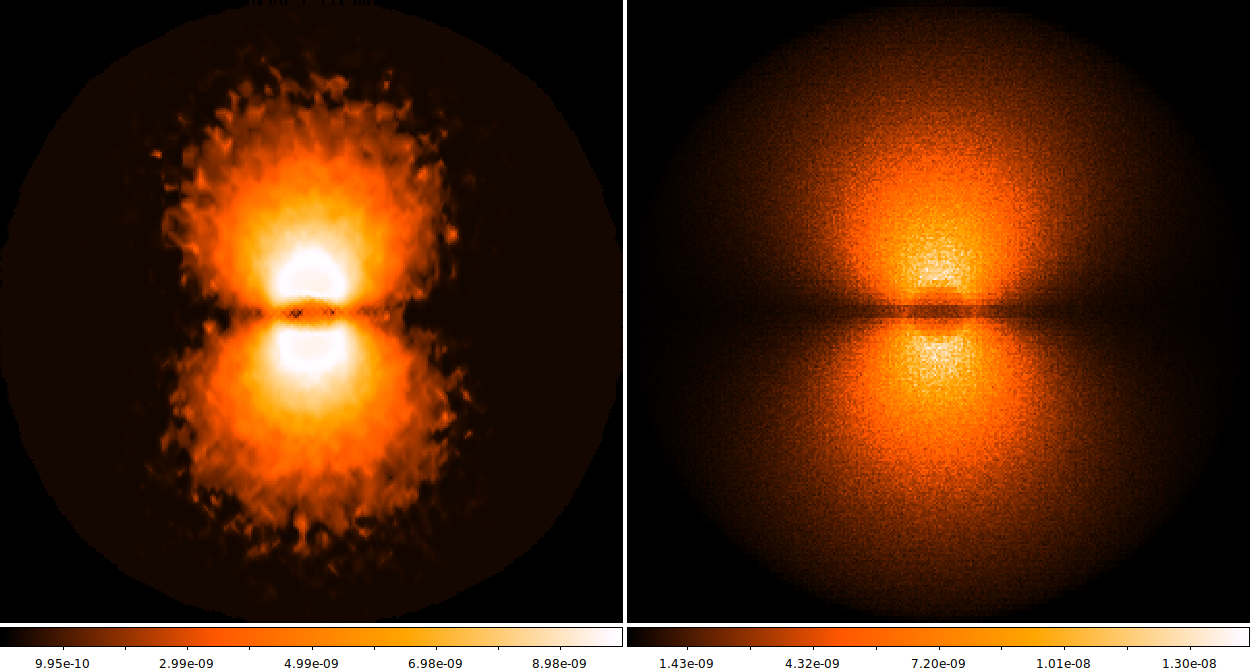}
        \caption{$\phi = 90^\circ$}
        \label{fig:edge01}
    \end{subfigure}
\caption{Emulation produced using as input a 25\% sample of the $N_p = 10^6$ realization, left, and HPN reference, right, for the case of dust shells with $\tau_{9.7} = 0.1$, at wavelengths 9.28 $\mu$m, with $\phi = 0^\circ$, (Fig. \ref{fig:face01}) and $\phi = 90^\circ$ (Fig. \ref{fig:edge01}).}
\label{fig:928_01}
\end{figure}

As with the $\tau_{9.7} = 1.0$ cases, for both tilt angle cases of $\tau_{9.7} = 0.1$ we observe a similar trend in the overall residuals. Fig. \ref{fig:928_01} shows that while the overall morphology of the spatial distribution is well recovered, the flux density value range for the emulations is underestimated, though not as considerably as with the $\tau_{9.7} = 1.0$ cases, and the contrast between the central and peripheral regions is higher than in the HPN references. This indicates a bias within the pipeline. Appendix \ref{app:F} further details into the likely source of such bias by comparing individual spaxels of emulations and HPN simulations.\par

\begin{figure}[ht]
\centering
    \begin{subfigure}[b]{\textwidth}
    \includegraphics[width =\textwidth]{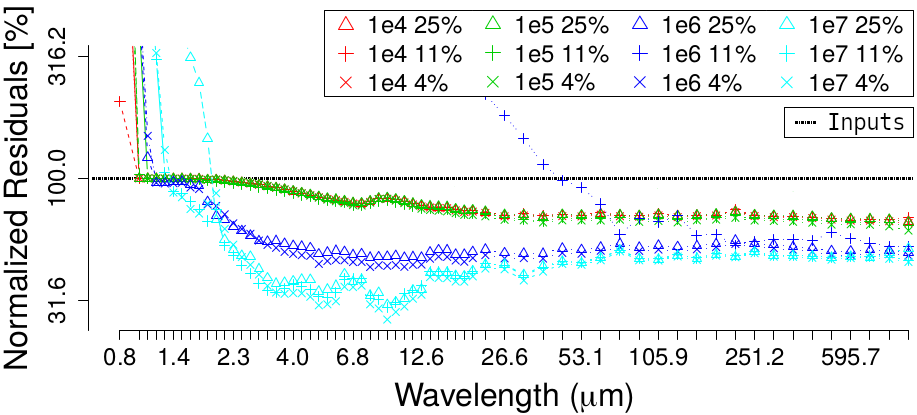}
    \caption{$\phi = 0^\circ$}
    \label{fig:sp_res_face01}
    \vspace{3mm}
    \end{subfigure}
    \begin{subfigure}[b]{\textwidth}
    \includegraphics[width =\textwidth]{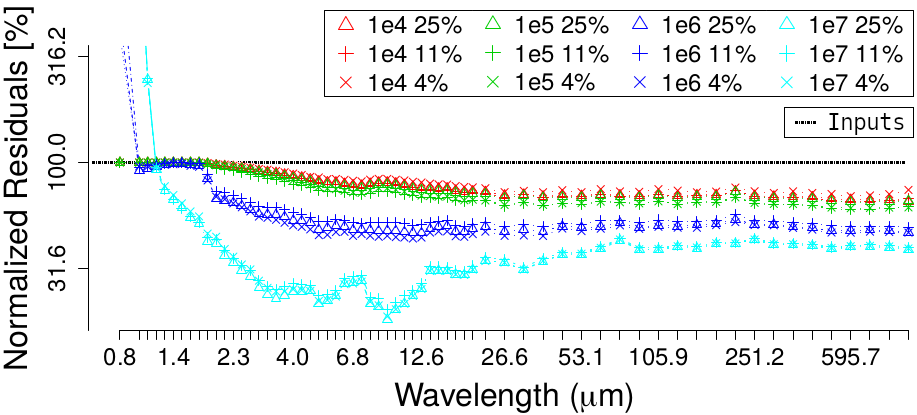}
    \caption{$\phi = 90^\circ$}
    \label{fig:sp_res_edge01}
    \end{subfigure}
    \caption{Median of the normalized residuals, at each wavelengths spatial map, for every emulation obtained with  $N_p = 10^4$ (red), $N_p = 10^5$ (green), $N_p = 10^6$ (blue) and $N_p = 10^7$ (cyan) realizations, for the case of a dust shell with $\tau_{9.7}=0.1$, with $\phi = 0^\circ$ (Fig. \ref{fig:sp_res_face01}) and $\phi = 90^\circ$ (Fig. \ref{fig:sp_res_edge01}). Emulations whose spatial inference was performed using 25\% of data are represented by ($\triangle$), 11\% by (+) and 4\% by ($\times$). The interrupted black line marks the same metric for the LPN inputs.}
    \label{fig:sp_res_01}
\end{figure}

\newpage
\setcounter{figure}{0}
\section{Single Spaxels}\label{app:F}
\begin{figure}[ht]
\centering
    \begin{subfigure}[b]{.47\textwidth}
        \centering
        \includegraphics[width =\textwidth]{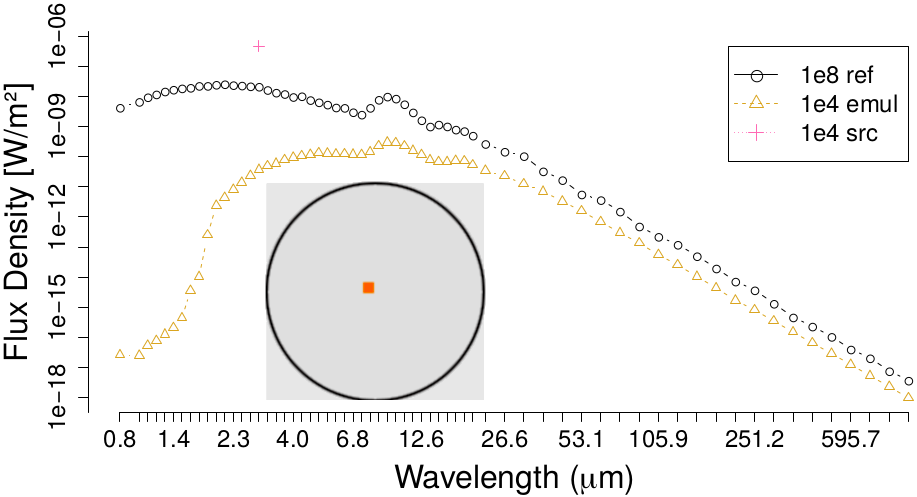}
        \subcaption{$N_p = 10^4$}
        \label{fig:sp41}
        \vspace{3mm}
    \end{subfigure}
    \hspace{5mm}
    \begin{subfigure}[b]{.47\textwidth}
        \centering
        \includegraphics[width =\textwidth]{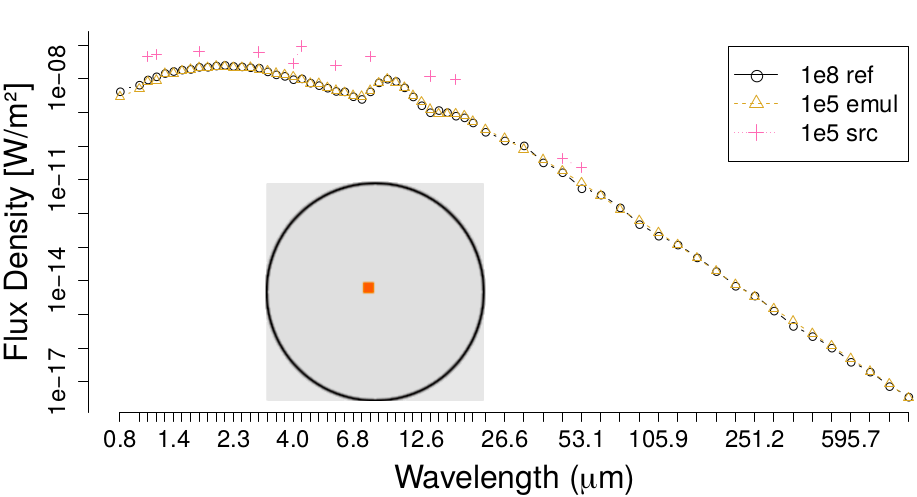}
        \subcaption{$N_p = 10^5$}
        \label{fig:sp51}
        \vspace{3mm}
    \end{subfigure}
    \begin{subfigure}[b]{.47\textwidth}
        \centering
        \includegraphics[width =\textwidth]{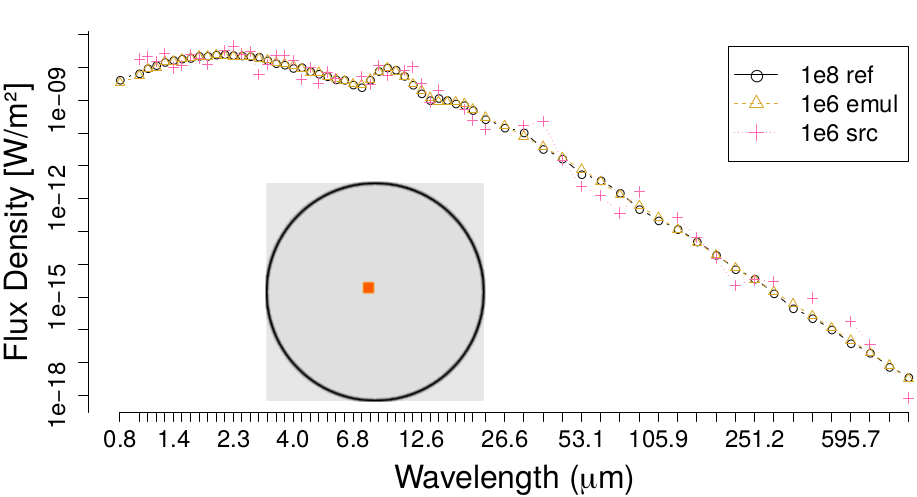}
        \subcaption{$N_p = 10^6$}
        \label{fig:sp61}
    \end{subfigure}
    \hspace{5mm}
    \begin{subfigure}[b]{.47\textwidth}
        \centering
        \includegraphics[width =\textwidth]{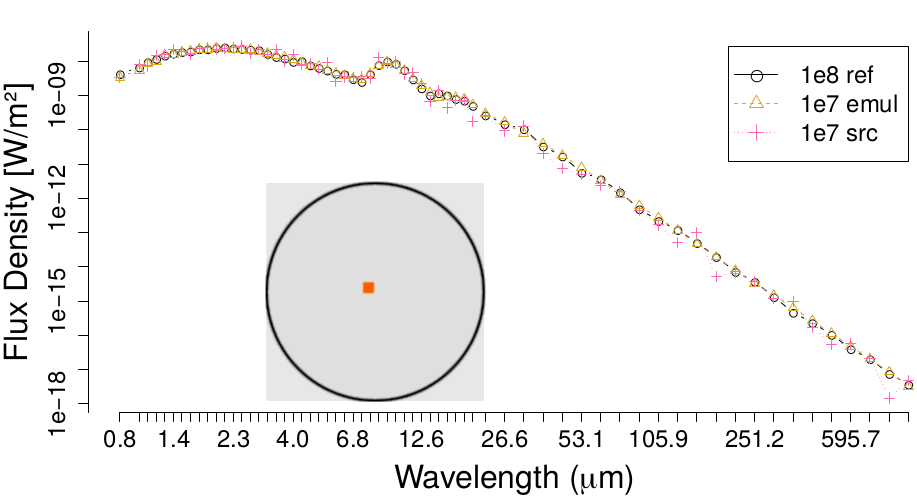}
        \subcaption{$N_p = 10^7$}
        \label{fig:sp71}
    \end{subfigure}
    \caption{Comparison of some emulations spaxels located in row = 150, col = 150, against the HPN references and the LPN input (indicated in each plot), for the case of a dust shell with $\tau_{9.7} = 0.05$ and $\phi = 0^\circ$. The emulations were performed using 25\% of the LPN input spatial information. HPN references spaxel in black ($\circ$), emulations in gold ($\triangle$) and LPN inputs, upon which the emulation was performed, in pink (+).}
    \label{fig:spaxels1}
\end{figure}

\begin{figure}[ht]
\centering
    \begin{subfigure}[b]{.47\textwidth}
        \centering
        \includegraphics[width =\textwidth]{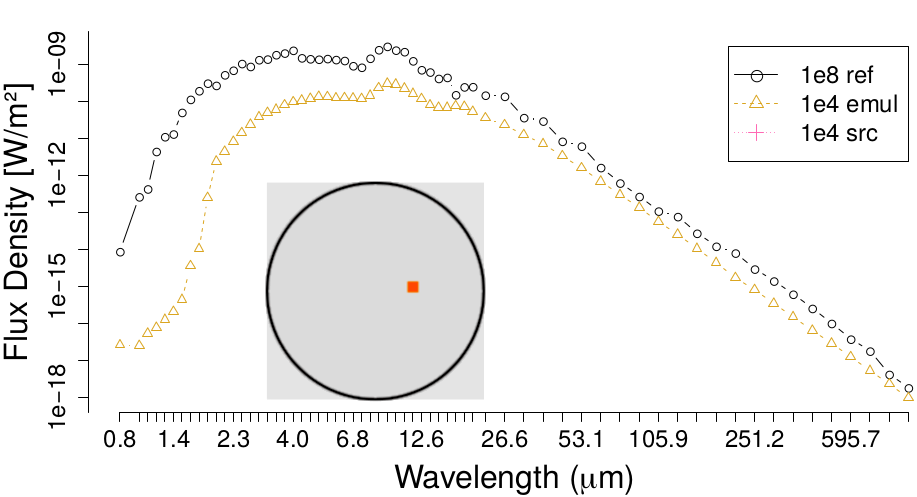}
        \subcaption{$N_p = 10^4$}
        \label{fig:sp42}
        \vspace{3mm}
    \end{subfigure}
    \hspace{5mm}
    \begin{subfigure}[b]{.47\textwidth}
        \centering
        \includegraphics[width =\textwidth]{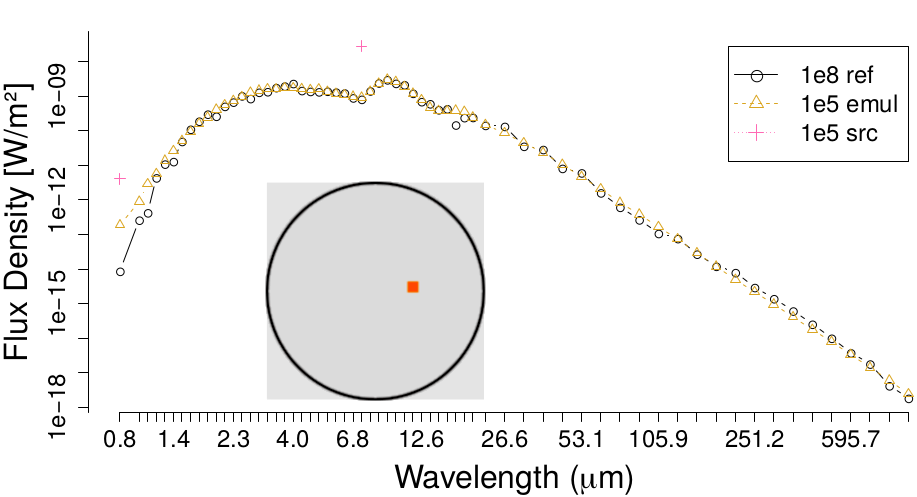}
        \subcaption{$N_p = 10^5$}
        \label{fig:sp52}
        \vspace{3mm}
    \end{subfigure}
    \begin{subfigure}[b]{.47\textwidth}
        \centering
        \includegraphics[width =\textwidth]{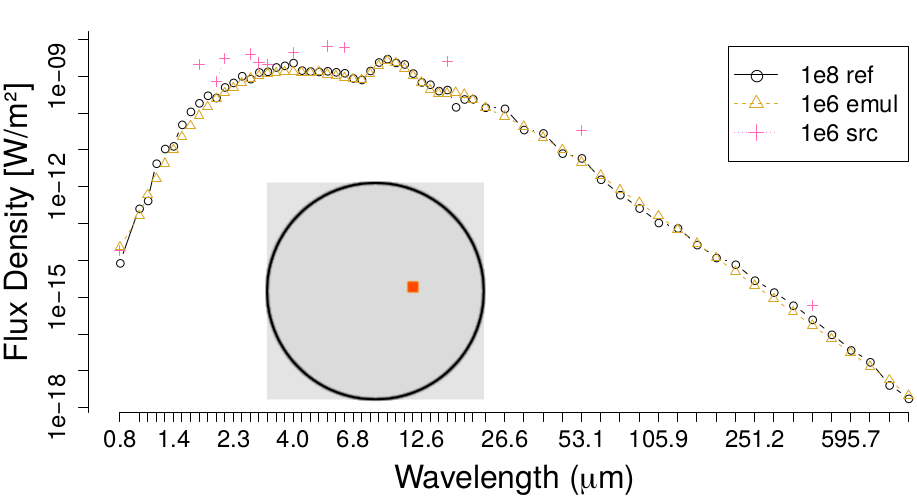}
        \subcaption{$N_p = 10^6$}
        \label{fig:sp62}
    \end{subfigure}
    \hspace{5mm}
    \begin{subfigure}[b]{.47\textwidth}
        \centering
        \includegraphics[width =\textwidth]{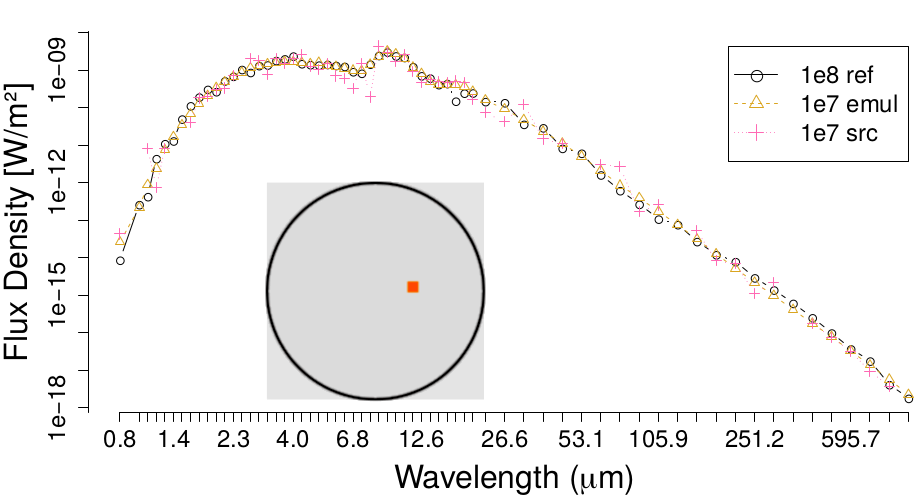}
        \subcaption{$N_p = 10^7$}
        \label{fig:sp72}
    \end{subfigure}
    \caption{Comparison of some emulations spaxels located in row = 150, col = 210, against the HPN references and the LPN input (indicated in each plot), for the case of a dust shell with $\tau_{9.7} = 0.05$ and $\phi = 0^\circ$. The emulations were performed using 25\% of the LPN input spatial information. HPN references spaxel in black ($\circ$), emulations in gold ($\triangle$) and LPN inputs, upon which the emulation was performed, in pink (+).}
    \label{fig:spaxels2}
\end{figure}

\begin{figure}[ht]
\centering
    \begin{subfigure}[b]{.47\textwidth}
        \centering
        \includegraphics[width =\textwidth]{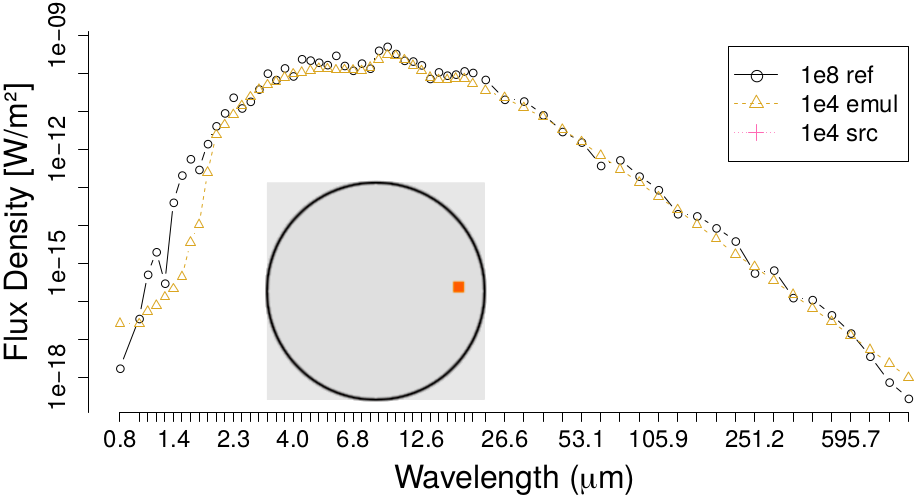}
        \subcaption{$N_p = 10^4$}
        \label{fig:sp43}
        \vspace{3mm}
    \end{subfigure}
    \hspace{5mm}
    \begin{subfigure}[b]{.47\textwidth}
        \centering
        \includegraphics[width =\textwidth]{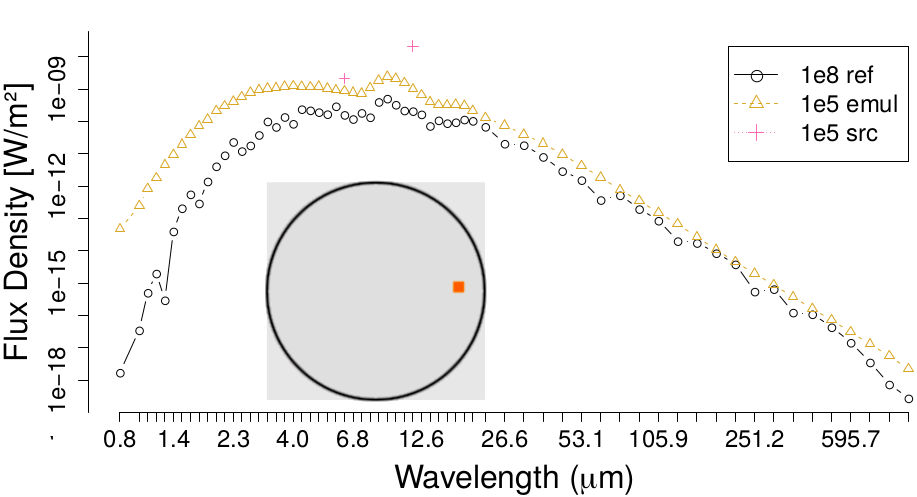}
        \subcaption{$N_p = 10^5$}
        \label{fig:sp53}
        \vspace{3mm}
    \end{subfigure}
    \begin{subfigure}[b]{.47\textwidth}
        \centering
        \includegraphics[width =\textwidth]{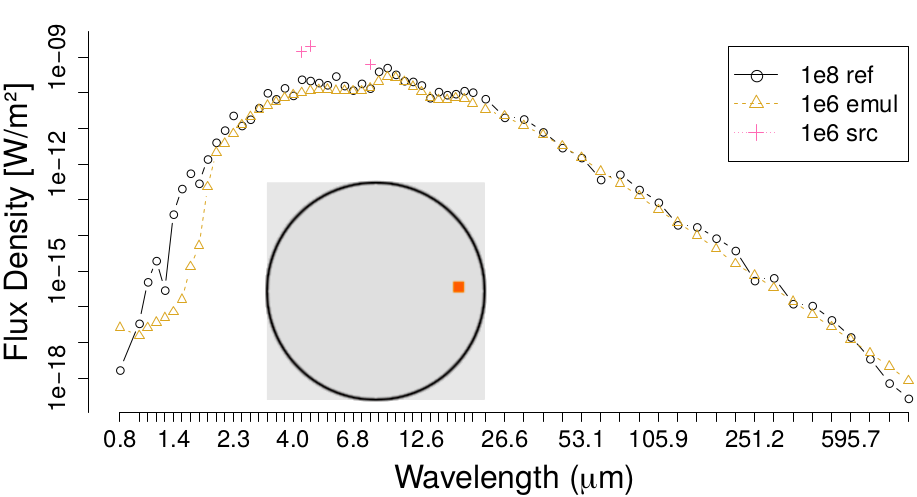}
        \subcaption{$N_p = 10^6$}
        \label{fig:sp63}
    \end{subfigure}
    \hspace{5mm}
    \begin{subfigure}[b]{.47\textwidth}
        \centering
        \includegraphics[width =\textwidth]{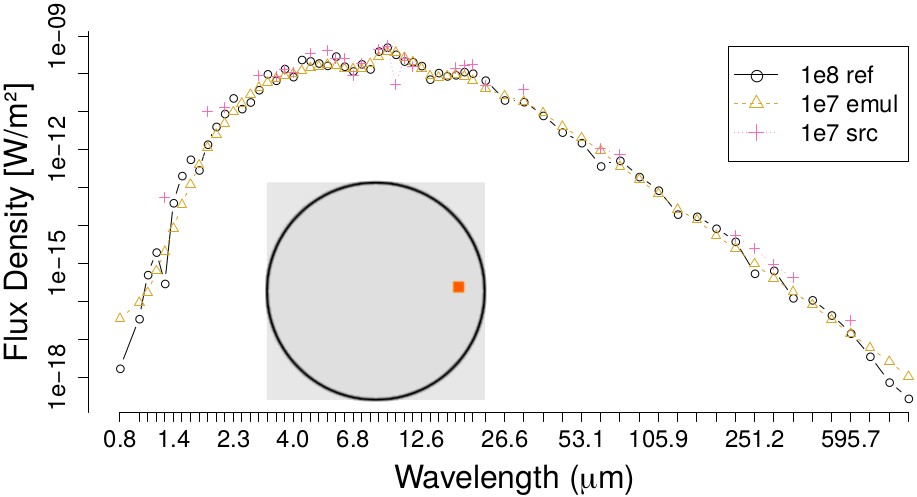}
        \subcaption{$N_p = 10^7$}
        \label{fig:sp73}
    \end{subfigure}
    \caption{Comparison of some emulations spaxels located in row = 150, col = 270, against the HPN references and the LPN input (indicated in each plot), for the case of a dust shell with $\tau_{9.7} = 0.05$ and $\phi = 0^\circ$. The emulations were performed using 25\% of the LPN input spatial information. HPN references spaxel in black ($\circ$), emulations in gold ($\triangle$) and LPN inputs, upon which the emulation was performed, in pink (+).}
    \label{fig:spaxels3}
\end{figure}

Figs. \ref{fig:spaxels1}, \ref{fig:spaxels2} and \ref{fig:spaxels3} compare 3 different spaxels between their versions in the LPN input, HPN reference and emulation based on the LPN input. These spaxels are radially aligned: one at the center of the object, another on the outskirts and another in-between those two. The emulations where achieved using 25\% of the corresponding source cube spaxels.\par
Some things easily noticeable are the underestimation of the shorter wavelengths for the versions of these spaxels taken from emulations based on $N_p = 10^4$ and $N_p = 10^5$ realizations. Nevertheless in those cases the LPN input spaxel is not complete, i.e. there are zero values at some wavelengths, if not all, meaning the pipeline is reaching what it was designed to do: take a sparse compressed model, infer the missing information and then decompress it.

\begin{figure}[ht]
\centering
    \begin{subfigure}[b]{.47\textwidth}
        \centering
        \includegraphics[width =\textwidth]{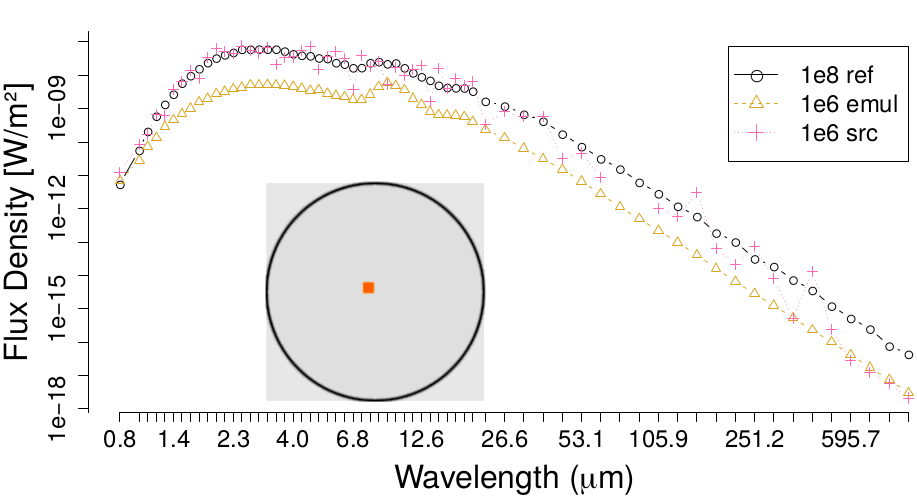}
        \subcaption{$\phi = 0^\circ$}
        \label{fig:sp1fa10}
    \end{subfigure}
    \hspace{5mm}
    \begin{subfigure}[b]{.47\textwidth}
        \centering
        \includegraphics[width =\textwidth]{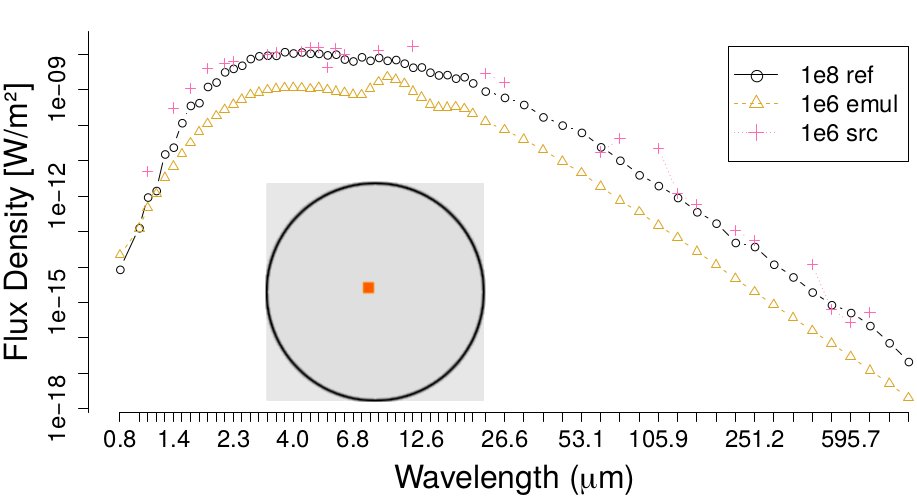}
        \subcaption{$\phi = 90^\circ$}
        \label{fig:sp1ed10}
    \end{subfigure}
    \caption{Comparison of emulations spaxels, located in row = 150, col = 150, against the HPN references and the $N_p = 10^6$ realization, for the case of a dust shell with $\tau_{9.7} = 1.0$, with $\phi = 0^\circ$ (Fig. \ref{fig:sp1fa10}) and $\phi = 90^\circ$ (Fig. \ref{fig:sp1ed10}). HPN references spaxel in black ($\circ$), emulations in gold ($\triangle$) and LPN inputs, upon which the emulation was performed, in pink (+).}
    \label{fig:sp110}
\end{figure}

\begin{figure}[ht]
\centering
    \begin{subfigure}[b]{.47\textwidth}
        \centering
        \includegraphics[width =\textwidth]{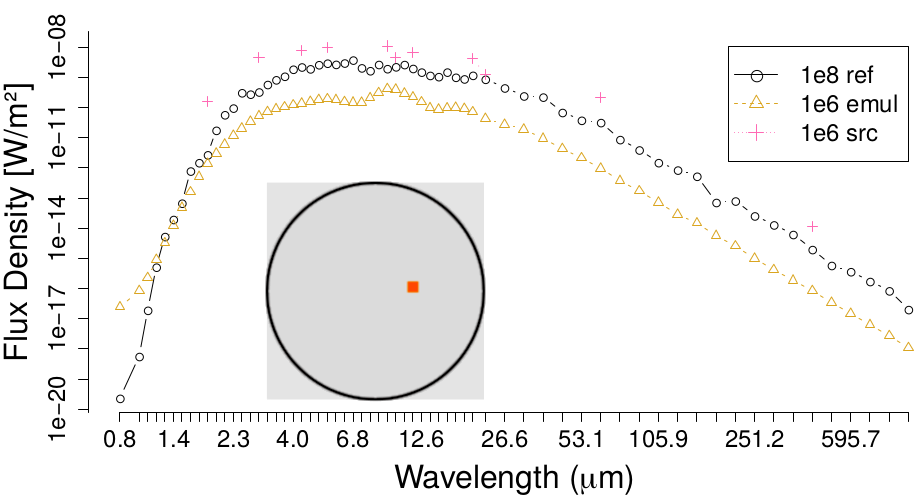}
        \subcaption{$\phi = 0^\circ$}
        \label{fig:sp2ed10}
    \end{subfigure}
    \hspace{5mm}
    \begin{subfigure}[b]{.47\textwidth}
        \centering
        \includegraphics[width =\textwidth]{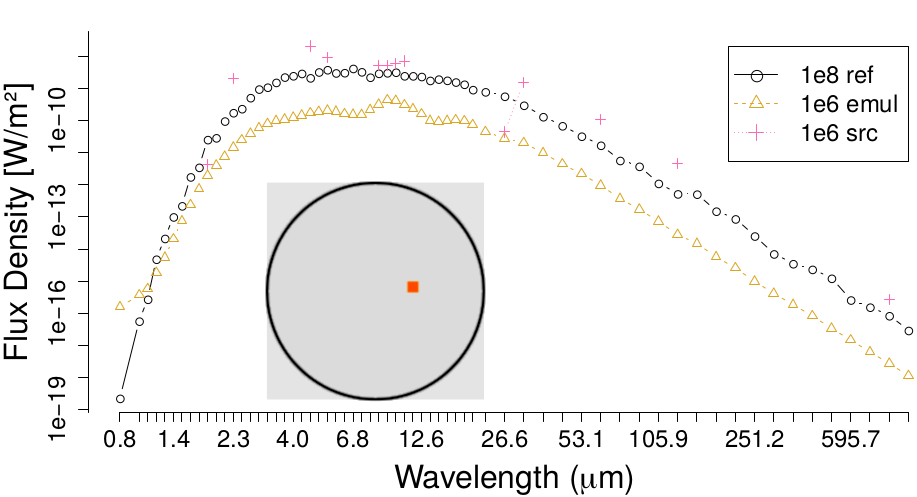}
        \subcaption{$\phi = 90^\circ$}
        \label{fig:sp2fa10}
    \end{subfigure}
    \caption{Comparison of emulations spaxels, located in row = 150, col = 210, against the HPN references and the $N_p = 10^6$ realization, for the cases of dust shells with $\tau_{9.7} = 1.0$, with $\phi = 0^\circ$ (Fig. \ref{fig:sp1fa10}) and $\phi = 90^\circ$ (Fig. \ref{fig:sp1ed10}). HPN references spaxel in black ($\circ$), emulations in gold ($\triangle$) and LPN inputs, upon which the emulation was performed, in pink (+).}
    \label{fig:sp210}
\end{figure}

\begin{figure}[ht]
\centering
    \begin{subfigure}[b]{.47\textwidth}
        \centering
        \includegraphics[width =\textwidth]{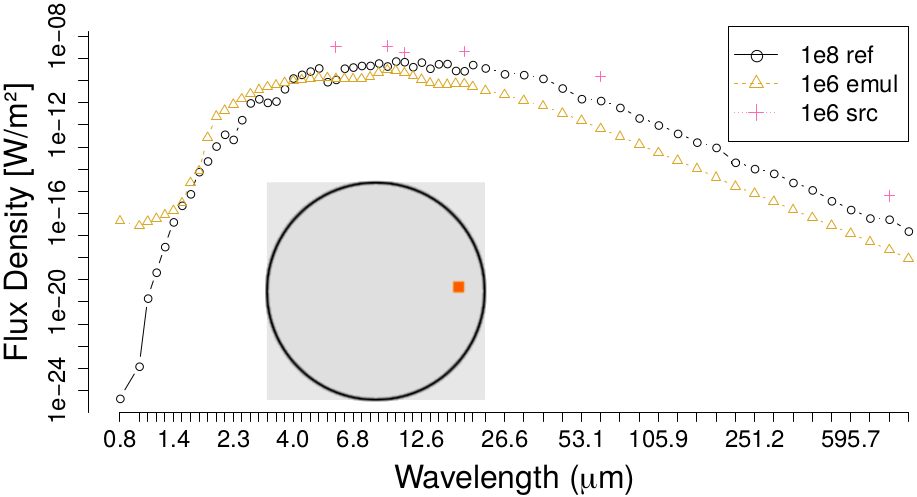}
        \subcaption{$\phi = 0^\circ$}
        \label{fig:sp3ed10}
    \end{subfigure}
    \hspace{5mm}
    \begin{subfigure}[b]{.47\textwidth}
        \centering
        \includegraphics[width =\textwidth]{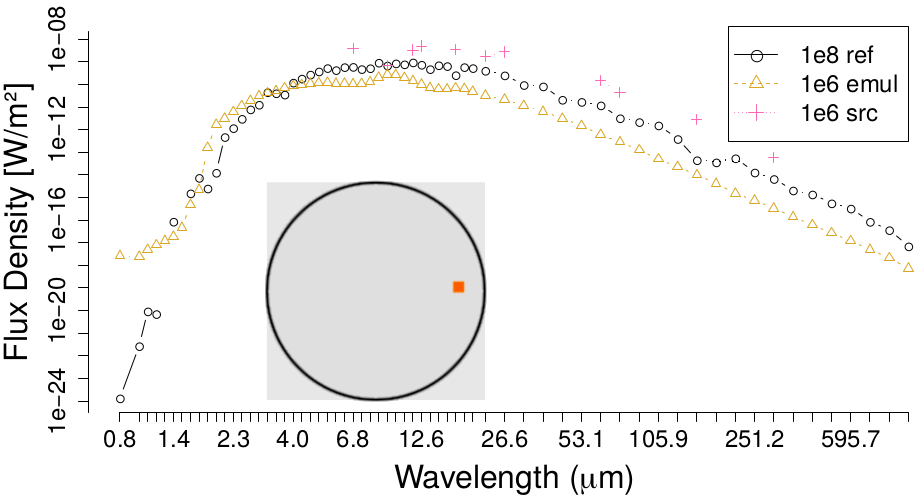}
        \subcaption{$\phi = 90^\circ$}
        \label{fig:sp3fa10}
    \end{subfigure}
    \caption{Comparison of emulations spaxels, located in row = 150, col = 270, against the HPN references and the $N_p = 10^6$ realization, for the cases of dust shells with $\tau_{9.7} = 1.0$, with $\phi = 0^\circ$ (Fig. \ref{fig:sp1fa10}) and $\phi = 90^\circ$ (Fig. \ref{fig:sp1ed10}). HPN references spaxel in black ($\circ$), emulations in gold ($\triangle$) and LPN inputs, upon which the emulation was performed, in pink (+).}
    \label{fig:sp310}
\end{figure}

Figs. \ref{fig:sp110}, \ref{fig:sp210} and \ref{fig:sp310} compare 3 different spaxels from the same spatial locations as before. The emulations were achieved using 11\% of the corresponding LPN input cube spaxels, which are the $\tau_{9.7} = 1.0$ and $\phi \in \{0^\circ, 90^\circ\}$ cases. Since the simulated bright central source has anisotropic emission at greater optical depths, $\tau_{9.7}$, we can expect that for different viewing angles, $\phi$, the shapes of spaxels near the central region of the spatial distribution\footnote{This is relevant because light coming from near the center of the spatial distribution is less likely to have been scattered towards the observer, unlike light coming from the edges which is mostly light that also came from the center, where the bright source is located, and was scattered towards the observer.} will present differences as can be seen in the reference curves of Fig. \ref{fig:sp110}.  \par
For both the face-on and edge-on cases, at all displayed positions, we can make the same assessment, that the shape of the emulated spaxels does not follow the shape of the HPN references. As the simulated models dust density increases so does the likelihood of scattering and absorption of light by the medium, consequently the peak emission features are flattened into the continuum. Figs. \ref{fig:sp110} and \ref{fig:sp210} show that the emulation of the spaxels in those regions is one to two orders of magnitude too low while the relative flux between the emission peak, at 9.7 $\mu$m, and the continuum is over-estimated, the allocation of flux along the spaxels does not match that of the reference. For the outer spaxels, \ref{fig:sp310}, the flux mismatch is greater for longer wavelengths but still the emulation spaxels present a peak emission feature while the references no longer show it. This is a clear sign that the DVAE model is biased.

\end{appendices}


\bibliography{sn-bibliography}


\end{document}